\documentclass{article}
\pdfoutput=1
\usepackage{graphicx}
\usepackage{subcaption}
\usepackage[includefoot,margin=1in]{geometry} 

\usepackage{tikz}
 
\usepackage{array} 
\usepackage{footnote}
 
\usepackage{amssymb}
\usepackage{amsmath}
\usepackage{amsfonts}
\usepackage{bm}

\usepackage{url}
\usepackage{hyperref}
\hypersetup{pdfstartview={XYZ null null 1.00}}

\newcommand{\beq}{\begin{equation}}
\newcommand{\eeq}{\end{equation}}

\newcommand{\beqr}{\begin{eqnarray}}
\newcommand{\eeqr}{\end{eqnarray}}
\def\bal#1\eal{\begin{align}#1\end{align}}
\def\bat#1#2\eat{\begin{alignat}{#1}#2\end{alignat}}

\begin{document}

\title{The 16-vertex model and its even and odd 8-vertex subcases on the square lattice}
\author{Michael Assis\footnote{School of Mathematics and Statistics, University of Melbourne, Carlton, VIC, Australia}}
\date{}
\maketitle

\begin{abstract}
We survey and enlarge the known mappings of the 16-vertex model, with emphasis on mappings between the even and odd 8-vertex subcases of the general model, also giving new mappings between these models, valid on finite toroidal lattices. In particular, we find new mappings between the models by using their algebraic invariants with respect to the $SL(2)\times SL(2)$ symmetry of the 16-vertex model; we also find a larger set of weak-graph transformations. We show many examples of models with negative weights which map to models with only positive weights. Using the algebraic invariant relations of the even and odd 8-vertex models, we find the complete set of points in the complex field plane of the square lattice Ising model in a field which map to the even or odd 8-vertex models; these points also correspond to the set of free-fermionic points of the model. We do not find any new integrable points, but we find a new mapping between the odd 8-vertex model and the square lattice Ising model at magnetic field $H= i\pi/(2\beta)$, valid on finite toroidal lattices. We also show directly through various examples that mappings via algebraic invariants do not fully exhaust the possible mappings a model may have with another model. We construct a new solution to the odd 8-vertex free-fermion model which is valid on the finite lattice, since the previous known solution resulted from a mapping valid only in the thermodynamic limit. Finally, we detail for the first time the phase transitions of the column staggered free-fermion 8-vertex model, and show that it can be mapped to the bi-partite staggered free-fermion model. 
\end{abstract}

\section{Introduction}
The 16-vertex model on the square lattice is a vertex model where each lattice bond can have one of two states, often represented in terms of arrows, leading to 16 Boltzmann weights, broken up into two sets of 8 variables $w_i=\exp(-\beta\epsilon_i)$ and $v_i=\exp(-\beta\xi_i)$, $i=1\ldots8$, assigned to each vertex of the lattice depending on the states of the four bonds incident to the vertex. Here $\epsilon_i$, $\xi_i$ are interaction energies, $\beta=(k_BT)^{-1}$, $k_B$ is the Boltzmann constant, and $T$ is the temperature. The partition function $Z$ is then defined by the following sum
\beq
Z = \sum_{\scriptscriptstyle\mathrm{configs}}\,\prod_{i,j=1}^{8} w_i^{n_i}v_j^{m_j} \label{partfn16def}
\eeq
where the sum is over all possible assignments of bond states on the lattice,  where $n_i$ is the total number of weights $w_i$, and where $m_i$ is the total number of weights $v_i$ in the lattice. In the thermodynamic limit, where the number of lattice sites $\mathcal{N}$ approaches infinity as the lattice grows proportionally in both directions, the free-energy $f$ is defined by the limit
\beq
-\beta\,f = \lim_{\mathcal{N}\to\infty}\frac{1}{\mathcal{N}}\ln(Z)
\eeq
We note that recently there has been direct experimental interest in the 16-vertex model in the context of artificial square ice nanomagnet systems~\cite{wang2006nflmclslcs,wang2007nflmclslcs,morgan2010slm,levis2012c,morgan2013aspelm,morgan2013bslm, levis2013c,levis2013cft}.

The 16-vertex model appears to have first been defined by Wu in~\cite{wu1969}. The full model, which has not been solved, encompasses many special subcases, several of which have been exactly solved. Among the solvable cases we list the square lattice close-packed dimer model~\cite{kasteleyn1961,temperley1961f,fisher1961,kasteleyn1963,lieb19675,fan1969,baxter1972,wu2004k}, the Ising model on the square, triangular and honeycomb lattices, all of which are subcases of the free-fermion model~\cite{fan1970w}, which is equivalent to the Ising model on the checkerboard~\cite{baxter19862,davies1987} or union-jack ~\cite{wu1987l} lattices, Baxter's symmetric even 8-vertex model~\cite{baxter1971,baxter1972,wu1986}, the symmetric even 8-vertex model on the Kagom{\'e} lattice~\cite{baxter19782}, the asymmetric even 6-vertex model~\cite{lieb1967,lieb19672,lieb19673,lieb19674,sutherland1967}, a 5-vertex model~\cite{motegi2013s}, the three-spin Ising model on the triangulated square~\cite{baxter1973w,baxter1974w,baxter1976e,wood1977p}, union-jack~\cite{hintermann1972m,wu1975,wood1977p,urumov1986}, dice~\cite{liu1974s,wood1977p}, honeycomb~\cite{liu1974s}, and bathroom-tile~\cite{wood1977p} lattices, two and three-spin Ising and lattice gas models on the Kagom{\'e} lattice~\cite{wu1989w3}, a 3-state IRF model~\cite{pearce1985}, an anisotropic generalized Kagom{\'e} lattice Ising model~\cite{debauche1991dag}, a staggered spin $1/2$ and spin $s$ model~\cite{strycharski1988c}, a two and three-spin Ising model in a field on the square lattice~\cite{merlini1972g}, a three and five-spin Ising model on the square lattice~\cite{lackova2003h}, vertical three-spin and horizontal two-spin Ising model on the square lattice~\cite{horiguchi1985}, a two and three-spin Ising model on the union-jack lattice~\cite{gitterman1980h}, a bond-decorated union-jack Ising model~\cite{jungling19763}, a two and four-spin Ising model on the square lattice~\cite{wu1971,kadanoff1971w}, a two and four-spin checkerboard Ising model~\cite{giacomini1985}, a nearest and next-next nearest Ising model on the square lattice~\cite{jungling1974o,jungling1975,horiguchi1985m}, the hard hexagon model~\cite{baxter1980,wood1980g}, the ferromagnetic self-dual $q$-state Potts model~\cite{baxter19823}, the self-dual, symmetric, and infinite-coupling-limit Ashkin-Teller models~\cite{wegner1972,fan1972,wu1977,kohmoto1981nk,ikhlef2012r,huang2013djs}, the generalized hard hexagon model~\cite{baxter19823}, and the three-coloring problem on the square lattice~\cite{baxter19702,truong1986s,pegg1982,pearce1989s}. Special cases which have resisted exact solutions are the monomer-dimer model~\cite{wood1980g}, the square lattice Ising model in a field~\cite{gaaff1974}, the ``symmetric" even 8-vertex model on the honeycomb lattice~\cite{wu19742}, the general even 8-vertex model on the square lattice and certain special cases~\cite{merlini19742,barber1985r}, and hard squares~\cite{wood1980g,takasaki2001nh}. Some of these unsolved models admit known integrable points, such as the Ising model with magnetic field $H=i\pi/(2\beta)$~\cite{gaaff1974} and hard squares at fugacity $z=-1$~\cite{fendley2005se,jonsson2006,jonsson2009,adamaszek2012}.

The vast majority of investigations have been done on the subcase of the 16-vertex model corresponding to an even rule at each vertex: only an even number of each type of bond state are allowed at each vertex, or alternatively, the number of incoming and outgoing arrows at each vertex must be even. This even condition requires that the odd weights $v_i=0$. We call this the ``even" 8-vertex model, to distinguish it from the ``odd" 8-vertex model where the even weights $w_i=0$. Baxter's symmetric 8-vertex model and the Ising model on the square lattice are examples. The odd 8-vertex model has been mostly unexplored, except for the close-packed dimer model, until 2004 when Wu and Kunz considered the odd 8-vertex free-fermion model~\cite{wu2004k} by mapping a staggered even 8-vertex model to an odd staggered 8-vertex model and then specializing to homogeneous odd variables. We are only aware of one solved case with mixed even and odd weights, the hard hexagon model~\cite{baxter1980,wood1980g}, although it was not solved as a vertex model. Other seemingly mixed models have non-mixed representations, such as the 11- and 16-vertex models of Deguchi~\cite{deguchi1991} and the symmetric 16-vertex model~\cite{wu1972,felderhof19734}.

\subsection{Notation}
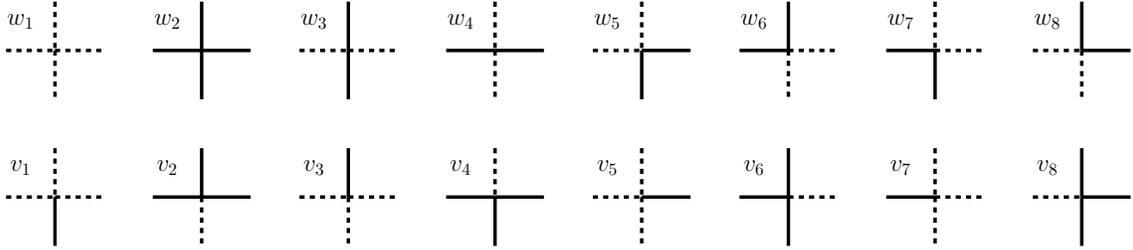
\begin{figure}[htpb]
\begin{center}
\scalebox{0.65}{
\begin{tikzpicture}
\draw[dashed,line width = 2pt] (0,1) -- (0,0); 
\draw[dashed,line width = 2pt] (0,0) -- (0,-1); 
\draw[dashed,line width = 2pt] (-1,0) -- (0,0); 
\draw[dashed,line width = 2pt] (0,0) -- (1,0); 
\node at (-.7,.6) {{\Large $w_1$}};
\begin{scope}[shift={(3,0)}]
\draw[line width = 2pt] (0,1) -- (0,0); 
\draw[line width = 2pt] (0,0) -- (0,-1); 
\draw[line width = 2pt] (-1,0) -- (0,0); 
\draw[line width = 2pt] (0,0) -- (1,0);
\node at (-.7,.6) {{\Large $w_2$}};
\end{scope}
\begin{scope}[shift={(6,0)}]
\draw[line width = 2pt] (0,1) -- (0,0); 
\draw[line width = 2pt] (0,0) -- (0,-1); 
\draw[dashed,line width = 2pt] (-1,0) -- (0,0); 
\draw[dashed,line width = 2pt] (0,0) -- (1,0);
\node at (-.7,.6) {{\Large $w_3$}};
\end{scope}
\begin{scope}[shift={(9,0)}]
\draw[dashed,line width = 2pt] (0,1) -- (0,0); 
\draw[dashed,line width = 2pt] (0,0) -- (0,-1); 
\draw[line width = 2pt] (-1,0) -- (0,0); 
\draw[line width = 2pt] (0,0) -- (1,0);
\node at (-.7,.6) {{\Large $w_4$}};
\end{scope}
\begin{scope}[shift={(12,0)}]
\draw[dashed,line width = 2pt] (0,1) -- (0,0); 
\draw[line width = 2pt] (0,0) -- (0,-1); 
\draw[dashed,line width = 2pt] (-1,0) -- (0,0); 
\draw[line width = 2pt] (0,0) -- (1,0);
\node at (-.7,.6) {{\Large $w_5$}};
\end{scope}
\begin{scope}[shift={(15,0)}]
\draw[line width = 2pt] (0,1) -- (0,0); 
\draw[dashed,line width = 2pt] (0,0) -- (0,-1); 
\draw[line width = 2pt] (-1,0) -- (0,0); 
\draw[dashed,line width = 2pt] (0,0) -- (1,0);
\node at (-.7,.6) {{\Large $w_6$}};
\end{scope}
\begin{scope}[shift={(18,0)}]
\draw[dashed,line width = 2pt] (0,1) -- (0,0); 
\draw[line width = 2pt] (0,0) -- (0,-1); 
\draw[line width = 2pt] (-1,0) -- (0,0); 
\draw[dashed,line width = 2pt] (0,0) -- (1,0);
\node at (-.7,.6) {{\Large $w_7$}};
\end{scope}
\begin{scope}[shift={(21,0)}]
\draw[line width = 2pt] (0,1) -- (0,0); 
\draw[dashed,line width = 2pt] (0,0) -- (0,-1); 
\draw[dashed,line width = 2pt] (-1,0) -- (0,0); 
\draw[line width = 2pt] (0,0) -- (1,0);
\node at (-.7,.6) {{\Large $w_8$}};
\end{scope}
\begin{scope}[shift={(0,-3)}]
\draw[dashed,line width = 2pt] (0,1) -- (0,0); 
\draw[line width = 2pt] (0,0) -- (0,-1); 
\draw[dashed,line width = 2pt] (-1,0) -- (0,0); 
\draw[dashed,line width = 2pt] (0,0) -- (1,0);
\node at (-.7,.6) {{\Large $v_1$}};
\begin{scope}[shift={(3,0)}]
\draw[line width = 2pt] (0,1) -- (0,0); 
\draw[dashed,line width = 2pt] (0,0) -- (0,-1); 
\draw[line width = 2pt] (-1,0) -- (0,0); 
\draw[line width = 2pt] (0,0) -- (1,0);
\node at (-.7,.6) {{\Large $v_2$}};
\end{scope}
\begin{scope}[shift={(6,0)}]
\draw[line width = 2pt] (0,1) -- (0,0); 
\draw[dashed,line width = 2pt] (0,0) -- (0,-1); 
\draw[dashed,line width = 2pt] (-1,0) -- (0,0); 
\draw[dashed,line width = 2pt] (0,0) -- (1,0);
\node at (-.7,.6) {{\Large $v_3$}};
\end{scope}
\begin{scope}[shift={(9,0)}]
\draw[dashed,line width = 2pt] (0,1) -- (0,0); 
\draw[line width = 2pt] (0,0) -- (0,-1); 
\draw[line width = 2pt] (-1,0) -- (0,0); 
\draw[line width = 2pt] (0,0) -- (1,0); 
\node at (-.7,.6) {{\Large $v_4$}};
\end{scope}
\begin{scope}[shift={(12,0)}]2
\draw[dashed,line width = 2pt] (0,1) -- (0,0); 
\draw[dashed,line width = 2pt] (0,0) -- (0,-1); 
\draw[dashed,line width = 2pt] (-1,0) -- (0,0); 
\draw[line width = 2pt] (0,0) -- (1,0);
\node at (-.7,.6) {{\Large $v_5$}};
\end{scope}
\begin{scope}[shift={(15,0)}]
\draw[line width = 2pt] (0,1) -- (0,0); 
\draw[line width = 2pt] (0,0) -- (0,-1); 
\draw[line width = 2pt] (-1,0) -- (0,0); 
\draw[dashed,line width = 2pt] (0,0) -- (1,0);
\node at (-.7,.6) {{\Large $v_6$}};
\end{scope}
\begin{scope}[shift={(18,0)}]
\draw[dashed,line width = 2pt] (0,1) -- (0,0); 
\draw[dashed,line width = 2pt] (0,0) -- (0,-1); 
\draw[line width = 2pt] (-1,0) -- (0,0); 
\draw[dashed,line width = 2pt] (0,0) -- (1,0);
\node at (-.7,.6) {{\Large $v_7$}};
\end{scope}
\begin{scope}[shift={(21,0)}]
\draw[line width = 2pt] (0,1) -- (0,0); 
\draw[line width = 2pt] (0,0) -- (0,-1); 
\draw[dashed,line width = 2pt] (-1,0) -- (0,0); 
\draw[line width = 2pt] (0,0) -- (1,0);
\node at (-.7,.6) {{\Large $v_8$}};
\end{scope}
\end{scope}
\end{tikzpicture}
}
\end{center}
\caption{The even (top row) and odd (bottom row) 8-vertex model weights, with bond states shown in terms of line type, dashed or solid.\label{fig:evenoddconfigsbonds}}
\end{figure}

\begin{figure}[htpb]
\begin{center}
\scalebox{0.65}{
\includegraphics{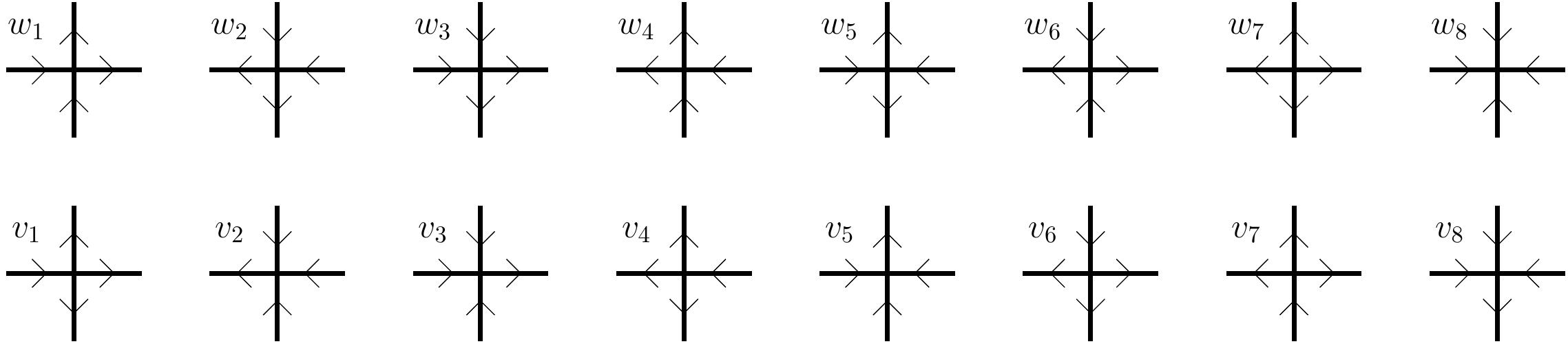}
}
\end{center}
\caption{The even (top row) and odd (bottom row) 8-vertex model weights, with bond states shown in terms of arrows.\label{fig:evenoddconfigsarrows}}
\end{figure}

There appears to be 6 and 5 different notation conventions in the literature for both the even and the odd 8-vertex models, respectively, which we list in tables in appendix~\ref{app:notation}. The choice we use in this paper for the even model is the most common choice in the literature, and the choice of odd weights $v_i$ is such that to go from an even weight $w_i$ to an odd weight $v_i$, the bottom bond is reversed. We show in figure~\ref{fig:evenoddconfigsbonds} our notation with bond states given in terms of solid or dashed line type, and in figure~\ref{fig:evenoddconfigsarrows} with bond states given in terms of arrows.

We also define a ``symmetric" model to be one where the weights $w_i$, $v_i$ are equal under reversal of all bond states. 
 We note that another use of ``symmetric" exists in the literature for vertex models~\cite{wu19742,wu1988w,wu1989w,perk1990ww,samaj1992k2,samaj1992k3,kolesik1993s,rollet1995w}, where it is taken to mean that all vertex weights which have the same number of solid (or dashed) bonds are equal. When describing such models we will use the terminology ``Wu's symmetric model", since this class of model was first defined by Wu in~\cite{wu19742}. We reserve the use of the term ``symmetric" to be in accordance with the use for Baxter's symmetric even 8-vertex model.

\subsection{Outline of paper}
In the next section we consider the topological constraints of the 16-vertex model and bring out for the first time the partition function's dependence on the bond occupation variables. In the following section we give for the first time in published literature the full set of lattice and bond-reversal symmetries of the 16-vertex model partition function. Since the only known solution at this point for the odd 8-vertex model is the free-fermion solution~\cite{wu2004k}, of which the close-packed dimers on the square lattice is a subcase, we focus in section~\ref{sec:ff} on free-fermion models in general, giving their historical overview as well as making the observation for the first time that all of the relevant critical phenomena of the free-fermion, union-jack and checkerboard lattice Ising models are all equivalent to the critical phenomena of the triangular Ising model. We also discuss the known mapping between the even and the odd free-fermion models, and how the mapping appears to only be valid in the thermodynamic limit, at least up until this point in time. We note that we haven't found a mapping between Baxter's symmetric even 8-vertex model and the odd 8-vertex model; under the weak-graph transformation, at least, Baxter's model is an invariant, as shown in section~\ref{sec:weakgraph}.

In section~\ref{sec:spinmaps} we give two different spin-model mappings for the 16-vertex model, with the Ising in a field as an example, then in section~\ref{sec:weakgraph} we consider in particular the weak-graph expansion transformation for the 16-vertex model, giving an enlarged set of weak-graph transformation matrices than previously considered. We also give new weak-graph transformations between the even and odd 8-vertex models and between the odd 8-vertex model and the 16-vertex model. 

In section~\ref{sec:alginvs} we calculate the polynomial algebraic invariants of the even and odd 8-vertex models with respect to the $SL(2)\times SL(2)$ symmetry of the 16-vertex model. In defining the relationships between the invariants for each model, we define the 16-vertex model generalization of each model which are equivalent to them. We then determine the valid mappings among the even and odd 8-vertex models which respect each of their algebraic invariants. However, in section~\ref{sec:ffmap} we show that the thermodynamic limit mapping between the even and odd free-fermion models does not respect each of their algebraic invariants, and in section~\ref{sec:dimermap} we show another mapping with disrepects the even and odd algebraic invariants, this time valid locally on finite toroidal lattices. 

In section~\ref{sec:isingfield} we consider mappings between the square lattice Ising model in a field to the even and odd 8-vertex models via their algebraic invariants, finding the full set of mappable points. We use both the spin-model mapping of section~\ref{sec:spinmaps} as well as a second one. These points are all free-fermionic points, and correspond to the known exact solutions of the model, at magnetic field $H=0$ and $H=i\pi/(2\beta)$. In particular, we find a new mapping between the Ising model at $H=i\pi/(2\beta)$ and the odd 8-vertex model, valid locally on finite toroidal lattices. 

In section~\ref{sec:stag} we consider both the column and the bi-partite staggered even and odd 8-vertex models, first giving spin-model mappings as well as showing for the first time that the column staggered model is a function of 11 variables in general. We then consider the mappings between the even and odd 8-vertex models, and give a new mapping between the column and the bi-partite staggered free-fermion models, closing the section with an analysis of the phase transition points of the column stgagered free-fermion 8-vertex model. 

Finally, in section~\ref{sec:discussion} we discuss boundary conditions as they relate to the odd 8-vertex model in general and its free-fermion subcase, we discuss the nature of free-fermion models in general, and we comment on 16-vertex model disorder lines relevant to the odd 8-vertex model. We finish in section~\ref{sec:conclusions} with a commentary on the benefits of using mappings in the analysis of the 16-vertex model, as well as a summary of the paper and results. 

In appendix~\ref{app:notation} we give tables of known even and odd 8-vertex weight notations, in appendix~\ref{app:parttrans} we list for the first time the full set of lattice and bond-reversal symmetries of the 16-vertex model partition function, and in appendix~\ref{app:genweakgraph} we show a $16\times 16$ matrix representation of the $SL(2)\times SL(2)$ symmetry transformation acting on the 16-vertex model weights, from which it can be shown that the weak-graph transformation is not a special case of this transformation. In appendix~\ref{app:dimerconst} we give full details for the construction of finite lattice even and odd 8-vertex partition functions on both the homogeneous and the column and bi-partite staggered lattices, listing the free-energy results in appendix~\ref{app:results}.

\section{16-vertex model topological constraints}
It is well known that toroidal boundary conditions induce topological constraints on the vertex weights, although it appears that only the even 8-vertex model constraints have appeared in print. When interpreting the bond states in terms of arrows, the number of horizontal and vertical sources and sinks must be equal in the lattice. For the even 8-vertex model, their counting produces the set of equations
\beq
n_5+n_8=n_6+n_7,\qquad n_5+n_7=n_6+n_8
\eeq
which can be re-expressed to give the known condition~\cite{baxter1982,baxter2007}
\beq
n_5=n_6,\qquad n_7=n_8
\eeq
so that for the even 8-vertex model, the weights $w_5$, $w_6$ always appear in the combination $w_5w_6$, and likewise for $w_7$, $w_8$, so that the partition function of the even 8-vertex model has the following form
\beq
Z_{8E} = \sum_{\scriptscriptstyle\mathrm{configs}}\, w_1^{n_1}w_2^{n_2}w_3^{n_3}w_4^{n_4}(w_5w_6)^{n_5}(w_7w_8)^{n_7}
\eeq
Therefore, it has been customary to set $w_5=w_6$, $w_7=w_8$ in the even 8-vertex model~\cite{baxter1982,baxter2007}. 

For the odd 8-vertex model, a similar counting of sources and sinks yields the set of equations
\beq
m_1+m_4=m_2+m_3,\qquad m_5+m_8=m_6+m_7 \label{oddtop}
\eeq
which does not have such a simple interpretation for the weights as for the even 8-vertex model. In section~\ref{sec:alginvs} we propose a set of 6 algebraically independent quantities for the odd 8-vertex model which eliminate the topological redundancy in the 8 variables $v_i$.

For the full 16-vertex model, we have the following set of equations
\beq
n_5+n_8+m_5+m_8=n_6+n_7+m_6+m_7,\qquad n_5+n_7+m_1+m_4=n_6+n_8+m_2+m_3 \label{16vtop}
\eeq

We will briefly consider these equations in the context of the hard hexagon model, which has the following mapping to the 16-vertex model~\cite{wood1980g}
\beq
w_1=1,\qquad w_7v_3v_5=z, \qquad w_i=v_i=0~~\mathrm{otherwise} \label{hhmap}
\eeq
where $z$ is the fugacity of a hard hexagon particle. 

The equations (\ref{16vtop}) give the constraint $n_7=m_3=m_5$, showing that the partition function is homogeneous in the combination of variables $(w_7v_3v_5)$, as it should be. The vertical and horizontal sources of $v_3$ and $v_5$, respectively, have a corresponding sink in $w_7$. If one were to swap $v_2$ for $v_3$, there continues to be a matching of vertical and horizontal sinks and sources, and the equations (\ref{16vtop}) continue to give a similar constraint $n_7=m_2=m_5$ even though physically the two models are quite different. In the hard hexagon mapping~(\ref{hhmap}) a combination of weights $w_7$, $v_3$, $v_5$ can be placed on the lattice in isolation, but in the second case a staircase pattern emerges that can potentially fill the toroidal lattice, as soon as one of the weights $w_7$, $v_2$, or $v_5$ are placed. This is because, though $v_2$ is not a horizontal source or sink, it still demands a matching bond on the right, which only $w_7$ can fill, which again demands a matching upward bond, which only $v_2$ can fill, and so on. So the total number of weights $n=n_7=m_2=m_5$ cannot be chosen arbitrarily, as was the case for hard hexagons, where $n$ can equal any integer up to the maximum allowed by the lattice size.

Under toroidal boundary conditions, we can count the number of horizontal and vertical solid and dashed bonds in the lattice. This gives the following combined set of both topological and geometric conditions
\beqr
& n_5+n_8+m_5+m_8=n_6+n_7+m_6+m_7 \label{topeqs1}\\
& n_5+n_7+m_1+m_4=n_6+n_8+m_2+m_3 \label{topeqs2}\\
& 2n_1+2n_4+n_5+n_6+n_7+n_8+m_1+m_2+m_3+m_4+2m_5+2m_7=2M_d \label{topeqs3}\\
& 2n_2+2n_3+n_5+n_6+n_7+n_8+m_1+m_2+m_3+m_4+2m_6+2m_8=2M_s \\
& 2n_1+2n_3+n_5+n_6+n_7+n_8+2m_1+2m_3+m_5+m_6+m_7+m_8=2N_d \\
& 2n_2+2n_4+n_5+n_6+n_7+n_8+2m_2+2m_4+m_5+m_6+m_7+m_8=2N_s \\
&\displaystyle \mathcal{N} = M_d+M_s=N_d+N_s = \sum_{i=1}^8n_i+\sum_{i=1}^8m_i \label{topeqs7}
\eeqr
where $\mathcal{N}$ is the total number of sites on the lattice, $M_d$ and $M_s$ are the number of vertical dashed and solid bonds, respectively, and $N_d$ and $N_s$ are the number of horizontal dashed and solid bonds, respectively, and where the factors of two on the right-hand sides of the equations~(\ref{topeqs3})--(\ref{topeqs7}) are due to double counting of bonds.

From equations (\ref{topeqs1})--(\ref{topeqs7}), we see that for any 16-vertex partition function under toroidal boundary conditions, its dependence on dashed or solid bonds can be explicitly added in as follows
\beq
Z_{16} = \sum_{\scriptscriptstyle\mathrm{configs}}\,\prod_{i,j=1}^{8} w_i^{n_i}v_j^{m_j}d^{N_d}_h d^{M_d}_v s^{N_s}_h s^{M_s}_v \label{partbondvars}
\eeq
This appears to be a new observation. We can use the sink and source equations~(\ref{topeqs1})--(\ref{topeqs2}) to eliminate any (apparent) square root dependence for each bond variable, in two ways each. We can choose, for example,
\beqr
N_d &=& n_1+n_3+n_5+n_8+m_1+m_3+m_5+m_8 \\
M_d &=& n_1+n_4+n_5+n_7+m_1+m_4+m_6+m_8 \\
N_s &=& n_2+n_4+n_5+n_8+m_2+m_4+m_5+m_8 \\
M_s &=& n_2+n_3+n_5+n_7+m_1+m_4+m_6+m_8 
\eeqr
so that the dependence on solid horizontal bonds can be added to the partition function~(\ref{partfn16def}) by the following set of variable transformations
\bat{4}
w_2&\to s_hw_2,&\qquad w_4&\to s_hw_4,&\qquad w_5&\to s_hw_5,&\qquad w_8&\to s_hw_8,\nonumber\\
v_2&\to s_hv_2,&\qquad v_4&\to s_hv_4,&\qquad v_5&\to s_hv_5,&\qquad v_8&\to s_hv_8 \label{bondvartransf1}
\eat
Similar transformations give the dependence of the partition function on the variables $s_v$, $d_h$, $d_v$.

Returning to the previous example, we can now see that for hard hexagons we have $n_7=m_3=m_5$ plus $M_s=N_s=n_7$, $M_d=N_d=n_1+2n_7$, $n_1+3n_7=\mathcal{N}$ so that as expected, every appearance of the combination of weights $w_7$, $v_3$, $v_5$ increases the number of $M_s=N_s$ by one, up to the maximum $(\mathcal{N}-n_1)/3$. By way of contrast, with non-zero weights $w_1$, $w_7$, $v_5$, $v_2$, we see again that $n_7=m_2=m_5$ but now we have $M_s=n_7$, $N_s=2n_7$, $M_d=n_1+2n_7$, $N_d=n_1+n_7$, $n_1+3n_7=\mathcal{N}$, and now we see that we cannot make $M_s=N_s$ unless $n_7=m_2=m_5=0$. 

\section{16-vertex model partition function symmetries}
The partition function of the 16-vertex model satisfies lattice and bond-reversal symmetries. It is invariant under rotations and reflections as well as horizontal and vertical bond reversals. It is invariant, therefore, under the group of transformations $C_2\times C_2\times D_8$ of order 32, where the $C_2$ are cyclic groups of order 2 and $D_8$ is the dihedral group of order 8. We list in appendix~\ref{app:parttrans} all 32 transformations, which can be specialized to either the even or odd 8-vertex models. We note that this list is larger than those found in~\cite{fan1969w,fan1970w,lieb1972w,wu2004k}.

It is apparent from the partition function symmetries that the variables permute among 3 distinct sets of variables. It would appear from these symmetries that there could be no mapping from the even to the odd 8-vertex models, but we will see below that this is not the case. It would be convenient to construct a new set of 16 variables which are invariant under all 32 partition function symmetry transformations; we are not aware of such a result.

\section{Free-fermion and dimer-solvable models} \label{sec:ff}
In a series of papers, Hurst and collaborators introduced an $S$-matrix approach to Ising model problems~\cite{hurst1960g,hurst1963,green1964h,hurst1964,hurst1966}, elaborated and modified by others (see~\cite{plechko1988} and references therein
), where they place fermion creation and annihilation operators on either end of each bond of the lattice. The model is solvable subject to the Lagrangian being a quadratic function of the fermion field operators~\cite{hurst1966,fan1969w}, rendering the problem one of non-interacting, free fermions. In order to map to the Ising model, they require the introduction of a ``terminal city" of extra ``internal" sites and bonds to replace each vertex of the original lattice. They then find conditions on the signs of the bond weights such that the partition function can be expressed as a Pfaffian, that is, the square root of the determinant of an anti-symmetric matrix. These conditions are such that an even element of an exterior algebra be expressible as an exponential~\cite{hurst1966}. These conditions were also considered much later in~\cite{viallet1997}. Separately, it is also known that every planar Ising model can also be solved via a close-packed dimer model on an equivalent lattice~\cite{fisher1966}, where the Pfaffian can be evaluated by choosing appropriate signs for the bond weights~\cite{mccoy1973w,mccoy2014w}. The restrictions Hurst finds on the bond weight signs are of the same nature as the sign restrictions required in solving planar close-packed dimer models, and explicitly, for the square and triangular lattices they are identical. 

Beyond Ising models, dimers and free-fermion methods can also be used to solve vertex models. Hurst first remarked on this point from the perspective of his free-fermionic method, noting the solution of what he called the generalized square lattice~\cite{hurst1963,green1964h}, but without consideration of its physical interpretation; in~\cite{mills1969h}, the triangular lattice is similarly considered. For the square lattice, the condition on the even weights such that it is solvable via this method is 
\beq
w_1w_2+w_3w_4=w_5w_6+w_7w_8 \label{ffcondeven}
\eeq 
Fan and Wu~\cite{fan1970w} later elaborated on the model, calling it the ``free-fermion model"~\cite{fan1969w}, with reference to Hurst, and relating it to other known models, though they also solve it via dimers\footnote{The earliest use of dimers to solve a vertex model was by Wu in the context of the modified KDP model~\cite{wu1968}, a subcase of the free-fermion model.}. Though no proof exists, as far as we are aware, it is reasonable to conjecture that both free-fermion and dimer-solvable models are exactly equivalent.

It is for these historical reasons that the even 8-vertex model with the constraint~(\ref{ffcondeven}) has been called the ``free-fermion model", even though it has been shown to be equivalent to the Ising model on the checkerboard lattice~\cite{davies1987,baxter19862} and the Ising model on the union-jack lattice~\cite{wu1987l}. Also, the partition function per site of the free-fermion model can be factored as a product of four quantities~\cite{bazhanov1985s,bazhanov1985s2,bazhanov1985s3}, which can be related to four anisotropic square lattice Ising models~\cite{baxter19862}, and furthermore it has another factorization as a product of two anisotropic triangular lattice Ising models~\cite{baxter19862}. This factorization property means that the free-fermion model can be expressed in terms of only four effective parameters. It would be interesting to see whether this factorization property, as well as a mapping to an Ising model on a different lattice, continue to hold for free-fermion vertex models defined on other lattices, for example the 32-vertex model on the triangular lattice~\cite{mills1969h,sacco1975w,sacco1977w,samuel1981} or the 128-vertex model on the union-jack lattice~\cite{samuel19812}.

The free-energy of the even free-fermion 8-vertex model is given by 
\beq
-\beta\,f = \frac{1}{8\pi^2}\int_0^{2\pi}\int_0^{2\pi}d\theta_1 d\theta_2 \ln(A+2B\cos(\theta_1)+2C\cos(\theta_2)+2D\cos(\theta_1-\theta_2)+2E\cos(\theta_1+\theta_2)  \label{even8vfreeenergy}
\eeq
where
\beqr
A &=& w_1^2+w_2^2+w_3^2+w_4^2 \\
B &=& w_1w_3-w_2w_4 \\
C &=& w_1w_4-w_2w_3 \\
D &=& w_3w_4-w_7w_8 = w_5w_6-w_1w_2 \\
E &=& w_3w_4-w_5w_6 = w_7w_8-w_1w_2
\eeqr
The even free-fermion model has many interesting subcases, such as the square, triangular, and honeycomb Ising models, close-packed dimers on the square lattice, various cases of the ice model, and the decoupling point of Baxter's symmetric even 8-vertex model~\cite{fan1970w,baxter1972}. In the analysis of the free-fermion model solutions, the transcendental relation between interaction energies $\epsilon_i$
\beq
e^{-(\epsilon_1+\epsilon_2)/k_BT}+e^{-(\epsilon_3+\epsilon_4)/k_BT}=
e^{-(\epsilon_5+\epsilon_6)/k_BT}+e^{-(\epsilon_7+\epsilon_8)/k_BT}
\eeq
is usually simplified~\cite{fan1970w,hsue1975lw,lin1977w} to two equalities $w_1w_2=w_5w_6$ and $w_3w_4=w_7w_8$ or else $w_1w_2=w_7w_8$ and $w_3w_4=w_5w_6$, or equivalently
\beq
w_1w_2w_3w_4=w_5w_6w_7w_8, \qquad\mathrm{or}\qquad \epsilon_1+\epsilon_2+\epsilon_3+\epsilon_4=\epsilon_5+\epsilon_6+\epsilon_7+\epsilon_8 \label{ffindcondeven}
\eeq
so that the free-fermion condition is temperature independent, that is, it is only a function of the interaction energies $\epsilon_i$ without any temperature dependence. The consequences of such a choice for the even 8-vertex model is that the model reduces to the triangular Ising model on the dual lattice~\cite{hurst1963,green1964h,felderhof19733}. This can also be seen from the integrand of the free-energy, where one term becomes identically zero~\cite{hurst1963,fan1970w} so that a mapping between the triangular Ising model and the free-fermion model free-energies can be carried out. 
In~\cite{hurst1963}, Hurst proves that for any integrand of the form of the even free-fermion model, the only singularities in the principal sheet of the model are those which respect the temperature independent free-fermion condition; singularities which occur only for the general free-fermion condition~(\ref{ffcondeven}) but not for~(\ref{ffindcondeven}) only appear in the analytic continuation of the model. Therefore, no critical phenomena is missed in choosing the temperature independent condition~(\ref{ffindcondeven}). Furthermore, in~\cite{davies1987}, the quantity $(w_1w_2w_3w_4-w_5w_6w_7w_8)$ is shown to be a normalization quantity in the problem, so that setting it to zero will not affect the model in a meaningful way. Therefore, the triangular, checkerboard, union-jack Ising models and the even free-fermion model all have the same essential physics, a statement which we believe has not appeared before in the literature. For staggered free-fermion models and free-fermion models on other lattices, the relevance of assuming temperature independent free-fermion conditions remains to be proven, however. 

We summarize the critical phenomena analysis of even free-fermion model using the analysis of the checkerboard lattice Ising model in~\cite{baxter19862}, where all of the phase transition and disorder conditions are given in terms of the quantity $\Omega^2$, expressed in terms of the even weights as
\beqr
\Omega^2 &=& 1+\frac{(w_1-w_2-w_3-w_4)(w_1-w_2+w_3+w_4)(w_1+w_2-w_3+w_4)(w_1+w_2+w_3-w_4)}{16\,w_5w_6w_7w_8} \\
&=& \frac{(w_1+w_2+w_3+w_4)(w_1+w_2-w_3-w_4)(w_1-w_2+w_3-w_4)(w_1-w_2-w_3+w_4)}{16\, w_1w_2w_3w_4} \label{Omega2}
\eeqr
where~(\ref{ffindcondeven}) was used going between the first and second lines. When $\Omega^2=1,-\infty$ the model is critical, at $\Omega^2=+\infty$ the model is completely ordered, and at $\Omega^2=0$ the model is at a disorder point. Also, at $\Omega^2=\pm\infty$ the model reduces to a six-vertex model; at $\Omega^2=+\infty$ it is in a frozen ferroelectric state while at $\Omega^2=-\infty$ it is an antiferroelectric~\cite{baxter19862}. The spontaneous magnetization $M_0$ of the checkerboard Ising model, and correspondingly, the free-fermion model is given by~\cite{baxter19862}
\beq
M_0 = \left\{ \begin{array}{lr}
(1-\Omega^{-2})^{1/8}, & \Omega^2>1 \\
0, & \Omega^2\leq 1
\end{array}
\right.
\eeq

While the even free-fermion 8-vertex model has been understood for a long while, the odd free-fermion 8-vertex model has not received much attention except for in~\cite{wu2004k}. In their work, Wu and Kunz found a mapping from the staggered even 8-vertex model to the staggered odd 8-vertex model, and by using the free-fermion solution of the staggered even model, they specialized to arrive at the homogeneous odd free-fermion 8-vertex model free-energy. The solution can also be found more directly using dimers, shown in appendix~\ref{app:dimerconst} for the first time, so that it is valid on finite lattices, and we give the free-energy here in our notation
\beq
-\beta\,f = \frac{1}{8\pi^2}\int_0^{2\pi}\int_0^{2\pi}d\theta_1 d\theta_2 \ln(A+2B\cos(\theta_1)+2C\cos(\theta_2)+2D\cos(\theta_1-\theta_2)+2E\cos(\theta_1+\theta_2) \label{odd8vfreeenergy}
\eeq
where
\beqr
A &=& (v_1v_2+v_3v_4)(v_5v_6+v_7v_8)+v_1^2v_4^2+v_2^2v_3^2+v_5^2v_7^2 +v_6^2v_8^2 \\
B &=& 2v_5v_6v_7v_8-v_1^2v_4^2-v_2^2v_3^2 \\
C &=& 2v_1v_2v_3v_4-v_5^2v_7^2-v_6^2v_8^2 \\
D &=& (v_1v_2-v_7v_8)(v_5v_6-v_3v_4) \\
E &=& (v_1v_2-v_5v_6)(v_7v_8-v_3v_4)
\eeqr
The phase transition points of the odd 8-vertex free-fermion model occur when one of the following hold~\cite{wu2004k}
\beqr
& (v_1v_2+v_3v_4)=(v_5v_6+v_7v_8) = 0 \\
& (v_1v_3+v_2v_4) = 0 \\
& (v_5v_7+v_6v_8) = 0 \\
& (v_1v_2+v_3v_4)(v_5v_6+v_7v_8)+(v_1v_3-v_2v_4)^2+(v_5v_7-v_6v_8)^2 = 0 
\eeqr

In the dimer construction of the solution, the internal sites enforce a constraint, 
\beq
v_1v_2+v_3v_4=v_5v_6+v_7v_8 \label{ffcondodd}
\eeq
which is equivalent to the even free-fermion condition. This can be seen on staggered lattices, where the free-fermion condition~(\ref{ffcondeven}) of the staggered even 8-vertex model gets mapped to this condition for the staggered odd 8-vertex model~\cite{wu2004k}, also shown below in section~\ref{sec:stag}. Again, temperature independent free-fermion conditions are equivalent to
\beq
v_1v_2v_3v_4=v_5v_6v_7v_8 \label{ffindcondodd}
\eeq
It will be shown in section~\ref{sec:weakgraph} below that under the weak-graph transformation, the temperature independent free-fermion condition of the even 8-vertex model maps to the general free-fermion condition of the odd 8-vertex model, and vice-versa, showing how closely related conditions~(\ref{ffcondeven}) and (\ref{ffcondodd}) are to (\ref{ffindcondeven}) and (\ref{ffindcondodd}), respectively.

As opposed to the even case, satisfying the appropriate bond weight signs in the equivalent dimer model requires a staggered lattice. A similar situation can be seen when attempting to find a spin model that is equivalent to the odd 8-vertex model --- a staggered spin model is necessary~\cite{wu2004k}. We are unaware of any method to solve the homogeneous odd 8-vertex model directly without recourse to an intermediate staggered lattice. Interestingly, we find a mapping in section~\ref{sec:isingfield} between the odd 8-vertex model and the Ising model in a field $H=i\pi/(2\beta)$~\cite{lee1952y,baxter1965,mccoy1967w,perk2006a}, whose partition function is known to be zero unless there are an even number of lattice sites~\cite{sherman1960,marshall1971,merlini1974,lin1988w}. 

The odd free-fermion constraint (\ref{ffcondodd}) at first sight seems to define a new free-fermion model as a subcase of the 16-vertex model, distinct from the even free-fermion model. From the partition function transformations of the 16-vertex model enumerated in appendix~\ref{app:parttrans}, it would appear that the even and the odd sectors of the 16-vertex model are independent of each other. Nevertheless, we consider several mappings below which map the 16-vertex model onto itself with mixing of the even and odd weights. Using the weak-graph transformation and mapping the algebraic invariant relations of each model to each other, we find exact mappings valid on the finite lattice between the two models, but subject to several constraints among each set of weights, which we outline below. One may be lead to believe that the even and the odd free-fermion models are different except where they overlap under certain restrictions of each set of weights. This appears to be true on the finite lattice for toroidal boundary conditions. In the thermodynamic limit, however, the free-energies of the even and the odd free-fermion models can be fully mapped to each other~\cite{wu2004k} by equating terms in the integrand of their double integral expressions, as can be easily seen above. Therefore, the two models have the same essential physics. 

To understand the mapping in the thermodynamic limit in contrast to the finite lattice, we can look at the construction of the free-energy solutions via dimers, given in appendix~\ref{app:dimerconst}, where the form of each free-energy integrand is simply the determinant whose square root gives each Pfaffian of the solution, the odd case involving two lattice units. For toroidal boundary conditions on the finite lattice, the partition function is a linear combination of 4 Pfaffians, so that the mapping found through equating the free-energy integrands is only approximate on the finite lattice. This is analogous to the case of the mappings between the checkerboard Ising model and the even free-fermion model, where the mapping in~\cite{davies1987} is valid on the finite toroidal lattice, but the mapping in~\cite{baxter19862} is only valid in the thermodynamic limit. It is possible that a boundary condition similar to those found by Brascamp and Kunz for the square lattice Ising model~\cite{brascamp1974k} could give the partition function of each model as a single double product, so that a finite lattice map between the partition functions of the even and odd free-fermion models could be found. We have not succeeded in generalizing those boundary conditions to the even and odd 8-vertex models, however. The checkerboard~\cite{davies1987} and union-jack ~\cite{wu1987l} lattice Ising models each have a local, finite toroidal lattice mapping to the even free-fermion model, and it would be satisfying to also find such a mapping between the even and odd free-fermion models. 

Furthermore, it is possible that this free-fermion, thermodynamic limit mapping is a special case of a more general mapping from the even to the odd 8-vertex models, since the mapping found in~\cite{wu2004k} automatically enforces the free-fermion constraints. Certainly, the mappings below found via the weak-graph transformations and the algebraic invariants continue to be valid in general for the full models, though subject to constraints among each set of weights. These mappings hint at a possible more general mapping among the general even and odd models. 

\section{Spin-model mappings}\label{sec:spinmaps}
\begin{figure}[htpb]
\begin{center}
\scalebox{1.2}{
\begin{tikzpicture}
\draw (0.01,0.01) grid (2.99,2.99);
\foreach \position in {(0.5,1),(0.5,2),(1.5,1),(1.5,2),(2.5,1),(2.5,2),(1,0.5),(2,0.5),(1,1.5),(2,1.5),(1,2.5),(2,2.5)}
\fill[black] \position circle (1.75pt);
\end{tikzpicture}
\hspace{0.5cm} 
\hspace{0.5cm}
\begin{tikzpicture}
\clip (0.1,0.1) rectangle (2.9,2.9);
\draw (0,1) -- (3,1);\draw (0,2) -- (3,2);
\draw (0,0.5) -- (3,3.5);\draw (0.5,0) -- (3,2.5);\draw (1.5,0) -- (3,1.5);\draw (0,1.5) -- (2,3.5);
\draw (1.5,0) -- (0,1.5);\draw (2.5,0) -- (0,2.5);\draw (3.5,0) -- (0,3.5);\draw (4.5,0) -- (0,4.5);
\draw (1,0.5) arc (-20:20:1.4);\draw (1,1.5) arc (-20:20:1.4);\draw (1,-0.5) arc (-20:20:1.4);\draw (1,2.5) arc (-20:20:1.4);\draw (2,0.5) arc (-20:20:1.4);\draw (2,1.5) arc (-20:20:1.4);\draw (2,-0.5) arc (-20:20:1.4);\draw (2,2.5) arc (-20:20:1.4);
\foreach \position in {(0.65,0.34),(0.65,1.34),(0.65,2.34),(1.65,0.34),(1.65,1.34),(1.65,2.34),(2.65,1.34),(2.65,2.34)}
\node at \position {\tiny{$J_1$}};
\foreach \position in {(0.3,0.6),(0.3,1.6),(1.3,0.6),(1.3,1.6),(1.3,2.6),(2.3,0.6),(2.3,1.6),(2.3,2.6)}
\node at \position {\tiny{$J_2$}};
\foreach \position in {(0.6,0.7),(0.6,1.7),(0.6,2.7),(1.6,0.7),(1.6,1.7),(1.6,2.7),(2.6,0.7),(2.6,1.7)}
\node at \position {\tiny{$J_3$}};
\foreach \position in {(0.32,1.35),(0.32,2.35),(1.32,0.35),(1.32,1.35),(1.32,2.35),(2.32,0.35),(2.32,1.35),(2.32,2.35)}
\node at \position {\tiny{$J_4$}};
\foreach \position in {(0.93,0.8),(0.93,1.8),(1.93,0.8),(1.93,1.8)}
\node at \position {\tiny{$J_5$}};
\foreach \position in {(0.85,1.1),(0.85,2.1),(1.85,1.1),(1.85,2.1)}
\node at \position {\tiny{$J_6$}};
\foreach \position in {(0.5,1),(0.5,2),(1.5,1),(1.5,2),(2.5,1),(2.5,2),(1,0.5),(2,0.5),(1,1.5),(2,1.5),(1,2.5),(2,2.5)}
\fill[black] \position circle (1.75pt);
\end{tikzpicture}
}
\caption{Mapping of the 16-vertex model to a staggered spin model, where the spin value corresponds to the bond state. On the left the vertex model lattice with spins placed at the center of each bond. On the right the staggered spin model; only two-spin interactions shown.\label{fig:spinbondmap}}
\end{center}
\end{figure}
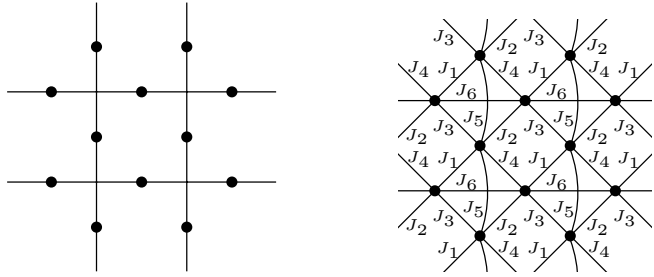

Three mappings to spin models have been introduced in the literature for the even 8-vertex model and the full 16-vertex model. One alternative is to introduce spins on the bonds of the lattice; this spin model was introduced in~\cite{suzuki1971f}. The 16 spin possibilities for the four spins on the bonds surrounding a vertex give 16 vertex weights, and introducing one, two, three, and four-spin interactions $J_i$ between spin variables, plus an overall constant, gives 16 spin interactions. The spin model, however, is inherently staggered, unless the only interactions allowed are one and two-spin interactions, shown in figure~\ref{fig:spinbondmap}.

The anisotropic square lattice Ising model in a magnetic field $H$ can be mapped to the 16-vertex model using this mapping~\cite{gaaff1974}, given the following weight assignments
\bat{4}
w_1 &= u_h^{-1}u_v^{-1}x^{-2}, &\qquad\qquad w_2 &= u_h^{-1}u_v^{-1}x^{2}, &\qquad\qquad w_3 &= u_hu_v, &\qquad\qquad w_4 &= u_hu_v, \nonumber\\
w_5 &= u_h^{-1}u_v, &\qquad\qquad w_6 &= u_h^{-1}u_v, &\qquad\qquad w_7 &= u_hu_v^{-1}, &\qquad\qquad w_8 &= u_hu_v^{-1}, \nonumber\\
v_1 &= x^{-1}, &\qquad\qquad v_2 &= x, &\qquad\qquad v_3 &= x^{-1}, &\qquad\qquad v_4 &= x, \nonumber\\
v_5 &= x^{-1}, &\qquad\qquad v_6 &= x, &\qquad\qquad v_7 &= x^{-1}, &\qquad\qquad v_8 &= x \label{isinginafield}
\eat
where $u_h=\exp(-2J_h/k_BT)$, $u_v=\exp(-2J_v/k_BT)$, $x=\exp(-H/k_BT)$. Because there are twice as many lattice sites for the Ising model compared to the vertex model, the relation of the free-energies is 
\beq
f=2\,f_{\mathrm{Ising}}(u_h,u_v,x)
\eeq
We give a second mapping between the 16-vertex model and the Ising model in a field in section~\ref{sec:isingfield}.

\begin{figure}[htpb]
\begin{center}
\scalebox{1.2}{
\begin{tikzpicture}
\draw (0.01,0.01) grid (2.99,2.99);
\foreach \position in {(0.5,0.5),(0.5,1.5),(0.5,2.5),(1.5,1.5),(1.5,2.5),(2.5,1.5),(2.5,2.5),(1.5,0.5),(2.5,0.5),(1.5,1.5),(2.5,1.5),(1.5,2.5),(2.5,2.5)}
\fill[black] \position circle (1.75pt);
\end{tikzpicture}
\hspace{0.5cm} 
\hspace{0.5cm}
\begin{tikzpicture}
\clip (0.1,0.1) rectangle (2.9,2.9);
\foreach \position in {(0.5,0.5),(0.5,1.5),(0.5,2.5),(1.5,1.5),(1.5,2.5),(2.5,1.5),(2.5,2.5),(1.5,0.5),(2.5,0.5),(1.5,1.5),(2.5,1.5),(1.5,2.5),(2.5,2.5)}
\fill[black] \position circle (1.75pt);
\draw (0.5,0) -- (0.5,3);\draw (1.5,0) -- (1.5,3);\draw (2.5,0) -- (2.5,3);
\draw (0,0.5) -- (3,0.5);\draw (0,1.5) -- (3,1.5);\draw (0,2.5) -- (3,2.5);
\draw (0,0) -- (3,3);\draw (1,0) -- (4,3);\draw (2,0) -- (5,3);
\draw (0,1) -- (3,4);\draw (0,2) -- (3,5);
\draw (-0.5,1.5) arc (65:25:2);\draw (0.5,1.5) arc (65:25:2);\draw (1.5,1.5) arc (65:25:2);\draw (2.5,1.5) arc (65:25:2);
\draw (-0.5,0.5) arc (65:25:2);\draw (0.5,0.5) arc (65:25:2);\draw (1.5,0.5) arc (65:25:2);\draw (2.5,0.5) arc (65:25:2);
\draw (-0.5,2.5) arc (65:25:2);\draw (0.5,2.5) arc (65:25:2);\draw (1.5,2.5) arc (65:25:2);\draw (2.5,2.5) arc (65:25:2);
\draw (-0.5,3.5) arc (65:25:2);\draw (0.5,3.5) arc (65:25:2);\draw (1.5,3.5) arc (65:25:2);\draw (2.5,3.5) arc (65:25:2);
\foreach \position in {(2,0.6),(1,1.6),(2,2.6),(1,0.6),(1,2.6),(2,1.6)}
\node at \position {\tiny{$J_1$}};
\foreach \position in {(0.35,1),(2.35,1),(1.35,2),(1.35,1),(0.35,2),(2.35,2)}
\node at \position {\tiny{$J_2$}};
\foreach \position in {(0.7,0.92),(1.7,1.92),(2.7,0.92),(1.7,0.92),(0.7,1.92),(2.7,1.92)}
\node at \position {\tiny{$J_3$}};
\foreach \position in {(0.75,1.2),(1.75,2.2),(1.75,0.2),(1.75,1.2),(0.75,2.2),(0.75,0.2)}
\node at \position {\tiny{$J_4$}};
\end{tikzpicture}
}
\caption{Mapping of the even 8-vertex model to a spin model. Bond states correspond to whether the spins across the bond are equal or not. On the left the vertex model lattice with spins placed at the center of each face. On the right the equivalent spin model; only two-spin interactions shown.\label{fig:spinfacemap}}
\end{center}
\end{figure}
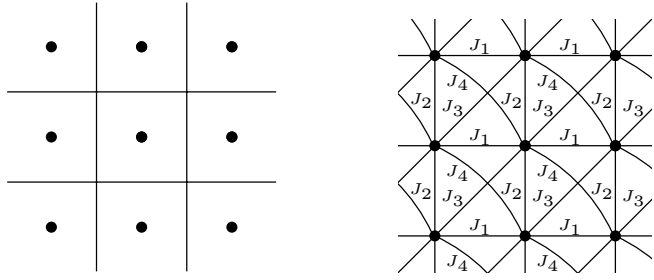

In another spin mapping, we assign spins to the faces of the lattice~\cite{wu1971,kadanoff1971w}; the two states of a bond separating two faces correspond to the spins on those those faces either being equal or opposite. Going around the faces surrounding a vertex and returning to the starting spin, the number of changes in spin values must be even in order to ensure that the upon returning to the initial spin it doesn't change its spin value. Therefore, this spin model only admits the even weights of the 16-vertex model. The mapping is 2:1 between spin states and vertex weights because the vertex weights are only determined after a spin somewhere in the lattice has been chosen to be $\pm1$. For toroidal boundary conditions, where $n_5=n_6$, $n_7=n_8$, we only need 6 spin interactions, for which we can use the four two-spin interactions show in figure~\ref{fig:spinfacemap}, a four-spin interaction $J_5$, and an overall constant $J_0$. The four-spin interactions can be removed in favor of only two-spin interactions if longer range next-next-nearest neighbor interactions are used~\cite{jungling1974o,jungling1975,horiguchi1985m}. The mapping is then given by
\bat{2}
\epsilon_1 &= J_0-J_1-J_2-J_3-J_4-J_5, &\qquad\qquad \epsilon_2 &=  J_0+J_1+J_2-J_3-J_4-J_5, \nonumber\\
\epsilon_3 &= J_0+J_1-J_2+J_3+J_4-J_5, &\qquad\qquad \epsilon_4 &=  J_0-J_1+J_2+J_3+J_4-J_5, \nonumber\\
\epsilon_5 &= \epsilon_6 = J_0-J_3+J_4+J_5, &\qquad\qquad \epsilon_7 &= \epsilon_8 = J_0+J_3-J_4+J_5, \label{homogspinmap}
\eat
and the inverse mapping
\bat{2}
8\,J_0 &= \epsilon_1+\epsilon_2+\epsilon_3+\epsilon_4+2\epsilon_5+2\epsilon_7, &\qquad\qquad 4\,J_1 &= -\epsilon_1+\epsilon_2+\epsilon_3-\epsilon_4, \nonumber\\
4\,J_2 &= -\epsilon_1-\epsilon_2+\epsilon_3+\epsilon_4, &\qquad\qquad 8\,J_3 &= -\epsilon_1-\epsilon_2+\epsilon_3+\epsilon_4-2\epsilon_5+2\epsilon_7, \nonumber\\
8\,J_4 &= -\epsilon_1-\epsilon_2+\epsilon_3+\epsilon_4+2\epsilon_5-2\epsilon_7, &\qquad\qquad 8\,J_5 &= -\epsilon_1-\epsilon_2-\epsilon_3-\epsilon_4+2\epsilon_5+2\epsilon_7,
\eat

On the finite lattice without toroidal boundary conditions, 8 spin interactions are needed to fully specify the vertex weights, so a magnetic field term (one-spin), and/or three-spin interactions are needed in addition. Choosing the upper-left spin around a vertex for the one-spin interaction $J_6$ and the upper-left, upper-right, and lower-right spins around a vertex for the three-spin interaction $J_7$ we have the following mapping
\beqr
e^{-\beta\epsilon_1} &=& 2\,e^{-\beta(J_0-J_1-J_2-J_3-J_4-J_5)}\cosh[\beta(J_6+J_7)], \\ 
e^{-\beta\epsilon_2} &=& 2\,e^{-\beta(J_0+J_1+J_2-J_3-J_4-J_5)}\cosh[\beta(J_6-J_7)], \\
e^{-\beta\epsilon_3} &=& 2\,e^{-\beta(J_0+J_1-J_2+J_3+J_4-J_5)}\cosh[\beta(J_6+J_7)], \\
e^{-\beta\epsilon_4} &=& 2\,e^{-\beta(J_0-J_1+J_2+J_3+J_4-J_5)}\cosh[\beta(J_6-J_7)], \\
e^{-\beta\epsilon_5} &=& 2\,e^{-\beta(J_0-J_3+J_4+J_5)}\cosh[\beta(J_6-J_7)], \\
e^{-\beta\epsilon_6} &=& 2\,e^{-\beta(J_0-J_3+J_4+J_5)}\cosh[\beta(J_6+J_7)], \\
e^{-\beta\epsilon_7} &=& 2\,e^{-\beta(J_0+J_3-J_4+J_5)}\cosh[\beta(J_6+J_7)], \\
e^{-\beta\epsilon_8} &=& 2\,e^{-\beta(J_0+J_3-J_4+J_5)}\cosh[\beta(J_6-J_7)]
\eeqr

We look at the third spin mapping, called the weak-graph expansion transformation, in the next section.

\section{Weak-graph expansion transformation}\label{sec:weakgraph}
Nagle introduced the so-called weak-graph expansion transformation for vertex models in~\cite{nagle1968,nagle1968t}, expanded on by Wegner~\cite{wegner1973}, for which Baxter gave a spin interpretation~\cite{baxter1982,baxter2007}. The weak-graph transformation has been previously applied in a couple of instances. Restricted to the even 8-vertex model, Wu used it to solve a modified Rys $F$ ice model~\cite{wu19692}, and Baxter used it to determine more relevant variables in his solution of the symmetric 8-vertex model~\cite{baxter1972}, for example. 

In the weak-graph transformation, each bond of the lattice is split in half; the partition function of the vertex model is then a sum over all vertex weights around all vertices, subject to the compatibility constraint that the two half-bonds must have the same state. 
\begin{figure}[htpb]
\begin{center}
\scalebox{0.75}{
\begin{tikzpicture}
\foreach \x in {2,4}
\draw (\x,0.4) -- (\x,5.6);
\foreach \y in {2,4}
\draw (0.4,\y) -- (5.6,\y);
\foreach \y in {1,3,5}
\draw[dashed] (1.5,\y) -- (2.5,\y);
\foreach \y in {1,3,5}
\draw[dashed] (3.5,\y) -- (4.5,\y);
\foreach \x in {1,3,5}
\draw[dashed] (\x,1.5) -- (\x,2.5);
\foreach \x in {1,3,5}
\draw[dashed] (\x,3.5) -- (\x,4.5);
\foreach \position in {(1.7,2.6),(3.7,4.6)}
\node at \position {$\sigma_A^{(1)}$};
\foreach \position in {(1.5,1.65),(3.5,3.65)}
\node at \position {$\sigma_B^{(1)}$};
\foreach \position in {(2.35,1.4),(4.35,3.4)}
\node at \position {$\sigma_C^{(1)}$};
\foreach \position in {(2.6,2.3),(4.6,4.3)}
\node at \position {$\sigma_D^{(1)}$};
\foreach \position in {(2.35,4.6),(4.35,2.6)}
\node at \position {$\sigma_C^{(2)}$};
\foreach \position in {(1.5,4.3),(3.5,2.3)}
\node at \position {$\sigma_D^{(2)}$};
\foreach \position in {(1.7,3.4),(3.7,1.4)}
\node at \position {$\sigma_A^{(2)}$};
\foreach \position in {(2.6,3.65),(4.6,1.65)}
\node at \position {$\sigma_B^{(2)}$};
\draw (1.9,1.9) rectangle (2.1,2.1);
\draw (3.9,3.9) rectangle (4.1,4.1);
\fill[black] (2,4) circle (0.115cm);
\fill[black] (4,2) circle (0.115cm);
\end{tikzpicture}
}
\caption{The weak-graph transformation. Spins on each bond are split into two, separated by the dashed lines, and a local map between spins variables and vertex weights is used around each vertex. The constraint that the split spin variables are equal is equivalent to the bond sharing of vertex weight variables.\label{fig:weakgraph}}
\end{center}
\end{figure}
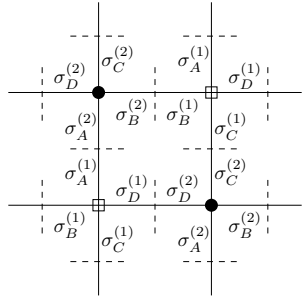

We can associate a spin variable to each half bond, distinguishing each of the spins around a vertex as in Figure~\ref{fig:weakgraph}. We then write each of the 16-vertex weights as functions of these four spin variables, so that the full expansion of the partition function can be written in terms of only the spin variables. Now when summing over the states of each spin, only those terms with an even number of spin variables for each bond will survive, that is, only those terms where the spins on either side of the split bond appear together survive the summation. The spin variables enforce the compatibility conditions among split bonds in the vertex weight summation, and we can replace the two spins on each bond by a single spin variable. Furthermore, there's a natural one-to-one mapping from spin variables to vertex weights. We choose that dashed bonds give $\sigma=1$ while solid bonds give $\sigma=-1$ below. Therefore, we can re-interpret these spin variables as new 16-vertex weights, since they fulfill all compatibility constraints among neighboring vertices and the mapping is one-to-one. We now show explicitly such a weak-graph transformation. 

{\scriptsize
\beq
G^{(1)} = \bordermatrix{
~ & w^{(1)}_{1}& w^{(1)}_{2}& w^{(1)}_{3}& w^{(1)}_{4}& w^{(1)}_{5}& w^{(1)}_{6}& w^{(1)}_{7}& w^{(1)}_{8}& v^{(1)}_{1}& v^{(1)}_{2}& v^{(1)}_{3}& v^{(1)}_{4}& v^{(1)}_{5}& v^{(1)}_{6}& v^{(1)}_{7}& v^{(1)}_{8}\cr
1 & 1 & 1 & 1 & 1 & 1 & 1 & 1 & 1 & 1 & 1 & 1 & 1 & 1 & 1 & 1 & 1 \cr
\sigma^{(1)}_A\sigma^{(1)}_B\sigma^{(1)}_C\sigma^{(1)}_D & 1 & 1 & 1 & 1 & 1 & 1 & 1 & 1 & -1 & -1 & -1 & -1 & -1 & -1 & -1 & -1 \cr
\sigma^{(1)}_A\sigma^{(1)}_C & 1 & 1 & 1 & 1 & -1 & -1 & -1 & -1 & -1 & -1 & -1 & -1 & 1 & 1 & 1 & 1 \cr
\sigma^{(1)}_B\sigma^{(1)}_D & 1 & 1 & 1 & 1 & -1 & -1 & -1 & -1 & 1 & 1 & 1 & 1 & -1 & -1 & -1 & -1 \cr
\sigma^{(1)}_C\sigma^{(1)}_D & 1 & 1 & -1 & -1 & 1 & 1 & -1 & -1 & -1 & -1 & 1 & 1 & -1 & -1 & 1 & 1 \cr
\sigma^{(1)}_A\sigma^{(1)}_B & 1 & 1 & -1 & -1 & 1 & 1 & -1 & -1 & 1 & 1 & -1 & -1 & 1 & 1 & -1 & -1 \cr
\sigma^{(1)}_B\sigma^{(1)}_C & 1 & 1 & -1 & -1 & -1 & -1 & 1 & 1 & -1 & -1 & 1 & 1 & 1 & 1 & -1 & -1 \cr
\sigma^{(1)}_A\sigma^{(1)}_D & 1 & 1 & -1 & -1 & -1 & -1 & 1 & 1 & 1 & 1 & -1 & -1 & -1 & -1 & 1 & 1 \cr
\sigma^{(1)}_C & 1 & -1 & 1 & -1 & -1 & 1 & -1 & 1 & -1 & 1 & 1 & -1 & 1 & -1 & 1 & -1 \cr
\sigma^{(1)}_A\sigma^{(1)}_B\sigma^{(1)}_D & 1 & -1 & -1 & 1 & -1 & 1 & -1 & 1 & 1 & -1 & -1 & 1 & -1 & 1 & -1 & 1 \cr
\sigma^{(1)}_A & 1 & -1 & -1 & 1 & 1 & -1 & 1 & -1 & 1 & -1 & -1 & 1 & 1 & -1 & 1 & -1 \cr
\sigma^{(1)}_B\sigma^{(1)}_C\sigma^{(1)}_D & 1 & -1 & -1 & 1 & 1 & -1 & 1 & -1 & -1 & 1 & 1 & -1 & -1 & 1 & -1 & 1 \cr
\sigma^{(1)}_D & 1 & -1 & 1 & -1 & -1 & 1 & 1 & -1 & 1 & -1 & 1 & -1 & -1 & 1 & 1 & -1 \cr
\sigma^{(1)}_A\sigma^{(1)}_B\sigma^{(1)}_C & 1 & -1 & 1 & -1 & -1 & 1 & 1 & -1 & -1 & 1 & -1 & 1 & 1 & -1 & -1 & 1 \cr
\sigma^{(1)}_B & 1 & -1 & 1 & -1 & 1 & -1 & -1 & 1 & 1 & -1 & 1 & -1 & 1 & -1 & -1 & 1 \cr
\sigma^{(1)}_A\sigma^{(1)}_C\sigma^{(1)}_D & 1 & -1 & 1 & -1 & 1 & -1 & -1 & 1 & -1 & 1 & -1 & 1 & -1 & 1 & 1 & -1 \cr
}\nonumber\label{weakgraphtrans}
\eeq
}
The transformation from vertex weights $w^{(1)}_i$, $v^{(1)}_i$ to spin variables $s^{(1)}_i$ is then
\beq
\mathbf{s}^{(1)} = G^{(1)}\mathbf{w}^{(1)}
\eeq
where the two vectors $\mathbf{s}^{(1)}$, $\mathbf{w}^{(1)}$ are as labeled on the border of the matrix $W^{(1)}$ defined above. Similarly, there's a relation $\mathbf{s}^{(2)} = G^{(2)}\mathbf{w}^{(2)}$ on alternate sites of the lattice, as shown in figure~\ref{weakgraphtrans}, where $G^{(2)}$ is determined from $G^{(1)}$ by swapping rows and columns under horizontal and vertical reflection of the bonds around a vertex, and where $\mathbf{w}^{(2)}$ is in principle a staggered set of weights.

When series expanding the partition function, the only non-zero terms correspond to the case where each half bond spin $\mathbf{s}^{(1)}$ is equal to $\mathbf{s}^{(2)}$. Demanding homogeneous spin variables $\mathbf{s}^{(2)}=\mathbf{s}^{(1)}$ requires the following compatibility condition on 32 variables
\beq
G^{(1)}\mathbf{w}^{(1)} = G^{(2)}\mathbf{w}^{(2)}
\eeq
which can be solved to show that simply $\mathbf{w}^{(2)}=\mathbf{w}^{(1)}$. Therefore the staggered set of variables $\mathbf{w}^{(2)}$ is not necessary, and the homogeneous weak-graph transformation is valid in general without constraints on the vertex weights. Furthermore, it is sufficient to only consider $G^{(1)}$, which we call $G$, and we similarly use the notation $\mathbf{s}$ and $\mathbf{w}$.

We now re-interpret the spin variables as new vertex weights $\mathbf{s}=\mathbf{w}^{(\mathrm{new})}$. This can be done in several ways. For example, the spin variable $\sigma_A\sigma_B\sigma_C\sigma_D$ may represent $w_1$, $w_2$, $w_3$, or $w_4$, according to how the spin variables represent either type of bond state. The horizontal or vertical bond state representations can be independently chosen, which leads to 4 different $G$ matrices, differing by swapped columns or rows according to the convention chosen. These extra matrices have two of which are related by a similarity transformation, a fact which has not been noted previously in the literature. The full set of weak-graph matrices $G$ will be what we call the weak-graph expansion transformation. We note that the $G$ matrices satisfy one of the following characteristic polynomials
\beqr
&(G^2-I)^8 = 0 \\
&(G^4-I)^4 = 0 \\
&(G^4-I)^2(G^2+I)^2(G-I)^4 = 0 
\eeqr
where $I$ is the $16\times16$ identity matrix.

\subsection{Mapping between the even and odd 8-vertex models}
Wu showed~\cite{wu1972} that the even 8-vertex model maps to the symmetric 16-vertex model with  pairs of weights equal, that is,
\beq
w_{2i}=w_{2i-1},\qquad v_{2i}=v_{2i-1},
\eeq
a point made later by Felderhof~\cite{felderhof19734} by looking directly at the transfer matrix. Wu further showed~\cite{wu1972} that when the even 8-vertex model satisfies the free-fermion condition, the condition on this symmetric 16-vertex model is 
\beq
w_1w_3+w_5w_7=v_1v_3+v_5v_7
\eeq
The only case of the even 8-vertex model that maps back onto the even 8-vertex model (the $v_i=0$ after the transformation) is precisely Baxter's symmetric 8-vertex model.

The case of the odd 8-vertex model has not been previously considered. The weak-graph transformation maps the odd 8-vertex model to the ``anti-symmetric" 16-vertex model, that is,
\beq
w_{2i}=-w_{2i-1},\qquad v_{2i}=-v_{2i-1},
\eeq
Once again, the free-fermion condition for the odd 8-vertex model is equivalent to the following condition on the anti-symmetric 16-vertex model
\beq
w_1w_3+w_5w_7=v_1v_3+v_5v_7
\eeq
The only case of the odd 8-vertex model which maps back onto the odd 8-vertex model is the anti-symmetric odd 8-vertex model. 

After applying the weak-graph transformation to either the even or the odd models, we can find the conditions on the sets of weights which map the two models to each other, yielding the following mapping using the explicit $G$ given in~(\ref{weakgraphtrans})
\beqr
&w_{2i}=-w_{2i-1},\qquad v_{2i}=v_{2i-1}, \nonumber\\
&2w_1 = v_1+v_3+v_5+v_7,\quad 2w_3 = -v_1-v_3+v_5+v_7 \nonumber\\ 
&2w_5 = -v_1+v_3-v_5+v_7,\quad 2w_7 = -v_1+v_3+v_5-v_7 \label{asetoso}
\eeqr
with the inverse map given by
\beqr
&w_{2i}=-w_{2i-1},\qquad v_{2i}=v_{2i-1}, \nonumber\\
&2v_1 = w_1- w_3-w_5- w_7,\quad 2v_3 = w_1- w_3+w_5+ w_7 \nonumber\\
&2v_5 = w_1+ w_3-w_5+ w_7,\quad 2v_7 = w_1+ w_3+w_5- w_7 \label{sotoase}
\eeqr
Other choices of $G$, or multiple applications $G^2$, $G^3$ give similar mappings. Note that in this mapping and its inverse, it is the anti-symmetric even and the symmetric odd 8-vertex models which are being mapped to each other.

From the mappings, we see that the anti-symmetric even 8-vertex model with the further constraint
\beq
w_1w_3=w_5w_7
\eeq
which is equivalent to the temperature independent even free-fermion condition, maps to the symmetric odd free-fermion model. Likewise, the symmetric odd 8-vertex model with the further constraint
\beq
v_1v_3=-v_5v_7
\eeq
which is also equivalent to the temperature independent odd free-fermion condition, maps to the anti-symmetric even free-fermion model. Here we see directly how intertwined the general free-fermion conditions~(\ref{ffcondeven}) and (\ref{ffcondodd}) are with the temperature independent free-fermion conditions~(\ref{ffindcondeven}) and (\ref{ffindcondodd}).

\section{Algebraic invariants mappings}\label{sec:alginvs}
In a series of papers, Gaaff, Hijmans, and Schram analyzed the full 16-vertex model, noting that the transfer matrix is invariant under an $SL(2)\times SL(2)$ transformation of the vertex weights for toroidal boundary conditions~\cite{gaaff1975h,gaaff1976h,gaaff1976h2,gaaff1978h,gaaff1979h,hijmans1983s, hijmans1984s,schram1984h,hijmans1985}. This analysis expanded an aside in~\cite{wegner1973} and was partially rediscovered in~\cite{davies1987}. They showed that with the 6 degrees of freedom available it is always possible to rewrite the partition function as a function of only 10 variables~\cite{gaaff1975h,gaaff1978h,gaaff1979h,hijmans1983s, hijmans1984s,hijmans1985}. Two particular subcases of the 16-vertex model can be shown to have the same partition function if their variables can be transformed into each other. They also constructed a set of 13 polynomial algebraic invariants which behave as scalars under the action of $SL(2)\times SL(2)$, such that the partition function of any 16-vertex model can be expressed as a function of only these 13 quantities~\cite{gaaff1976h,gaaff1976h2,schram1984h}. As a practical matter, we can use these invariants to look for mappings between two particular subcases of the 16-vertex model by constraining their invariants to be equal to each other. A null result, however, does not prevent a mapping between the models, as will be shown in sections~\ref{sec:ffmap} and \ref{sec:dimermap}. This set of transformations has been called a ``generalized weak-graph transformation", e.g.~\cite{samaj1991k,johnston1999}, even though it does not reduce, for any choice of the 6 arbitrary parameters, to the weak-graph transformation (see appendix~\ref{app:genweakgraph}). 
In this section we study the use of the algebraic invariants in order to find direct mappings between the even and odd 8-vertex model. We follow closely the notation defined in~\cite{gaaff1975h}. 

We note that algebraic invariants have also been studied in detail for the special case of the 16-vertex model we call Wu's symmetric 16-vertex model~\cite{perk1990ww,wu1990w,rollet1995w}, where the weights $w_i$, $v_i$ are equal if they have the same number of solid bonds. This model has been shown, under an extra  constraint on the weights, to be equal to the Ising model in a field; see section~\ref{sec:isingfield} for more details. For this model, the symmetry group of the transformation is $O(2)$, leading to only 5 algebraic invariants, with one syzygie between them~\cite{perk1990ww}. 

\subsection{Algebraic invariants of the 16-vertex model}
The start of the method is the recognition that the partition function is invariant under the following $SL(2)\times SL(2)$ transformation of the vertex weights~\cite{gaaff1975h,gaaff1978h,gaaff1979h,hijmans1983s, hijmans1984s,hijmans1985}. If we define the matrices
\beq
M = \begin{pmatrix}
w_1 & v_4 & v_6 & w_8 \\
v_1 & w_4 & w_6 & v_7 \\
v_8 & w_5 & w_3 & v_2 \\
w_7 & v_5 & v_3 & w_2
\end{pmatrix},\qquad S = \begin{pmatrix}
s_1 & s_2 \\
s_3 & s_4
\end{pmatrix},\qquad T = \begin{pmatrix}
t_1 & t_2 \\
t_3 & t_4
\end{pmatrix}
\eeq
\beq
V = S\otimes T,\qquad V^{-1}=S^{-1}\otimes T^{-1}
\eeq
for arbitrary complex constants $s_i$, $t_i$, and where the matrices $S$, $T$ can have unit determinant without loss of generality, then the partition function of the 16-vertex model is invariant under the transformation of the weights given by 
\beq
M = V^{-1}\,M\,V\label{genweakgraphtrans}
\eeq
In appendix~\ref{app:genweakgraph} we rewrite this transformation as a $16\times16$ matrix acting on the vector of 16 weights $\mathbf{w}$ defined in section~\ref{sec:weakgraph}, where it can be shown that there's no choice of $s_i$, $t_i$ which specialize to the weak-graph transformation matrix~(\ref{weakgraphtrans}). 

Under this transformation, the weights transform as either scalar $W_0$, vector $\mathbf{u}$, $\mathbf{v}$, or divector $W$ quantities
\beqr
W_0 &=& \frac{1}{4}~\big(w_1+w_2+w_3+w_4\big) \\
\textbf{u}^{\mathrm{T}} &=& \frac{1}{4}~\Big(v_1+v_2+v_3+v_4,i(-v_1+v_2+v_3-v_4),w_1-w_2-w_3+w_4\Big) \\
\textbf{v}^{\mathrm{T}} &=& \frac{1}{4}~\Big(v_5+v_6+v_7+v_8,i(v_5-v_6-v_7+v_8),w_1-w_2+w_3-w_4\Big) 
\eeqr
\beq
W = \frac{1}{4}\begin{pmatrix}
w_5+w_6+w_7+w_8 & i(w_5-w_6+w_7-w_8) & v_1-v_2+v_3-v_4 \\
i(-w_5+w_6+w_7-w_8) & w_5+w_6-w_7-w_8 & i(-v_1-v_2+v_3+v_4) \\
v_5-v_6+v_7-v_8 & i(v_5+v_6-v_7-v_8) & w_1+w_2-w_3-w_4
\end{pmatrix}
\eeq
with transformations given by
\beq
W_0\to W_0,\qquad \mathbf{u}\to R^{-1}(S)\,\mathbf{u},\qquad \mathbf{v}\to R^{-1}(T)\,\mathbf{v},\qquad W\to R^{-1}(S)\,WR(T)
\eeq
where the matrix R(S) is defined as 
\beq
R(S) = \begin{pmatrix}
\displaystyle \frac{1}{2}(s_1^2-s_2^2-s_3^2+s_4^2) & -\frac{i}{2}(s_1^2+s_2^2-s_3^2-s_4^2) & -s_1s_2+s_3s_4\vspace{0.05in} \\
\displaystyle \frac{i}{2}(s_1^2-s_2^2+s_3^2-s_4^2) & \frac{1}{2} (s_1^2+s_2^2+s_3^2+s_4^2) & -i(s_1s_2+s_3s_4) \\
-s_1s_3+s_2s_4 & i(s_1s_3+s_2s_4) & s_1s_4+s_2s_3
\end{pmatrix}
\eeq
so that
\beq
R^{-1}(S) = R^{\mathrm{T}}(S)
\eeq
and the matrix $R(T)$ is defined equivalently.

Therefore given an arbitrary 16-vertex model, it is always possible to rewrite the vertex weights in terms of only 10 new variables. This procedure is carried out in the series of papers~\cite{gaaff1975h,gaaff1978h,gaaff1979h,hijmans1983s, hijmans1984s,hijmans1985}, where they choose to diagonalize the $W$ matrix. It is not easy in practice, however, to use the transformation directly to seek mappings between two different cases of the 16-vertex model. To proceed further, we use the scalar polynomial invariants of this transformation, of which there are a total of 13~\cite{gaaff1976h,gaaff1976h2,schram1984h}. Though there are more polynomial invariants than the minimal set of variables, leading to three syzygies, they are nevertheless useful because they are unchanged by the $SL(2)\times SL(2)$ transformation. 

The 13 scalar polynomial invariants $I_i$ are defined as follows
\bat{2}
I_1 &= W_0 \qquad\qquad\qquad &  I_8 &= \textbf{u}^{\mathrm{T}}WW^{\mathrm{T}}W\textbf{v}  \\
I_2 &= \textbf{u}^{\mathrm{T}} \textbf{u}  \qquad\qquad\qquad &  I_9 &= \mathrm{Trace}\left[(WW^{\mathrm{T}})^2 \right]  \\
I_3 &= \textbf{v}^{\mathrm{T}} \textbf{v}   \qquad\qquad\qquad &  I_{10} &= \textbf{u}^{\mathrm{T}}(WW^{\mathrm{T}})^2\textbf{u}  \\
I_4 &= \textbf{u}^{\mathrm{T}}W \textbf{v} \qquad\qquad\qquad &  I_{11} &= \textbf{v}^{\mathrm{T}}(W^{\mathrm{T}}W)^2\textbf{v}  \\
I_5 &= \mathrm{Trace}\left[WW^{\mathrm{T}} \right]  \qquad\qquad\qquad &  I_{12} &= \textbf{u}^{\mathrm{T}}(WW^{\mathrm{T}})^2W\textbf{v}  \\
I_6 &= \textbf{u}^{\mathrm{T}}WW^{\mathrm{T}}\textbf{u}  \qquad\qquad\qquad &  I_{13} &=  \mathrm{Trace}\left[(WW^{\mathrm{T}})^3 \right]
 \\
I_7 &= \textbf{v}^{\mathrm{T}}WW^{\mathrm{T}}\textbf{v}  \qquad\qquad\qquad &  
\eat
For a given set of weights $w_j$ and $v_j$, the partition function of the model can be written as a function of only these 13 invariants. Therefore, for any mapping between models that preserves these invariants, the two partition functions will be equal to each other. We note that this is a sufficient but not necessary condition for two models to be equal, as exemplified in sections~\ref{sec:ffmap} and \ref{sec:dimermap}. 

For the odd 8-vertex model, with $w_i=0$, the invariants are as follows
\beqr
I_1 &=& I_4 ~=~ I_8 ~=~ I_{12} ~= 0 \label{invodd1}\\
I_2 &=& \frac{1}{4}~(v_1+v_4)(v_3+v_2) \\
I_3 &=& \frac{1}{4}~(v_5+v_8)(v_7+v_6) \\
I_5 &=& \frac{1}{4}~(v_1-v_4)(v_3-v_2)+\frac{1}{4}~(v_5-v_8)(v_7-v_6) \\
I_6 &=& \frac{1}{16}~(v_1v_3-v_2v_4)^2 \\
I_7 &=& \frac{1}{16}~(v_5v_7-v_6v_8)^2 \\
I_9 &=& \frac{1}{16}~(v_1-v_4)^2(v_3-v_2)^2+\frac{1}{16}~(v_5-v_8)^2(v_7-v_6)^2 \\
I_{10} &=& \frac{1}{64}~(v_1-v_4)(v_3-v_2)(v_1v_3-v_2v_4)^2 \\
I_{11} &=& \frac{1}{64}~(v_5-v_8)(v_7-v_6)(v_5v_7-v_6v_8)^2 \\
I_{13} &=& \frac{1}{64}~(v_1-v_4)^3(v_3-v_2)^3+\frac{1}{64}~(v_5-v_8)^3(v_7-v_6)^3 \label{invodd13}
\eeqr
where we see that there are 6 algebraically independent quantities, 
\beq
(v_1+v_4)(v_3+v_2),\quad (v_5+v_8)(v_7+v_6),\quad (v_5v_7-v_6v_8),\quad (v_1v_3-v_2v_4),\quad (v_1-v_4)(v_3-v_2),\quad (v_5-v_8)(v_7-v_6)\label{odd6algind}
\eeq
so that the odd 8-vertex model is a function of only these 6 quantities. There are 9 non-zero invariants and only 6 algebraically independent quantities, leading to relations between the invariants, listed below
\beqr
I_1 &=& I_4~=~I_8~=~I_{12}~=~0 \label{odddef1a}\\
I_5 &=& \left(\frac{I_{10}}{I_6}\right)+ \left(\frac{I_{11}}{I_7}\right) \\
I_9 &=& \left(\frac{I_{10}}{I_6}\right)^2+ \left(\frac{I_{11}}{I_7}\right)^2 \\
I_{13} &=& \left(\frac{I_{10}}{I_6}\right)^3 + \left(\frac{I_{11}}{I_7}\right)^3  \label{odddef1b}
\eeqr
or alternatively
\beqr
& I_1 ~=~ I_4~=~I_8~=~I_{12}~=~0 \label{odddef2a}\\
& I_5\,I_6\,I_7 = I_6\,I_{11}+I_7\,I_{10} \\
& I_6\,I_7\,(I_5^2-I_9) = 2\,I_{10}\,I_{11} \\
& I_6\,I_7\,(I_5I_9 - I_{13}) = I_5\,I_{10}\,I_{11} \label{odddef2b}
\eeqr
Any general 16-vertex model that has these same relations among its invariants will be fully equivalent to the odd 8-vertex model. We therefore define the class of odd 8-vertex models, invariant under the $SL(2)\times SL(2)$ symmetry, in terms of equations (\ref{odddef1a}--\ref{odddef1b}) or (\ref{odddef2a}--\ref{odddef2b}).

The odd 8-vertex model free-fermion case is given by the following extra condition on the invariants
\beq
I_2-\frac{I_{10}}{I_6} = I_3 - \frac{I_{11}}{I_7} \label{oddinvff}
\eeq

For the even 8-vertex model, with $v_i=0$, the invariants are as follows
\beqr
I_1 &=& \frac{1}{4}~(w_1+w_2+w_3+w_4) \label{inveven1}\\
I_2 &=& \frac{1}{16}~(w_1-w_2-w_3+w_4)^2 \\
I_3 &=& \frac{1}{16}~(w_1-w_2+w_3-w_4)^2 \\
I_4 &=& \frac{1}{64}~(w_1-w_2+w_3-w_4)(w_1-w_2-w_3+w_4)(w_1+w_2-w_3-w_4) \\
I_5 &=& \frac{1}{16}~(w_1+w_2-w_3-w_4)^2+\frac{1}{2}~(w_5 w_6+w_7 w_8) \\
I_6 &=& \frac{1}{256}~(w_1+w_2-w_3-w_4)^2(w_1-w_2-w_3+w_4)^2 \\
I_7 &=& \frac{1}{256}~(w_1+w_2-w_3-w_4)^2(w_1-w_2+w_3-w_4)^2 \\
I_8 &=& \frac{1}{1024}~(w_1-w_2+w_3-w_4)(w_1-w_2-w_3+w_4)(w_1+w_2-w_3-w_4)^3 \\
I_9 &=& \frac{1}{256}~(w_1+w_2-w_3-w_4)^4+\frac{1}{8}~(w_5 w_6+w_7 w_8)^2+\frac{1}{2}~(w_5w_6w_7w_8) \\
I_{10} &=& \frac{1}{4096}~(w_1-w_2-w_3+w_4)^2(w_1+w_2-w_3-w_4)^4 \\
I_{11} &=& \frac{1}{4096}~(w_1-w_2+w_3-w_4)^2(w_1+w_2-w_3-w_4)^4 \\
I_{12} &=& \frac{1}{16384}~(w_1-w_2+w_3-w_4)(w_1-w_2-w_3+w_4)(w_1+w_2-w_3-w_4)^5 \\
I_{13} &=& \frac{1}{4096}~(w_1+w_2-w_3-w_4)^6+\frac{1}{32}~(w_5 w_6+w_7 w_8)^3+\frac{3}{8}~(w_5w_6w_7w_8)(w_5 w_6+w_7 w_8) \label{inveven13}
\eeqr
where we see that there are also only 6 algebraically independent quantities, 
\beqr
& (w_1+w_2+w_3+w_4),\quad (w_1-w_2-w_3+w_4),\quad (w_1-w_2+w_3-w_4),\quad (w_1+w_2-w_3-w_4),\\
& (w_5 w_6+w_7 w_8),\quad (w_5w_6w_7w_8)
\eeqr
leading to 7 relations among the invariants. These are
\beqr
& I_4^2 = I_2\,I_7 = I_3\,I_6 \label{evendef1a}\\
& I_4^4 = I_2\,I_3^2\,I_{10} = I^2_2\,I_3\,I_{11} \\
& I_4^6 = I_2^2\,I_3^2\,I_8^2 \\
& I_4^{10} = I_2^4\,I_3^4\,I_{12}^2 \\
& I_2\,I_3\,(I_5^3+2I_{13})+6\,I_5\,I_6\,I_7+3\,I_4^2\,I_9=3\,(I_2\,I_3\,I_9+I_4^2\,I_5)\,I_5+6\,I_8^2 \label{evendef1b}
\eeqr
or alternatively
\beqr
& I_4^2 = I_2\,I_7 = I_3\,I_6 \label{evendef2a}\\
& I_6\,I_7 = I_2\,I_{11} = I_3\,I_{10} \\
& I_6\,I_{11} = I_7\,I_{10} = I_8^2 \\
& I_6\,I_7\,I_8^2 = I_2\,I_3\,I_{12}^2 \\
& I_2\,I_3\,(I_5^3+2I_{13})+6\,I_5\,I_6\,I_7+3\,I_4^2\,I_9=3\,(I_2\,I_3\,I_9+I_4^2\,I_5)\,I_5+6\,I_8^2 \label{evendef2b}
\eeqr
Any general 16-vertex model that satisfies these same invariant relations will be fully equivalent to the even 8-vertex model. We therefore define the class of even 8-vertex models, invariant under the $SL(2)\times SL(2)$ symmetry, in terms of equations (\ref{evendef1a}--\ref{evendef1b}) or (\ref{evendef2a}--\ref{evendef2b}).

The even 8-vertex model free-fermion case is given by the following extra condition on the invariants
\beq
I_1^2+\frac{2\,I_4^2}{I_2\,I_3} = I_2+I_3+I_5 \label{eveninvff}
\eeq

For the full 16-vertex model we list the first seven invariants
\beqr
I_1 &=& I_1^{\mathrm{even}} \label{full1a}\\
I_2 &=& I_2^{\mathrm{odd}}+I_2^{\mathrm{even}} \\
I_3 &=& I_3^{\mathrm{odd}}+I_3^{\mathrm{even}} \\
I_4 &=& I_4^{\mathrm{even}} + \frac{1}{16}~(w_1-w_2-w_3+w_4)(v_5v_7-v_6v_8)+\frac{1}{16}~(w_1-w_2+w_3-w_4)(v_1v_3-v_2v_4) \nonumber\\
&&+\frac{1}{16}~(v_7+v_6)(v_1w_8+v_2w_5+v_3w_5+v_4w_8)+\frac{1}{16}~(v_8+v_5)(v_1w_6+v_2w_7+v_3w_7+v_4w_6)  \\
I_5 &=& I_5^{\mathrm{odd}}+I_5^{\mathrm{even}} \\
I_6 &=& I_6^{\mathrm{odd}}+I_6^{\mathrm{even}} + \frac{1}{16}~[w_5(v_2+v_3)+w_8(v_1+v_4)][w_6(v_1+v_4)+w_7(v_2+v_3)] \nonumber\\
&&+\frac{1}{32}~(w_1-w_2-w_3+w_4)(v_1+v_4)[w_6(v_5-v_8)+w_8(v_7-v_6)] \nonumber\\
&& +\frac{1}{32}~(w_1-w_2-w_3+w_4)(v_2+v_3)[w_7(v_5-v_8)+w_5(v_7-v_6)] \nonumber\\
&& +\frac{1}{32}~(w_1-w_2-w_3+w_4)(w_1+w_2-w_3-w_4)(v_1v_3-v_2v_4) \nonumber\\
&& +\frac{1}{64}~(w_1-w_2-w_3+w_4)^2(v_5-v_8)(v_7-v_6)
\eeqr
\beqr
I_7 &=& I_7^{\mathrm{odd}}+I_7^{\mathrm{even}} + \frac{1}{32}~(v_5v_7-v_6v_8)[(w_1-w_4)^2-(w_2-w_3)^2]\nonumber\\
&& +\frac{1}{16}~[w_5(v_6+v_7)+w_7(v_8+v_5)][w_6(v_5+v_8))+w_8(v_7+v_6)]\nonumber\\
&& +\frac{1}{32}~(w_1-w_2+w_3-w_4)(v_1-v_4)[w_6(v_5+v_8))+w_8(v_7+v_6)] \nonumber\\
&& +\frac{1}{32}~(w_1-w_2+w_3-w_4)(v_3-v_2)[w_7(v_5+v_8)+w_5(v_7+v_6)] \nonumber\\
&& +\frac{1}{64}~(w_1-w_2+w_3-w_4)^2(v_1-v_4)(v_3-v_2)\hspace{2.6in} \label{full1b}
\eeqr
The sizes of the invariants beyond the seventh invariant are very large, though they presumably simplify somewhat after some effort. Unfortunately, the relations~(\ref{oddinvff}) and (\ref{eveninvff}) corresponding to the odd and even free-fermion conditions, respectively, do not either by themselves or in combination determine a simple means of recognizing a free-fermion model for the general 16-vertex model. We note that the invariant $I_1$ can be interpreted as the partition function of a single vertex under toroidal boundary conditions; we have not found similar interpretations of the other invariants, however.  

\subsection{Topological constraints}
As noted above, for toroidal boundary conditions, counting sink and sources constrains the even weights to have $n_5=n_6$ and $n_7=n_8$, so that the even 8-vertex model is a function of only 6 variables, for example $w_1$, $w_2$, $w_3$, $w_4$, $w_5w_6$, and $w_7w_8$, as can be seen clearly in its algebraic invariants. But in defining the even 8-vertex model in terms of the 7 relations~(\ref{evendef1a})--(\ref{evendef1b}) among its algebraic invariants, it no longer is necessary that $n_5=n_6$ and $n_7=n_8$, as can be seen by comparing with the invariants of the full 16-vertex model in (\ref{full1a})--(\ref{full1b}). Nevertheless, there must remain only 6 algebraically independent quantities in the model, though it is not clear how best to define these six quantities other than through the invariants themselves and their relations. 

For the odd 8-vertex model, counting sinks and sources under toroidal boundary conditions leads to $m_1+m_4=m_2+m_3$, $m_5+m_8=m_6+m_7$, which does not have a simple interpretation in terms of the odd weights, as is the case for the even 8-vertex model. Nevertheless, the set of 6 algebraically independent quantities~(\ref{odd6algind}) in the odd 8-vertex model invariants~(\ref{invodd1})--(\ref{invodd13}) provides natural variables which eliminate the redundancy in the 8 vertex weights by proper accounting of the topological constraints. Again, defining the odd 8-vertex model in terms of the relations among algebraic invariants~(\ref{odddef1a})--(\ref{odddef1b}) leads to a more general set of variables which respect the counting of sources and sinks, so that it remains a 6-variable model, though a convenient choice of variable definitions is not obvious, beyond using the invariants themselves along with their relations.

In the full 16-vertex model, the counting of horizontal and vertical sinks and sources leads to more general topological constraints, given in (\ref{16vtop}). Using the methods in~\cite{gaaff1975h,gaaff1978h,gaaff1979h,hijmans1983s, hijmans1984s,hijmans1985}, the full model can always be reduced to 10 variables, more than eliminating the redundancy caused by the model respecting the topological constraints. 

\subsection{Invariant mappings between even and odd 8-vertex models}
We now set the even and odd invariants given in equations (\ref{invodd1})--(\ref{invodd13}) and (\ref{inveven1})--(\ref{inveven13}) equal to each other in order to construct mappings between the two models, valid on the finite torus. We see that since $I_1$, $I_4$, $I_8$, $I_{12}$ are identically zero for the odd 8-vertex model, that on the even side we must have at least one of the following constraints, which leads to three separate mappings, considered below
\beq
\left.\substack{\displaystyle \vphantom{w^2_2}(w_1+w_2+w_3+w_4) ~=~ (w_1+w_2-w_3-w_4)~=~0\vspace{0.1in}\\\displaystyle (w_1+w_2+w_3+w_4) ~=~ (w_1-w_2+w_3-w_4)~=~0\vspace{0.1in}\\\displaystyle \vphantom{w_2^2}(w_1+w_2+w_3+w_4) ~=~ (w_1-w_2-w_3+w_4)~=~0}~~\right\}\quad \Rightarrow\quad \left\{~~\substack{\displaystyle \vphantom{w^2_2} w_1=-w_2,~~ w_3=-w_4  \vspace{0.1in}\\\displaystyle w_1=-w_3,~~ w_2=-w_4 \vspace{0.1in}\\\displaystyle \vphantom{w_2^2} w_1=-w_4,~~ w_2=-w_3}\right.\label{mappingcases}
\eeq 

\subsubsection{First mapping}\label{sec:firstmapping}
We start with the first case in~(\ref{mappingcases}). This leads to the following mapping, for which the only non-zero invariants are $I_2$, $I_3$, $I_5$, $I_9$, $I_{13}$
\beqr
&\displaystyle w_1=-w_2, \\
&\displaystyle w_3=-w_4, \\
&\displaystyle v_5v_7=v_6v_8, \\
&\displaystyle v_1v_3=v_2v_4, \\
&\displaystyle (w_1+w_3)^2 = \frac{v_5}{v_6}(v_7+v_6)^2 \\
&\displaystyle (w_1-w_3)^2 = \frac{v_2}{v_1}(v_1+v_4)^2 \\
&\displaystyle (w_5w_6+w_7w_8) = -\frac{1}{2}\left[\frac{v_2}{v_1}(v_1-v_4)^2+\frac{v_5}{v_6}(v_7-v_6)^2 \right] \\
&\displaystyle w_5w_6w_7w_8 = \frac{1}{16}\left[\frac{v_2}{v_1}(v_1-v_4)^2-\frac{v_5}{v_6}(v_7-v_6)^2 \right]^2 
\eeqr

We see that we can recover the weak-graph transformation results when setting
\beqr
w_{2i}=-w_{2i-1},\qquad v_{2i}=v_{2i-1} \label{firstmapping1}
\eeqr
such that
\beqr
&\displaystyle (w_1+w_3)^2 = (v_5+v_7)^2 \\
&\displaystyle (w_1-w_3)^2 = (v_1+v_3)^2 \\
&\displaystyle (w_5^2+w_7^2) = \frac{1}{2}\left[(v_1-v_3)^2+(v_5-v_7)^2 \right] \\
&\displaystyle w_5^2w_7^2 = \frac{1}{16}\left[(v_1-v_3)^2-(v_5-v_7)^2 \right]^2 \label{firstmapping2}
\eeqr
giving a set of 32 weak-graph transformations.

When the even 8-vertex model satisfies the free-fermion condition~(\ref{ffcondeven}), the odd 8-vertex model under this mapping~(\ref{firstmapping1})--(\ref{firstmapping2}) has the following further restrictions
\beq
v_1v_3=v_2v_4=-v_5v_7=-v_6v_8
\eeq
so that trivially,
\beq
v_1v_2v_3v_4=v_5v_6v_7v_8
\eeq
which is the temperature independent free-fermion condition, and when the odd 8-vertex model satisfies the free-fermion condition~(\ref{ffcondodd}), the even 8-vertex model under this mapping~(\ref{firstmapping1})--(\ref{firstmapping2}) has the following further restriction
\beq
w_1w_2w_3w_4=w_5w_6w_7w_8
\eeq
which is the temperature independent free-fermion condition.

\subsubsection{Second mapping}\label{sec:secondmapping}
From the second case in~(\ref{mappingcases}), for which the only non-zero invariants are $I_2$, $I_5$, $I_6$, $I_9$, $I_{10}$, $I_{13}$, we have by the following mapping
\beqr
&\displaystyle w_1=-w_3, \\
&\displaystyle w_2=-w_4, \\
&\displaystyle v_5=-v_8, \\
&\displaystyle v_6=-v_7, \\
&\displaystyle v_1v_2 = v_3v_4 \\
&\displaystyle (w_1+w_2)^2 = \frac{v_1}{v_3}(v_3-v_2)^2 \\
&\displaystyle (w_1-w_2)^2 = \frac{v_1}{v_3}(v_3+v_2)^2 \\
&\displaystyle w_5w_6=w_7w_8=v_5v_7
\eeqr

When the even 8-vertex model satisfies the free-fermion condition, the odd 8-vertex model under this mapping has the following further restrictions
\beq
v_1v_2=v_3v_4=v_5v_6=v_7v_8
\eeq
so that trivially 
\beq
v_1v_2v_3v_4=v_5v_6v_7v_8
\eeq
which is the temperature-independent odd free-fermion condition, and when the odd 8-vertex model satisfies the free-fermion condition, the even 8-vertex model under this mapping has the following further restrictions
\beq
w_1w_2=w_3w_4=-w_5w_6=-w_7w_8
\eeq
so that also
\beq
w_1w_2w_3w_4=w_5w_6w_7w_8
\eeq
which is the temperature independent free-fermion condition.

\subsubsection{Third mapping}\label{sec:thirdmapping}
The third case in~(\ref{mappingcases}), for which the only non-zero invariants are $I_3$, $I_5$, $I_7$, $I_9$, $I_{11}$, $I_{13}$, is equivalent to the second mapping under the following interchange of variables
\beq
w_3 \leftrightarrow w_4,\quad v_1 \leftrightarrow v_5,\quad v_2 \leftrightarrow v_6,\quad v_3 \leftrightarrow v_7,\quad v_4 \leftrightarrow v_8
\eeq

\section{Free-fermion free-energy mapping}\label{sec:ffmap}
From the free-energy expressions of the even and the odd free-fermion 8-vertex models in~(\ref{even8vfreeenergy}) and (\ref{odd8vfreeenergy}), a direct mapping~\cite{wu2004k} from one model to the other can be found which is more general than those found through the weak-graph transformation or the algebraic invariants. It is a mapping constructed in the thermodynamic limit, but is only approximate on the finite lattice with toroidal boundary conditions. This is because the mapping is from one Pfaffian determinant to another, and on the finite lattice the partition function under toroidal boundary conditions is a linear combination of 4 Pfaffians~\cite{mccoy1973w,mccoy2014w}. This is analogous to the case of the mapping from the free-fermion model to the checkerboard lattice Ising model, where the mapping of~\cite{davies1987} is valid on the finite toroidal lattice, but the mapping of~\cite{baxter19862} is only valid in the thermodynamic limit. For the square lattice Ising model, special boundary conditions have been found by Brascamp and Kunz~\cite{brascamp1974k} such that the partition function on the finite lattice has a simple double product form. A similar boundary condition would allow a direct mapping between the even and odd 8-vertex models, but we have not succeeded in finding similar boundary conditions relevant to these models. 

Furthermore, because the known constructions of the odd free-fermion model all use a staggered lattice as a starting point, the mapping is formally from two staggered odd lattice units to a single even unit, that is, the free-energies satisfy the relation
\beq
f_{8O} = \tfrac{1}{2}\,f_{8E}
\eeq
Any finite lattice mapping could presumably be a mapping from two lattice units in the odd model to one lattice unit in the even model. Another way to account for the factor of 2 in the free-energies is to map to the dual lattice. When mapping between the checkerboard Ising model and free-fermion model, there's a factor of 2 difference between the mapping of~\cite{davies1987} and the mapping of~\cite{baxter19862}, where the latter mapping used the dual lattice. Nevertheless, for the odd 8-vertex model, it appears a staggered lattice is still necessary even when considering the dual lattice, so that it appears that two odd lattice units must be mapped to one even lattice unit.

We now give the mapping explicitly. Let
\beqr
A &=& \pm \sqrt{B^2+\bar{C}^2+\bar{D}^2} \label{freemap1a}\\
B &=& \pm(v_1v_2+v_3v_4) \\
C &=& \pm (v_5v_7+v_6v_8) \\
\bar{C} &=& (v_5v_7-v_6v_8) \\
D &=& \pm (v_1v_3+v_2v_4) \\
\bar{D} &=& (v_1v_3-v_2v_4) \\
E &=& (v_1v_2-v_3v_4)(v_5v_6-v_7v_8)
\eeqr
then we have the following mapping
\bat{2}
2\,w_1 &= -A+B+C+D, &\qquad\qquad 2\,w_2 &= A-B+C+D \\
2\,w_3 &= A+B-C+D, &\qquad\qquad 2\,w_4 &= A+B+C-D \\
2\,w_5w_6 &= AB+CD-E &\qquad\qquad 2\,w_7w_8 &= AB+CD+E \label{freemap1b}
\eat
Note that this is a more general mapping between the even and odd 8-vertex models than those found via the weak-graph transformation and the algebraic invariants, since it does not reduce the number of algebraically independent quantities for either model, although it is only valid for free-fermion cases. In particular, this mapping does not respect the algebraic invariants of each model, that is, purely from the perspective of the algebraic invariants there should be no such general mapping between the even and odd models.

\section{An even-odd mapping of Baxter}\label{sec:dimermap}
The free-energy mapping~(\ref{freemap1a})--(\ref{freemap1b}) between the even and odd free-fermion models does not respect the algebraic invariants of the models, although it is only valid in the thermodynamic limit. Baxter, however, has found another mapping~\cite{baxter1972} which notably does not respect the algebraic invariants of the two models but also is defined locally so that it is valid on the finite toroidal lattice. The mapping is between the close-packed dimer model on the square lattice, with horizontal and vertical fugacities $z_h$ and $z_v$, respectively, and the even free-fermion 8-vertex model. Defined as an odd 8-vertex model, it has the following weights
\beq
v_1=v_3=z_v,\qquad v_5=v_7=z_h,\qquad v_2=v_4=v_6=v_8=0
\eeq
By considering the ``superbond" lattice, Baxter forms the following mapping to the free-fermion case of his symmetric even 8-vertex model~\cite{baxter1972}
\beq
w_1=w_2=z_h,\qquad w_3=w_4=z_v,\qquad w_5w_6=0,\qquad w_7w_8=z_h^2+z_v^2
\eeq
We can see that it does not respect the algebraic invariants because the first invariant $I_1$ is not zero in the even case. Clearly, having equal algebraic invariants is only a sufficient condition for the equality of two 16-vertex models. Furthermore, we see that this mapping is not equivalent to the free-energy free-fermion mapping in section~\ref{sec:ffmap}, which gives for the close-packed dimer model the following new mapping
\beq
2\,w_1 = 2\,w_2 = z_h^2+z_v^2,\qquad 2\,w_3=z_v^2-z_h^2,\qquad 2\,w_4 = z_h^2-z_v^2,\qquad 2\,w_5w_6=2\,w_7w_8=z_h^2z_v^2
\eeq

\section{Ising model in a field as even and odd 8-vertex models}\label{sec:isingfield}
The square lattice Ising model at $H=0$ is clearly a special case of the even free-fermion model~\cite{fan1970w}, and Gaaff showed in~\cite{gaaff1974} that the Ising model at $H=i\pi/(2\beta)$ is also a special case of the even free-fermion model. We have sought for mappings between the general even and odd 8-vertex models and the full Ising model in a field, which has the 16-vertex model weight representation given in (\ref{isinginafield}). Its algebraic invariants are given as follows

\beqr
I_1 &=& \frac{1+2\,u_h^2u_v^2x^2+x^4}{4\,u_hu_vx^2} \\
I_2 &=& I_3 ~=~ \frac{(x^2+1)^2(1-2\,x^2+4\,u_h^2u_v^2x^2+x^4)}{16\,u_h^2u_v^2x^4} ~=~ I_1^2-\frac{(u_h^2u_v^2-1)^2}{4\,u_h^2u_v^2}  \\
I_4 &=& \frac{(x^2+1)^2\,p_{(4,4,8)}}{64\,u_h^3u_v^3x^6} \\
I_5 &=& \frac{q_{(4,4,8)}}{16\,u_h^2u_v^2x^4} ~=~ I_2 +\frac{1+2\,u_h^4+2\,u_v^4-6u_h^2u_v^2+u_h^4u_v^4}{4\,u_h^2u_v^2} \\
I_6 &=& I_7 ~=~ \frac{(x^2+1)^2\,p_{(6,6,12)}}{256\,u_h^4u_v^4x^8} \\
I_8 &=& \frac{(x^2+1)^2\,p_{(8,8,16)}}{1024\,u_h^5u_v^5x^{10}} \\
I_9 &=& \frac{q_{(8,8,16)}}{256\,u_h^4u_v^4x^8} \\
I_{10} &=& I_{11} ~=~ \frac{(x^2+1)^2\,p_{(10,10,20)}}{4096\,u_h^6u_v^6x^{12}} \\
I_{12} &=& \frac{(x^2+1)^2\,p_{(12,12,24)}}{16384\,u_h^7u_v^7x^{14}} \\
I_{13} &=&  \frac{q_{(12,12,24)}}{4096\,u_h^6u_v^6x^{12}}
\eeqr
where the $p_{(n,m,o)}$, $q_{(n,m,o)}$ are polynomials of degree $n$ in $u_h$, degree $m$ in $u_v$, and degree $o$ in $x$. 

It can be shown that the Ising model in a field can only satisfy the algebraic invariant relations (\ref{evendef1a})--(\ref{evendef1b}) of the general even 8-vertex model if the magnetic field variable is $x^2=\pm1$, at which points it also automatically satisfies the even free-fermion invariant relation (\ref{eveninvff}). For the odd 8-vertex model, the Ising model in a field always satisfies the odd free-fermion invariant relation (\ref{oddinvff}) but it can only satisfy the general odd 8-vertex model invariant relations (\ref{odddef1a})--(\ref{odddef1b}) trivially, if the magnetic field variable is $x^2=-1$ and $u_h,u_v=\pm1$. These are all exact mappings on the finite toroidal lattice. In the thermodynamic limit, however, the free-fermion mapping between the even and the odd model can be used, so that the odd free-fermion model can fully map to the Ising model at $H=\{0,i\pi/(2\beta)\}$.

There is a second mapping between Wu's symmetric 16-vertex model and the isotropic Ising model in a field, given in~\cite{wu19722,wu1974,wu1990w,wu1989w} where the 16-vertex model is first mapped to a lattice gas on the square lattice, and then to the Ising model in a field. An alternate version of the mapping was also considered in~\cite{samaj1991k}, though it is equivalent to the mapping of~\cite{wu19722,wu1974} after applying the $SL(2)\times SL(2)$ transformation~(\ref{genweakgraphtrans}). We mention that this Ising in a field mapping can be generalized to lattices with general coordination numbers~\cite{samaj1992k2,johnston1999} and the $(q-1)/2$-spin Ising model in a field can be mapped to $q$-state vertex models generalizing this mapping~\cite{samaj1992k3}. Here we restrict ourselves to the square lattice and the spin $1/2$ Ising model in a field. The mapping in~\cite{samaj1991k} has weights given by
\beqr
& w_1=a,\qquad w_2=e,\qquad w_3=w_4=w_5=w_6=w_7=w_8=c,\nonumber\\
& v_1=v_3=v_5=v_7=b,\qquad v_2=v_4=v_6=v_8=d
\eeqr
and is valid under the following constraint of the weights
\beq
ace-ad^2-b^2e+2bcd-c^3=0
\eeq
We prefer the simplicity of the treatment in~\cite{wu19722,wu1974} and will only consider it below, since the results are equivalent. The weights in the mapping of~\cite{wu19722,wu1974} are given as follows
\beqr
& w_1=1,\qquad w_2=v,\qquad w_3=w_4=w_5=w_6=w_7=w_8=w^2,\nonumber\\
& v_1=v_3=v_5=v_7=w,\qquad v_2=v_4=v_6=v_8=w^3
 \label{isinginafield2}
\eeqr
where the variables $w$ and $v$ are related to the isotropic variable $u=u_h=u_v$ and the field variable $x$ by
\beq
u^2=\frac{w^2}{1+w^2},\qquad x^2=\frac{v-w^4}{(1+w^2)^2}
\eeq
and
\beq
w = \frac{u}{\sqrt{1-u^2}},\qquad v = \frac{x^2+u^4}{(1-u^2)^2}
\eeq
and the partition function of this 16-vertex model is related to the Ising model in a field on a toroidal lattice with $\mathcal{N}$ sites by 
\beq
Z = \left[\frac{ux}{(1-u^2)^2}\right]^{\mathcal{N}} Z_{\mathrm{Ising}}(u,x)
\eeq
so that the free-energy is given by
\beq
f = f_{\mathrm{Ising}}(u,x) + \ln\left[\frac{ux}{(1-u^2)^2}\right]
\eeq
The algebraic invariants are now of the form
\beqr
I_1 &=& \frac{(x^2+1)}{4\,(u^2-1)^2} \\
I_2 &=& I_3 ~=~ \frac{1-2\,x^2+x^4+4\,u^2x^2}{16\,(u^2-1)^4} \\
I_4 &=& \frac{(x^2+1)(1-2\,x^2+x^4+4\,u^4x^2)}{64\,(u^2-1)^6} \\
I_5 &=& \frac{1+2\,x^2+x^4-8\,u^2x^2+8\,u^4x^2}{16\,(u^2-1)^4} \\
I_6 &=& I_7 ~=~ \frac{p_{(6,8)}}{256\,(u^2-1)^8} \\
I_8 &=& \frac{(x^2+1)p_{(8,8)}}{1024\,(u^2-1)^{10}} \\
I_9 &=& \frac{(x^2+1)q_{(8,8)}}{256\,(u^2-1)^{8}} \\
I_{10} &=& I_{11} ~=~ \frac{p_{(10,12)}}{4096\,(u^2-1)^{12}} \\
I_{12} &=& \frac{(x^2+1)p_{(12,12)}}{16384\,(u^2-1)^{14}} \\
I_{13} &=& \frac{(x^2+1)q_{(12,12)}}{4096\,(u^2-1)^{12}}
\eeqr
where the $p_{(n,m)}$, $q_{(n,m)}$ are polynomials of degree $n$ in $u$ and degree $m$ in $x$.

In this case, the Ising model in a field can only satisfy the algebraic invariant relations (\ref{evendef1a})--(\ref{evendef1b}) of the general even 8-vertex model if the magnetic field is $H=0,\infty$, in which case it also automatically satisfies the even free-fermion invariant relation (\ref{eveninvff}). For the odd 8-vertex model, the Ising model in a field always satisfies the odd free-fermion invariant relation (\ref{oddinvff}) but it can only satisfy the general odd 8-vertex model invariant relations (\ref{odddef1a})--(\ref{odddef1b}) if the magnetic field is $x^2=-1$. We give the following new mappings from the odd 8-vertex model to the Ising model with field $H=i\pi/(2\beta)$
\beq
v_1v_2=v_3v_4=v_5v_6=v_7v_8=\frac{1}{4(1-u^2)^2},\qquad \frac{v_1}{v_4} =  \frac{v_3}{v_2} = \frac{v_5}{v_8} = \frac{v_7}{v_6} = \frac{(1+u)}{(1-u)} 
\eeq
\beq
v_1v_2=v_3v_4=v_5v_6=v_7v_8=\frac{1}{4(1-u^2)^2},\qquad \frac{v_1}{v_4} =  \frac{v_3}{v_2} = \frac{v_8}{v_5} = \frac{v_6}{v_7} = \frac{(1+u)}{(1-u)} 
\eeq
There are also two alternate mappings given by making $u$ negative. The above mappings are new for the odd 8-vertex model, and generalizes what was known previously~\cite{auyang1984p,perk2006a}, that at criticality $T_c=\infty$ the Ising model at $H=i\pi/(2\beta)$ is equivalent to the square lattice close-packed dimer model, which is a special case of the odd 8-vertex free-fermion model. These are exact mappings on the finite toroidal lattice. In the thermodynamic limit, the free-fermion free-energy mapping between the even and the odd models can be used to map back and forth between the even and odd cases, however.

From the analysis of the algebraic invariants of the Ising model in a field using both mappings, we conclude that unless another mapping is found, the only magnetic field points which map to either the even and odd 8-vertex models are those described above, which also are simultaneously free-fermion points of the model. From the perspective of algebraic invariants of the 16-vertex model we have found all of the free-fermionic points of the Ising model in a field.

\section{Staggered 8-vertex models}\label{sec:stag}
Most exactly solved models in statistical mechanics use homogeneous variables, which are equal at all lattice sites, but occasionally staggered lattices have also been studied. In~\cite{wu1975l,hsue1975lw}, different weights were assigned to each bi-partite sublattice of the square lattice, in~\cite{tanaka1992m} both column and bi-partite staggering were considered, and in~\cite{lin1977w} Lin and Wu considered a staggering units of four lattice sites. By specializing from the results of Lin and Wu~\cite{lin1977w}, the column or bi-partite staggered results can be recovered. Also, in~\cite{wolff1981z}, the Ising model on the square lattice is solved for abitrary staggering unit sizes in the diagonal direction, which can be re-cast as a staggered vertex model. In each of these cases, the restriction to the free-fermion condition was used. 

There are very few known solvable cases of staggered 8-vertex models which are not free-fermion models. In~\cite{truong1985}, sufficient conditions on the weights for the solubility of staggered symmetric 6-vertex models are given. Bariev solved, however, a staggered 6-vertex model outside of those conditions~\cite{bariev19802,bariev1981,bariev1984}. Baxter in~\cite{baxter19712} also gave conditions such that a general inhomogeneous 6-vertex model can be solvable by Bethe's ansatz. For the bi-partite staggered lattice, Baxter showed in~\cite{baxter19824} that the equivalent critical-temperature antiferromagnetic $q$-state Potts model on the square lattice admits two solutions; for one, the second set of staggered weights is simply proportional to the other set of weights; the second solution is non-trivial.  

As was first noted in~\cite{wu2004k}, the staggered even and odd 8-vertex models are fully equivalent, since changing the convention of the bond state of one (or three) bond(s) per staggering unit of two vertices will convert each even vertex weight on the lattice to an odd one, and vice-versa. We will make this precise below. Therefore, the critical phenomena of the staggered even and odd 8-vertex models are equivalent. 

We use the notation $\bar{w}_i$ and $\bar{v}_i$ for the second set of vertex weights on the alternate vertices of the staggering units. We can also define staggered solid and dashed bond variables, in anology with~(\ref{partbondvars}), which can be introduced into any result by a simple variable transformation of the form~(\ref{bondvartransf1}).

\subsection{Spin model mappings}
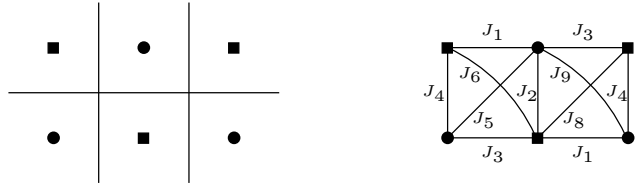
\begin{figure}[htpb]
\begin{center}
\scalebox{1.2}{
\begin{tikzpicture}
\draw (1,0) -- (1,2); \draw (2,0) -- (2,2); \draw (0,1) -- (3,1);
\foreach \position in {(0.5,0.5),(1.5,1.5),(2.5,0.5)}
\fill[black] \position circle (2pt);
\begin{scope}[shift={(-.06,-0.06)}]
\foreach \position in {(0.5,1.5),(1.5,0.5),(2.5,1.5)}
\filldraw \position rectangle ++(3.2pt,3.2pt);
\end{scope}
\end{tikzpicture}
\hspace{0.5cm} 
\hspace{0.5cm}
\begin{tikzpicture}
\draw (0.5,0.5) -- (0.5,1.5) -- (1.5,1.5) -- (1.5,0.5) -- (0.5,0.5);
\draw (0.5,0.5) -- (1.5,1.5);
\draw (1.5,1.5) -- (2.5,1.5) -- (2.5,0.5) -- (1.5,0.5) -- (2.5,1.5);
\draw (0.5,1.5) arc (65:25:2);
\draw (1.5,1.5) arc (65:25:2);
\draw[white] (0,0) -- (0,2);
\foreach \position in {(0.5,0.5),(1.5,1.5),(2.5,0.5)}
\fill[black] \position circle (2pt);
\begin{scope}[shift={(-.06,-0.06)}]
\foreach \position in {(0.5,1.5),(1.5,0.5),(2.5,1.5)}
\filldraw \position rectangle ++(3.2pt,3.2pt);
\end{scope}
\node at (1,1.67) {\tiny{$J_1$}};
\node at (1.38,1) {\tiny{$J_2$}};
\node at (1,0.32) {\tiny{$J_3$}};
\node at (0.35,1) {\tiny{$J_4$}};
\node at (0.9,0.7) {\tiny{$J_5$}};
\node at (0.75,1.2) {\tiny{$J_6$}};
\begin{scope}[shift={(1,0)}]
\node at (1,1.67) {\tiny{$J_3$}};
\node at (1.38,1) {\tiny{$J_4$}};
\node at (1,0.32) {\tiny{$J_1$}};
\node at (0.9,0.7) {\tiny{$J_8$}};
\node at (0.75,1.2) {\tiny{$J_9$}};
\end{scope}
\end{tikzpicture}
}
\caption{Mapping of the bi-partite staggered even 8-vertex model to a staggered spin model. Bond states correspond to whether the spins across the bond are equal or not. On the left the vertex model lattice with spins placed at the center of each face. On the right the equivalent spin model; only two-spin interactions are shown.\label{fig:bistagspinfacemap}}
\end{center}
\end{figure}

In~\cite{hsue1975lw} the homogeneous spin-model mapping in~(\ref{homogspinmap}) was generalized for a bi-partite staggered even 8-vertex model, as shown in figure~\ref{fig:bistagspinfacemap}. 

We define in figure~\ref{fig:bistagspinfacemap} the two-spin interactions for both units. In each unit there is also a four-spin interaction $J_7$ and $J_{10}$, plus an overall constant $J_0$. For the left unit, we have the mapping
\beqr
\epsilon^{(1)}_1 &=& J_0 - J_1 - J_2 - J_3 - J_4 - J_5 - J_6 - J_7 \label{bipspinmap1}\\
\epsilon^{(1)}_2 &=& J_0 + J_1 + J_2 + J_3 + J_4 - J_5 - J_6 - J_7 \\
\epsilon^{(1)}_3 &=& J_0 + J_1 - J_2 + J_3 - J_4 + J_5 + J_6 - J_7 \\
\epsilon^{(1)}_4 &=& J_0 - J_1 + J_2 - J_3 + J_4 + J_5 + J_6 - J_7 \\
\epsilon^{(1)}_5 &=& J_0 - J_1 + J_2 + J_3 - J_4 - J_5 + J_6 + J_7 \\
\epsilon^{(1)}_6 &=& J_0 + J_1 - J_2 - J_3 + J_4 - J_5 + J_6 + J_7 \\
\epsilon^{(1)}_7 &=& J_0 - J_1 - J_2 + J_3 + J_4 + J_5 - J_6 + J_7 \\
\epsilon^{(1)}_8 &=& J_0 + J_1 + J_2 - J_3 - J_4 + J_5 - J_6 + J_7 \label{bipspinmap2}
\eeqr
and similarly for the right unit.

For this bi-partite staggered lattice under toroidal boundary conditions, there are only 11 independent quantities $u_i$, where our definitions differ slightly from~\cite{hsue1975lw}
\bat{6}
u_1 &= w_1 \bar{w}_1,&\qquad u_2 &= w_2 \bar{w}_2,&\qquad u_3 &= w_3 \bar{w}_3,&\qquad u_4u_9 &= w_4 \bar{w}_4,&\qquad u_5 &= w_5 \bar{w}_6,&\qquad u_6u_{10} &= w_6 \bar{w}_5 \nonumber\\
u_7 &= w_7 \bar{w}_8,&\qquad u_8u_{11} &= w_8 \bar{w}_7,&\qquad u_9 &= \frac{\bar{w}_3\bar{w}_4}{\bar{w}_1\bar{w}_2},&\qquad u_{10} &= \frac{\bar{w}_5\bar{w}_6}{\bar{w}_1\bar{w}_2},&\qquad u_{11} &= \frac{\bar{w}_7\bar{w}_8}{\bar{w}_1\bar{w}_2}&& \label{stagbnotation}
\eat

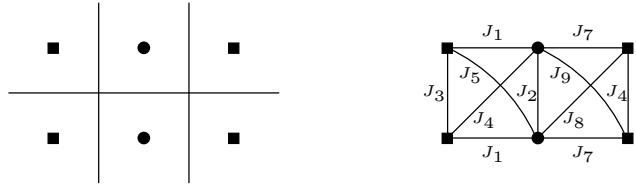
\begin{figure}[htpb]
\begin{center}
\scalebox{1.2}{
\begin{tikzpicture}
\draw (1,0) -- (1,2); \draw (2,0) -- (2,2); \draw (0,1) -- (3,1);
\foreach \position in {(1.5,0.5),(1.5,1.5)}
\fill[black] \position circle (2pt);
\begin{scope}[shift={(-.06,-0.06)}]
\foreach \position in {(0.5,0.5),(0.5,1.5),(2.5,0.5),(2.5,1.5)}
\filldraw \position rectangle ++(3.2pt,3.2pt);
\end{scope}
\end{tikzpicture}
\hspace{0.5cm} 
\hspace{0.5cm}
\begin{tikzpicture}
\draw (0.5,0.5) -- (0.5,1.5) -- (1.5,1.5) -- (1.5,0.5) -- (0.5,0.5);
\draw (0.5,0.5) -- (1.5,1.5);
\draw (1.5,1.5) -- (2.5,1.5) -- (2.5,0.5) -- (1.5,0.5) -- (2.5,1.5);
\draw (0.5,1.5) arc (65:25:2);
\draw (1.5,1.5) arc (65:25:2);
\draw[white] (0,0) -- (0,2);
\foreach \position in {(1.5,0.5),(1.5,1.5)}
\fill[black] \position circle (2pt);
\begin{scope}[shift={(-.06,-0.06)}]
\foreach \position in {(0.5,0.5),(0.5,1.5),(2.5,0.5),(2.5,1.5)}
\filldraw \position rectangle ++(3.2pt,3.2pt);
\end{scope}
\node at (1,1.67) {\tiny{$J_1$}};
\node at (1.38,1) {\tiny{$J_2$}};
\node at (1,0.32) {\tiny{$J_1$}};
\node at (0.35,1) {\tiny{$J_3$}};
\node at (0.9,0.7) {\tiny{$J_4$}};
\node at (0.75,1.2) {\tiny{$J_5$}};
\begin{scope}[shift={(1,0)}]
\node at (1,1.67) {\tiny{$J_7$}};
\node at (1.38,1) {\tiny{$J_4$}};
\node at (1,0.32) {\tiny{$J_7$}};
\node at (0.9,0.7) {\tiny{$J_8$}};
\node at (0.75,1.2) {\tiny{$J_9$}};
\end{scope}
\end{tikzpicture}
}
\caption{Mapping of the column staggered even 8-vertex model to a staggered spin model. Bond states correspond to whether the spins across the bond are equal or not. On the left the vertex model lattice with spins placed at the center of each face. On the right the equivalent spin model; only two-spin interactions are shown.\label{fig:colstagspinfacemap}}
\end{center}
\end{figure}
Similarly for the column staggered lattice, shown in figure~\ref{fig:colstagspinfacemap}, there is a mapping similar to~(\ref{bipspinmap1})--(\ref{bipspinmap2}). Under toroidal boundary conditions there are only 11 independent quantities $t_i$, which we give for the first time below
\bat{6}
t_1 &= w_1\bar{w}_1,&\qquad t_2 &= w_2\bar{w}_2,&\qquad t_3 &= w_3\bar{w}_3,&\qquad t_4 &= w_4\bar{w}_2,&\qquad t_5 &= w_5w_6\bar{w}_1\bar{w}_2,&\qquad t_6 &= \frac{w_6\bar{w}_8}{w_8\bar{w}_6} \nonumber\\
t_7 &= w_7w_8\bar{w}_1\bar{w}_2,&\qquad t_8 &= \frac{\bar{w}_3}{\bar{w}_1},&\qquad t_9t_8 &= \frac{\bar{w}_3\bar{w}_4}{\bar{w}_1\bar{w}_2},&\qquad t_{10} &= \frac{\bar{w}_5\bar{w}_6}{\bar{w}_1\bar{w}_2},&\qquad t_{11}t_6 &= \frac{\bar{w}_7\bar{w}_8}{\bar{w}_1\bar{w}_2}
\eat

\subsection{Mappings between the even and odd staggered 8-vertex models}
\begin{figure}[htpb]
\begin{center}
\begin{subfigure}{0.3\textwidth}
\centering
\scalebox{1.}{
\includegraphics{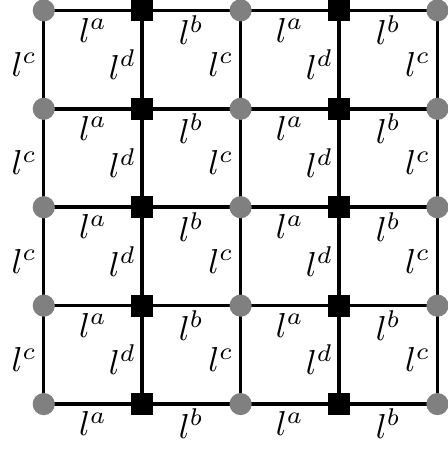}
}
\end{subfigure}%
\begin{subfigure}{0.3\textwidth}
\centering
\scalebox{1.}{
\includegraphics{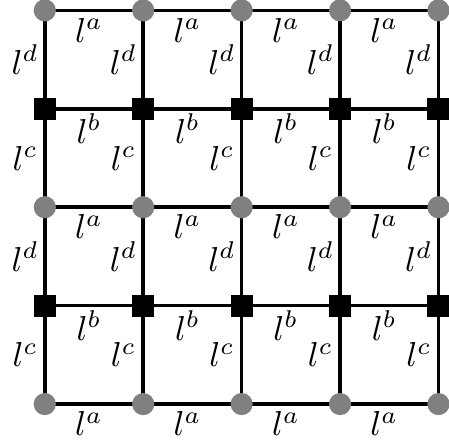}
}
\end{subfigure}%
\begin{subfigure}{0.3\textwidth}
\centering
\scalebox{1.}{
\includegraphics{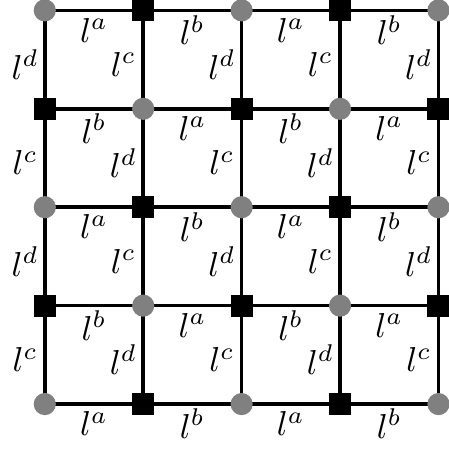}
}
\end{subfigure}%
\end{center}
\caption{The square lattice with edge variables $l=\pm 1$ on each edge. From the edge values a choice of correspondence determines the vertical and horizontal arrows or, alternatively, the solid and dashed bonds of the vertex model. Each sub-lattice is identified by either a square or a circle at each site, from which we identify four sets of edges $l^a$, $l^b$, $l^c$, $l^d$. By changing the interpretation of one edge set, an even staggered 8-vertex model can be converted to an odd staggered 8-vertex model and vice-versa. On the left a column staggered lattice, in the center a row staggered lattice, and on the right a bi-partite staggered lattice.\label{fig:staggered}}
\end{figure}

In Figure~\ref{fig:staggered}, we draw the square lattice, identifying two of its sub-lattices with either circles or squares at each site. We also divide the set of edges into four categories, which we denote by edge variables $l^a$, $l^b$, $l^c$, $l^d$, which take the values $\pm1$. From the edge variables, we can then choose a correspondence between the values $\pm1$ and vertical and horizontal arrow directions, or alternatively, solid and dashed bonds. For arbitrary sets of edge values on the lattice, the resulting vertex model will be described by the most general 16-vertex model. However, this general 16-vertex model will reduce to an even or odd 8-vertex model if an appropriate constraint is applied to all four edge variables around each vertex. 

Once the edge variables are constrained to form an 8-vertex model, either even or odd, changing the interpretation of one set of edges, say $l^a$, will take the even 8-vertex model into the odd 8-vertex model and vice-versa. Since the underlying edge variables remain unchanged and only the interpretation of the bond values changes, the two models are equal under this transformation. Similarly, upon transforming to a staggered spin model, the spin interactions remain unchanged even if the bond interpretation changes.


By swapping the interpretation of the edge variables on a horizontal subset of edges, say $l^a$, we have the following mapping between column staggered lattices or bi-partite staggered lattices
\bat{4}
w_1 &\leftrightarrow v_7 \quad\quad\quad\quad & w_5 &\leftrightarrow v_4 \quad\quad\quad\quad & \bar{w}_1 &\leftrightarrow \bar{v}_5 \quad\quad\quad\quad & \bar{w}_5 &\leftrightarrow \bar{v}_1 \label{stagevenoddmap1a}\nonumber\\
w_2 &\leftrightarrow v_8 \quad\quad\quad\quad & w_6 &\leftrightarrow v_3 \quad\quad\quad\quad & \bar{w}_2 &\leftrightarrow \bar{v}_6 \quad\quad\quad\quad & \bar{w}_6 &\leftrightarrow \bar{v}_2 \nonumber\\
w_3 &\leftrightarrow v_6 \quad\quad\quad\quad & w_7 &\leftrightarrow v_1 \quad\quad\quad\quad & \bar{w}_3 &\leftrightarrow \bar{v}_8 \quad\quad\quad\quad & \bar{w}_7 &\leftrightarrow \bar{v}_4 \nonumber\\
w_4 &\leftrightarrow v_5 \quad\quad\quad\quad & w_8 &\leftrightarrow v_2 \quad\quad\quad\quad & \bar{w}_4 &\leftrightarrow \bar{v}_7 \quad\quad\quad\quad & \bar{w}_8 &\leftrightarrow \bar{v}_3 
\eat
while swapping the interpretation of the edge variables on a vertical subset of edges, say $l^c$, we have the following mapping between row staggered lattices or bi-partite staggered lattices
\bat{4}
w_1 &\leftrightarrow v_1 \quad\quad\quad\quad & w_5 &\leftrightarrow v_5 \quad\quad\quad\quad & \bar{w}_1 &\leftrightarrow \bar{v}_3 \quad\quad\quad\quad & \bar{w}_5 &\leftrightarrow \bar{v}_8 \nonumber\\
w_2 &\leftrightarrow v_2 \quad\quad\quad\quad & w_6 &\leftrightarrow v_6 \quad\quad\quad\quad & \bar{w}_2 &\leftrightarrow \bar{v}_4 \quad\quad\quad\quad & \bar{w}_6 &\leftrightarrow \bar{v}_7 \nonumber\\
w_3 &\leftrightarrow v_3 \quad\quad\quad\quad & w_7 &\leftrightarrow v_7 \quad\quad\quad\quad & \bar{w}_3 &\leftrightarrow \bar{v}_1 \quad\quad\quad\quad & \bar{w}_7 &\leftrightarrow \bar{v}_6 \nonumber\\
w_4 &\leftrightarrow v_4 \quad\quad\quad\quad & w_8 &\leftrightarrow v_8 \quad\quad\quad\quad & \bar{w}_4 &\leftrightarrow \bar{v}_2 \quad\quad\quad\quad & \bar{w}_8 &\leftrightarrow \bar{v}_5 \label{stagevenoddmap1b}
\eat
Furthermore, because it is arbitrary which unit cell is used for each staggered model, it is possible to interchange the $\bar{v}_i$ and $\bar{w}_i$ sets for each model independently, yielding 6 more mappings, for a total of 8. The bi-partite staggered lattice respects all 8 mappings while the column staggered or row staggered lattices only respects 4 of the mappings.  While these mappings are valid in general, we note in particular that in each of these mappings, the free-fermion condition in one model maps to the free-fermion condition in the other model. 

The column staggered even 8-vertex model can map to the homogeneous odd 8-vertex model with the following specialization
\bat{6}
w_1 &= \bar{w}_4 &= v_5 \quad\quad\quad & w_2 &= \bar{w}_3 &= v_6 \quad\quad\quad & w_3 &= \bar{w}_2 &= v_8 \quad\quad\quad & w_4 &= \bar{w}_1 &= v_7 
\nonumber\\
w_5 &= \bar{w}_7 &= v_1 \quad\quad\quad & w_6 &= \bar{w}_8 &= v_2 \quad\quad\quad & w_7 &= \bar{w}_5 &= v_4 \quad\quad\quad & w_8 &= \bar{w}_6 &= v_3
\eat
and the column staggered odd 8-vertex model can map to the homogeneous even 8-vertex model with the following specialization
\bat{6}
v_1 &= \bar{v}_4 &= w_5 \quad\quad\quad & v_2 &= \bar{v}_3 &= w_6 \quad\quad\quad & v_3 &= \bar{v}_2 &= w_8 \quad\quad\quad & v_4 &= \bar{v}_1 &= w_7 
\nonumber\\
v_5 &= \bar{v}_7 &= w_1 \quad\quad\quad & v_6 &= \bar{v}_8 &= w_2 \quad\quad\quad & v_7 &= \bar{v}_5 &= w_4 \quad\quad\quad & v_8 &= \bar{v}_6 &= w_3
\eat

Likewise, the row staggered even 8-vertex model can map to the homogeneous odd 8-vertex model with the following specialization
\bat{6}
w_1 &= \bar{w}_3 &= v_3 \quad\quad\quad & w_2 &= \bar{w}_4 &= v_4 \quad\quad\quad & w_3 &= \bar{w}_1 &= v_1 \quad\quad\quad & w_4 &= \bar{w}_2 &= v_2 
\nonumber\\
w_5 &= \bar{w}_8 &= v_8 \quad\quad\quad & w_6 &= \bar{w}_7 &= v_7 \quad\quad\quad & w_7 &= \bar{w}_6 &= v_6 \quad\quad\quad & w_8 &= \bar{w}_5 &= v_5
\eat
and the row staggered odd 8-vertex model can map to the homogeneous even 8-vertex model with the following specialization
\bat{6}
v_1 &= \bar{v}_3 &= w_3 \quad\quad\quad & v_2 &= \bar{v}_4 &= w_4 \quad\quad\quad & v_3 &= \bar{v}_1 &= w_1 \quad\quad\quad & v_4 &= \bar{v}_2 &= w_2 
\nonumber\\
v_5 &= \bar{v}_8 &= w_8 \quad\quad\quad & v_6 &= \bar{v}_7 &= w_7 \quad\quad\quad & v_7 &= \bar{v}_6 &= w_6 \quad\quad\quad & v_8 &= \bar{v}_5 &= w_5
\eat
For the bi-partite staggered either set of mappings above holds.

We note the following typos in~\cite{wu2004k} with regards to these mappings. The second line of the Theorem, in the first set of weights, $u_3$ and $u_4$ are swapped; see the second line of equation (10) of~\cite{wu2004k}. In equation (15) of~\cite{wu2004k}, $w'_7$ and $w'_8$ are swapped, $w'_5$ and $w'_6$ are swapped, and the bottom left equation should read $w_7=w'_6=u_7$; see the first line of equation (10).

\subsubsection{Mixed even and odd 8-vertex models}
In staggering the lattice it is also possible to consider having only even weights on one sublattice and only odd weights on the other sublattice. In this case, the model maps into itself when re-interpreting the edge variables on one set of edges. These mixed models can be seen as specializations of the more general staggered model with a staggering unit of 4 sites: by changing the interpretation of one or two edges per staggering unit, an all even or all odd model can be changed to a mixed even-odd column or bi-partite staggered model, respectively. In the free-fermion case, the column or bi-partite staggered mixed models can be specialized from the work on Lin and Wu in~\cite{lin1977w}.

\subsection{Mapping between column and bi-partite staggered free-fermion 8-vertex models}
In appendix~\ref{app:results} we give expressions for the free-energies of the even and odd column and bi-partite staggered free-fermion 8-vertex models. The even bi-partite staggered free-fermion model was analyzed in~\cite{hsue1975lw}. The phase transitions occur at the points given by
\beqr
-w_1\bar{w}_1-w_2\bar{w}_2+w_3\bar{w}_3+w_4\bar{w}_4+w_5\bar{w}_6+w_6\bar{w}_5+w_7\bar{w}_8+w_8\bar{w}_7&=&0 \\
w_1\bar{w}_1+w_2\bar{w}_2-w_3\bar{w}_3-w_4\bar{w}_4+w_5\bar{w}_6+w_6\bar{w}_5+w_7\bar{w}_8+w_8\bar{w}_7&=&0 \\
w_1\bar{w}_1+w_2\bar{w}_2+w_3\bar{w}_3+w_4\bar{w}_4-w_5\bar{w}_6-w_6\bar{w}_5+w_7\bar{w}_8+w_8\bar{w}_7&=&0 \\
w_1\bar{w}_1+w_2\bar{w}_2+w_3\bar{w}_3+w_4\bar{w}_4+w_5\bar{w}_6+w_6\bar{w}_5-w_7\bar{w}_8-w_8\bar{w}_7&=&0 
\eeqr
where each condition allows up to 3 phase transitions. At these points the model has second order, logarithmic phase transitions, except under certain subcases, where the phase transition has exponent $\alpha=1/2$ going to a frozen state below $T_c$~\cite{hsue1975lw}.

We have already given mappings between the even and odd cases of the staggered models in (\ref{stagevenoddmap1a})--(\ref{stagevenoddmap1b}). It is also possible to map the column and bi-partite staggered models to each other. The column staggered free-fermion model has not received much attention in the past. Looking at the integrands of the respective free-energy expressions in appendix~\ref{app:results}, we would like to set terms equal to each other. This can be accomplished if the $G$ term in the bi-partite staggered free-energy~(\ref{evenbstagint}) vanishes, along with the $E$ and $I$ terms for the column staggered free-energy~(\ref{evencstagint}), which correspond to
\beq
(v_3v_4-v_7v_8)=(\bar{v}_3\bar{v}_4-\bar{v}_7\bar{v}_8)=0 \label{biptocol1}
\eeq
for the odd case or 
\beq
(w_3w_4-w_7w_8)=(\bar{w}_3\bar{w}_4-\bar{w}_7\bar{w}_8)=0 \label{biptocol2}
\eeq
for the even case. These conditions are equivalent to temperature independent free-fermion conditions for each model. Alternatively, one could require the $F$ term in the bi-partite staggered free-energy~(\ref{evenbstagint}) and the $D$ and $H$ terms for the column staggered free-energy~(\ref{evencstagint}) to vanish, which corresponds to
\beq
(w_3w_4-w_5w_6)=(\bar{w}_3\bar{w}_4-\bar{w}_5\bar{w}_6)=0 \label{biptocol3}
\eeq
for the even case, and similarly for the odd case. Again this corresponds to the temperature independent free-fermion conditions of each model.

For the homogeneous model, all of the phase transition and disorder points happen only under temperature independent free-fermion conditions~\cite{hurst1963}. If that continues to be the case for staggered models, then there are no extra phase transition or disorder points in the bi-partite staggered model compared to the column staggered model that is missed in choosing~(\ref{biptocol1})--(\ref{biptocol2}) or (\ref{biptocol3}). Therefore, for the purposes of analyzing the column staggered model, we can use all of the knowledge from the bi-partite staggered analysis in~\cite{hsue1975lw}. Unfortunately a direct mapping of integrand terms does not appear to lead to enlightening expressions, so that we consider a direct approach below in order to study the critical points of the column staggered model. We point out also that the critical phenomena of both models are included as special cases of the model of Lin and Wu~\cite{lin1977w}, though the specialization to column staggering was not considered.

\subsection{Analysis of the column staggered free-fermion 8-vertex model}
We now look at the column staggered even free-fermion model, first considered in~\cite{tanaka1992m}, but whose phase transitions have not been previously analyzed. Following~\cite{green1964h,hurst1963}, we can find a set of singularities of the free-energy by setting $\{\theta_1,\theta_2\}=\{0,\pi\}$ in (\ref{evencstagint}), which leads to four phase transition conditions, given as follows
\beqr
(w_1+w_3)(\bar{w}_1+\bar{w}_3)\pm(w_2-w_4)(\bar{w}_2-\bar{w}_4) &=& 0 \\
(w_2+w_4)(\bar{w}_2+\bar{w}_4)\pm(w_1-w_3)(\bar{w}_1-\bar{w}_3) &=& 0 
\eeqr
While these conditions are implicit in~\cite{lin1977w}, they have not been detailed before. The only singularities which could possibly still occur would require strong constraints on the terms of the integrand, which at least in the case of the homogeneous free-fermion model leads to singularities only in the analytic continuation of the model~\cite{hurst1963}. A careful analysis of possible extra singularities which only obey the general free-fermion condition~(\ref{ffcondeven}) but not the temperature independent free-fermion condition~(\ref{ffindcondeven}) has never been performed for staggered free-fermion models or for free-fermion models on other lattices, as far as we are aware.

From the mapping of the column to bi-partite staggered free-fermion model outlined above, since the number of singularities in each case match, we can conclude that these phase transitions for the column staggered free-fermion model correspond to temperature-independent free-fermion conditions. Furthermore, we can conclude from the analysis of the bi-partite staggered free-fermion model in~\cite{hsue1975lw} that these phase transition points for the column staggered free-fermion model correspond to second order, logarithmic phase transitions, except under certain subcases, where the phase transition has exponent $\alpha=1/2$ going to a frozen state below $T_c$.

\section{Discussion} \label{sec:discussion}
\subsection{Boundary condition considerations}
Under toroidal boundary conditions, the number of horizontal and vertical arrow sources and sinks in the 16-vertex model must be equal, which gives nontrivial global constraints on the number of particular vertex weights $w_i$, $v_i$. Under free boundary conditions, however, there are no topological constraints on sources and sinks and there are no corresponding constraints on the vertex weights. In particular, domain boundary conditions could be considered for the odd 8-vertex model, analogous to those of the even 6-vertex model~\cite{brascamp1973kw}. A sufficient condition for boundary condition independence of the 16-vertex model in the thermodynamic limit, given in~\cite{brascamp1973kw}, is that 
\beq
w_i>0,\quad v_i\geq0,\qquad i=1\ldots8
\eeq
which does not constrain the odd 8-vertex model. We will note that in the case of close-packed dimers on the square lattice, which is a subcase of the odd 8-vertex model, exact results show boundary condition independence in the thermodynamic limit for free, cylindrical, and toroidal boundary conditions~\cite{kasteleyn1961,mccoy1973w,mccoy2014w}.

Under toroidal boundary conditions the partition function on the finite lattice is given by the sum of four Pfaffians~\cite{mccoy1973w,mccoy2014w}. In the thermodynamic limit, however, only a single Pfaffian is necessary to be considered, since the others are degenerate. In appendix~\ref{app:dimerconst} we only consider one Pfaffian in order to construct the free-energy solution in the thermodynamic limit, although the correct linear combination of four Pfaffians could be derived following the methods in~\cite{mccoy1973w,mccoy2014w}. It would presumably also be possible to solve the model on a finite cylinder, using the methods in~\cite{mccoy1967w2,mccoy1973w,mccoy2014w}, or perhaps also under free boundary conditions, as done recently for the Ising model~\cite{baxter2016}. 

As discussed in section~\ref{sec:ffmap}, the current full mapping~(\ref{freemap1a})--(\ref{freemap1b}) from the even to the odd free-fermion models is only valid in the thermodynamic limit, because on the finite torus a linear combination of four Pfaffians is necessary to give the partition function, so that the mapping is only approximate between the models. Furthermore, the mapping is between two lattice units in the odd model to one lattice unit in the even model. Both these issues are analogous to the mapping from the free-fermion model to the checkerboard lattice Ising model, where the mapping of~\cite{davies1987} is valid on the finite toroidal lattice, but the mapping of~\cite{baxter19862} is only valid in the thermodynamic limit and has an extra factor of 2 in the free energy compared to the free energy of~\cite{davies1987}. It would be convenient to find a boundary condition similar to that found by Brascamp and Kunz~\cite{brascamp1974k} for the square lattice Ising model, which would give the finite lattice partition functions of the even and odd free-fermion models in a single double product form, so that a finite lattice mapping can be defined. Perhaps also special fixed boundary conditions like in~\cite{baxter1976e} could be found to allow a direct mapping. We have not succeeded in finding such suitable boundary conditions.

\subsection{Free-fermion models}
While a mapping remains to be found between the general even and odd free-fermion models on the finite lattice, it holds in special cases, such as Baxter's map from close-packed dimers to the even free-fermion model in section~\ref{sec:dimermap}. Nevertheless, from the perspective of critical phenomena, only the thermodynamic limit is relevant, and they are fully equivalent in this limit, from~(\ref{freemap1a})--(\ref{freemap1b}). From the perspective of staggered lattices, the even and odd free-fermion conditions are fully equivalent, since their partition functions are fully equivalent. But using the weak-graph transformations in section~\ref{sec:weakgraph}, valid on the finite torus, we see that in mapping between the anti-symmetric even model and the symmetric odd homogeneous models, the general free-fermion condition in one model maps to the temperature-independent free-fermion condition in the other model, showing the close association between the free-fermion conditions~(\ref{ffcondeven}) and~(\ref{ffcondodd}) and the temperature-independent equivalents~(\ref{ffindcondeven}) and~(\ref{ffindcondodd}), respectively.  

We believe that it hasn't been so directly stated that because all of the phase transitions of the free-fermion model occur for temperature independent free-fermion conditions, all of the essential physics of the free-fermion model and the union-jack and checkerboard lattice Ising models can already be seen in the triangular lattice Ising model, contrary to expectations based on counting the number of interactions in these models. Moving to staggered lattices, the column and bi-partite square lattice staggered 8-vertex models have a richer phase transition structure than the homogeneous free-fermion model, but the homogeneous triangular lattice free-fermion model~\cite{sacco1975w,sacco1977w} appears to be a larger model compared to the square lattice staggered free-fermion models given in appendix~\ref{app:results}. It would appear that longer range, next-nearest-neighbor interactions define larger models than staggered shorter range nearest-neighbor models. It would also be interesting to compare the triangular free-fermion model with the union-jack or checkerboard free-fermion model, as well as to larger staggered Ising models.

\subsection{Disorder lines}
The study of disorder lines in vertex models is well developed, starting with Stephenson's characterization of the disorder point $T_D$ of the anisotropic antiferromagnetic triangular Ising model~\cite{stephenson1969,stephenson19702,stephenson19703, stephenson19704,stephenson19705}. This disorder point corresponds to the condition $\Omega^2=0$ in~(\ref{Omega2}) and can be mapped to the odd 8-vertex model using the free-energy free-fermion mapping~(\ref{freemap1a})--(\ref{freemap1b}). Peschel's disorder condition for the general even 8-vertex model~\cite{peschel1982r}, with $w_3=w_4$, $w_5=w_6$, $w_7=w_8$, is given as
\beq
2\,(w_3+w_5)-(w_1+w_2)=\pm\sqrt{4\,w_7^2+(w_1-w_2)^2}
\eeq
Under the free-fermion condition~(\ref{ffcondeven}), this expression can be translated to the odd free-fermion model using~(\ref{freemap1a})--(\ref{freemap1b}), but for the general odd 8-vertex model it cannot be done, since the mappings in section~\ref{sec:firstmapping} from the even to the odd 8-vertex model require an anti-symmetric even 8-vertex model, which violates Peschel's starting assumptions.
Other disorder conditions are given in~\cite{rujan19822,diep1992dg} for the even 8-vertex model, and in~\cite{hansel1987jm} for the staggered symmetric even 8-vertex model and the 16-vertex model. In~\cite{rujan1984}, disorder conditions are given for the general 8-vertex model, the staggered 8-vertex model, which does not have a disorder point for all positive weights, and the general 16-vertex model. Rae~\cite{rae19732} gives the following disorder conditions for the general 16-vertex model, for an arbitrary real number $z$ satisfying the following
\beqr
& w_i,~v_i \neq 0,\quad i=1\ldots8 \nonumber\\
& v_7\,z^2+(w_4-w_1)z-v_5 = 0 \nonumber\\
& v_6\,z^2+(w_2-w_3)z -v_8 = 0 \nonumber\\
& w_7\,z^2+(v_4-v_1)z -w_5 = 0 \nonumber\\
& w_6\,z^2+(v_2-v_3)z-w_8 = 0 
\eeqr
or else the condition
\beq
w_5=w_8=v_5=v_8=0
\eeq
or the condition
\beq
w_6=w_7=v_6=v_7=0
\eeq
In all these cases, the disorder conditions correspond to a dimensional reduction of the models.

\section{Conclusions} \label{sec:conclusions}
We have collected together and expanded many mappings of the 16-vertex model, in particular between the even and the odd 8-vertex models, and have given new mappings as well. We have also collected and expanded many transformations of the 16-vertex model, giving for the first time a complete list of partition function symmetries in appendix~\ref{app:parttrans}, and giving for the first time variable transformations which bring out the dependence of the partition function on bond states in (\ref{partbondvars}). We have also enlarged the number of known weak-graph transformation matrices in section~\ref{sec:weakgraph}. 

Historically mappings have allowed certain models to be solved, by relating a model to a known exact solution. We have also seen in sections~\ref{weakgraphtrans}, \ref{sec:alginvs}, \ref{sec:ffmap} many examples of models with negative weights being mapped to models with only positive weights, showing the physicality of seemingly unphysical models. Further, work on seemingly different models can be simplified once a mapping is provided. For example, the work on special cases of the symmetric 16-vertex model in~\cite{stilck1983,stilck1984os,ahn1994hs,wang2006nflmclslcs,wang2007nflmclslcs,morgan2010slm,levis2012c, morgan2013aspelm,morgan2013bslm,levis2013c,levis2013cft,foini2013ltc} can all be related via the weak-graph transformation~(\ref{weakgraphtrans}) to the even 8-vertex model~\cite{wu1972}. In particular, the models in~\cite{stilck1983,stilck1984os} map to the general even 8-vertex model with $w_5=w_6=w_7=w_8$, which has recently been studied in~\cite{krcmar2016s}, the models in~\cite{levis2012c,levis2013c,levis2013cft,foini2013ltc} map to the general even 8-vertex model with $w_3=w_4$, $w_5=w_6$, $w_7=w_8$, and the models in~\cite{wang2006nflmclslcs,wang2007nflmclslcs,morgan2010slm,morgan2013aspelm,morgan2013bslm} map to the general even 8-vertex model with $w_3=w_4$, $w_5=w_6=-w_7=w_8$. The models solved in~\cite{ahn1994hs} map, via the weak-graph transformation~(\ref{weakgraphtrans}) and with the partition function symmetries of appendix~\ref{app:parttrans}, to particular cases of the general 6-vertex model; their candidate points for integrability are particular cases of the general 8-vertex model with $w_3=w_4$, $w_5=w_6$, $w_7=w_8$. Also, in~\cite{garel1983m} it is shown that a mapping can be found to take a fully frustrated model with critical point at $T_c=0$ to an equivalent model with a finite temperature critical point. It would be interesting to apply this technique to the odd 8-vertex model.

We have made use for the first time of the theory of algebraic invariants of the 16-vertex model as given by Gaaff and Hijmans~\cite{gaaff1975h,gaaff1976h}, although we have found that using the invariants to construct mappings between models leads to only sufficient conditions. Through other mappings, such as the free-energy free-fermion mapping in section~\ref{sec:ffmap} and Baxter's mapping in section~\ref{sec:dimermap}, we see that mappings can be found which do not respect the algebraic invariants of the model, even locally on the finite toroidal lattice. Consequently, the set of algebraic invariants for a model and their relationships do not fully specify a model's equivalency class. We have used the algebraic invariants, however, to find new mappings in section~\ref{sec:alginvs} between the even and odd 8-vertex models, valid on finite toroidal lattices, as well as a new mapping between the square lattice Ising model in a field at $H=i\pi/(2\beta)$ and the odd 8-vertex model in section~\ref{sec:isingfield}. At least with regards to the algebraic invariants of the Ising model in a field, mapped to the 16-vertex model via (\ref{isinginafield}) or (\ref{isinginafield2}), we have found the complete list of complex magnetic field points which can be mapped to the even or odd 8-vertex models,  which also automatically correspond to free-fermionic points of the general model. Unless another mapping from the Ising model in a field to the 16-vertex model is found or else another means of specializing the 16-vertex model to the even or odd 8-vertex models is found, our list of free-fermionic points, also corresponding to 8-vertex model points, of the Ising model in a field is complete.

We have given complete derivation information for the dimer constructions of the solutions to the even and odd staggered free-fermion 8-vertex models in appendix~\ref{app:dimerconst}, from which the homogeneous results can be specialized. These are new solutions to the odd 8-vertex free-fermion model, valid on finite lattices, since the only previously known solution was via the free-energy mapping of~(\ref{freemap1a})--(\ref{freemap1b}), which is only valid in the thermodynamic limit.

On the staggered lattices, we consider a new mapping in section~\ref{sec:stag} between the column and bi-partite staggered free-fermion 8-vertex models, valid under temperature-independent free-fermion conditions for each model. In analogy with the homogeneous models, we conjecture that no critical phenomena is missed by choosing temperature independent free-fermion conditions in the mapping; for homogeneous free-fermion models the general and temperature independent free-fermion conditions are interlinked, as explained in sections~\ref{sec:ff}, \ref{sec:weakgraph}, and~\ref{sec:alginvs}. Analyzing the column staggered model separately for the first time, we have provided new details of the phase transitions of the model, which was only implicit in~\cite{lin1977w}. A proof equivalent to that in~\cite{hurst1963} that no further singularities occur in staggered free-fermion models beyond the temperature independent free-fermion conditions would complete the analysis of these models.

\section*{Acnowledgements}
We gratefully acknowledge numerous helpful conversations with Nathan Clisby and Iwan Jensen during the preparation of this work. We would like to thank the Australian Research Council
 for supporting this work under the Discovery Project scheme (project number DP140101110).

\newpage
\appendix
\section{Notation conventions}\label{app:notation}
We collect various notation conventions used in the literature for the even and odd weights in tables~\ref{tableeven} and \ref{tableodd}, respectively.
\begin{table}[htpb]
\begin{center}
\begin{tabular}{|cccccccc|c|}
\hline
$w_1$ & $w_2$ & $w_3$ & $w_4$ & $w_5$ & $w_6$ & $w_7$ & $w_8$ & References \\\hline
\begin{tikzpicture} \node at (0,-1) {}; \node at (0,1) {}; \draw[line width = 1pt, dashed] (0,0)--(-0.5,0); \draw[line width = 1pt, dashed] (0,0)--(0.5,0); \draw[line width = 1pt, dashed] (0,0)--(0,0.5); \draw[line width = 1pt, dashed] (0,0)--(0,-0.5); \end{tikzpicture} &
\begin{tikzpicture} \node at (0,-1) {}; \node at (0,1) {}; \draw[line width = 1pt] (0,0)--(-0.5,0); \draw[line width = 1pt] (0,0)--(0.5,0); \draw[line width = 1pt] (0,0)--(0,0.5); \draw[line width = 1pt] (0,0)--(0,-0.5); \end{tikzpicture} &
\begin{tikzpicture} \node at (0,-1) {}; \node at (0,1) {}; \draw[line width = 1pt, dashed] (0,0)--(-0.5,0); \draw[line width = 1pt, dashed] (0,0)--(0.5,0); \draw[line width = 1pt] (0,0)--(0,0.5); \draw[line width = 1pt] (0,0)--(0,-0.5); \end{tikzpicture} &
\begin{tikzpicture} \node at (0,-1) {}; \node at (0,1) {}; \draw[line width = 1pt] (0,0)--(-0.5,0); \draw[line width = 1pt] (0,0)--(0.5,0); \draw[line width = 1pt, dashed] (0,0)--(0,0.5); \draw[line width = 1pt, dashed] (0,0)--(0,-0.5); \end{tikzpicture} &
\begin{tikzpicture} \node at (0,-1) {}; \node at (0,1) {}; \draw[line width = 1pt, dashed] (0,0)--(-0.5,0); \draw[line width = 1pt] (0,0)--(0.5,0); \draw[line width = 1pt, dashed] (0,0)--(0,0.5); \draw[line width = 1pt] (0,0)--(0,-0.5); \end{tikzpicture} &
\begin{tikzpicture} \node at (0,-1) {}; \node at (0,1) {}; \draw[line width = 1pt] (0,0)--(-0.5,0); \draw[line width = 1pt, dashed] (0,0)--(0.5,0); \draw[line width = 1pt] (0,0)--(0,0.5); \draw[line width = 1pt, dashed] (0,0)--(0,-0.5); \end{tikzpicture} &
\begin{tikzpicture} \node at (0,-1) {}; \node at (0,1) {}; \draw[line width = 1pt] (0,0)--(-0.5,0); \draw[line width = 1pt, dashed] (0,0)--(0.5,0); \draw[line width = 1pt, dashed] (0,0)--(0,0.5); \draw[line width = 1pt] (0,0)--(0,-0.5); \end{tikzpicture} &
\begin{tikzpicture} \node at (0,-1) {}; \node at (0,1) {}; \draw[line width = 1pt, dashed] (0,0)--(-0.5,0); \draw[line width = 1pt] (0,0)--(0.5,0); \draw[line width = 1pt] (0,0)--(0,0.5); \draw[line width = 1pt, dashed] (0,0)--(0,-0.5); \end{tikzpicture} &
\begin{tikzpicture} \node at (0,-1) {}; \node at (0,1) {}; \node at (0,0.25) {\cite{wu1969,suzuki1971f,temperley1971l,lieb1972w,wu1972,rae1973}}; \node at (0,-0.25) {\cite{hsue1975lw}\footnotemark[1],\cite{gaaff1974,gaaff1975h,stilck1983,wu1989w,wu2004k}\footnotemark[1]};
\end{tikzpicture}  \\\hline
\begin{tikzpicture} \node at (0,-1) {}; \node at (0,1) {}; \draw[line width = 1pt, dashed] (0,0)--(-0.5,0);
\draw[line width = 1pt, dashed] (0,0)--(0.5,0); \draw[line width = 1pt, dashed] (0,0)--(0,0.5); \draw[line width = 1pt, dashed] (0,0)--(0,-0.5); \end{tikzpicture} &
\begin{tikzpicture} \node at (0,-1) {}; \node at (0,1) {}; \draw[line width = 1pt] (0,0)--(-0.5,0); \draw[line width = 1pt] (0,0)--(0.5,0); \draw[line width = 1pt] (0,0)--(0,0.5); \draw[line width = 1pt] (0,0)--(0,-0.5); \end{tikzpicture} &
\begin{tikzpicture} \node at (0,-1) {}; \node at (0,1) {}; \draw[line width = 1pt, dashed] (0,0)--(-0.5,0); \draw[line width = 1pt, dashed] (0,0)--(0.5,0); \draw[line width = 1pt] (0,0)--(0,0.5); \draw[line width = 1pt] (0,0)--(0,-0.5); \end{tikzpicture} &
\begin{tikzpicture} \node at (0,-1) {}; \node at (0,1) {}; \draw[line width = 1pt] (0,0)--(-0.5,0); \draw[line width = 1pt] (0,0)--(0.5,0); \draw[line width = 1pt, dashed] (0,0)--(0,0.5); \draw[line width = 1pt, dashed] (0,0)--(0,-0.5); \end{tikzpicture} &
\begin{tikzpicture} \node at (0,-1) {}; \node at (0,1) {}; \draw[line width = 1pt, dashed] (0,0)--(-0.5,0); \draw[line width = 1pt] (0,0)--(0.5,0); \draw[line width = 1pt, dashed] (0,0)--(0,0.5); \draw[line width = 1pt] (0,0)--(0,-0.5); \end{tikzpicture} &
\begin{tikzpicture} \node at (0,-1) {}; \node at (0,1) {}; \draw[line width = 1pt] (0,0)--(-0.5,0); \draw[line width = 1pt, dashed] (0,0)--(0.5,0); \draw[line width = 1pt] (0,0)--(0,0.5); \draw[line width = 1pt, dashed] (0,0)--(0,-0.5); \end{tikzpicture} &
\begin{tikzpicture} \node at (0,-1) {}; \node at (0,1) {}; \draw[line width = 1pt, dashed] (0,0)--(-0.5,0); \draw[line width = 1pt] (0,0)--(0.5,0); \draw[line width = 1pt] (0,0)--(0,0.5); \draw[line width = 1pt, dashed] (0,0)--(0,-0.5); \end{tikzpicture} &
\begin{tikzpicture} \node at (0,-1) {}; \node at (0,1) {}; \draw[line width = 1pt] (0,0)--(-0.5,0); \draw[line width = 1pt, dashed] (0,0)--(0.5,0); \draw[line width = 1pt, dashed] (0,0)--(0,0.5); \draw[line width = 1pt] (0,0)--(0,-0.5); \end{tikzpicture} &
\begin{tikzpicture} \node at (0,-1) {}; \node at (0,1) {}; \node at (0,0.25) {\cite{baxter1971,baxter1972,felderhof1973,felderhof19732,felderhof19733, baxter1982}}; \node at (0,-0.25) {\cite{baxter19862,wu1987l,baxter2007,tracy1985,tracy1987}}; \end{tikzpicture}  \\\hline
\begin{tikzpicture} \node at (0,-1) {}; \node at (0,1) {}; \draw[line width = 1pt, dashed] (0,0)--(-0.5,0);
\draw[line width = 1pt, dashed] (0,0)--(0.5,0); \draw[line width = 1pt, dashed] (0,0)--(0,0.5); \draw[line width = 1pt, dashed] (0,0)--(0,-0.5); \end{tikzpicture} &
\begin{tikzpicture} \node at (0,-1) {}; \node at (0,1) {}; \draw[line width = 1pt] (0,0)--(-0.5,0); \draw[line width = 1pt] (0,0)--(0.5,0); \draw[line width = 1pt] (0,0)--(0,0.5); \draw[line width = 1pt] (0,0)--(0,-0.5); \end{tikzpicture} &
\begin{tikzpicture} \node at (0,-1) {}; \node at (0,1) {}; \draw[line width = 1pt, dashed] (0,0)--(-0.5,0); \draw[line width = 1pt, dashed] (0,0)--(0.5,0); \draw[line width = 1pt] (0,0)--(0,0.5); \draw[line width = 1pt] (0,0)--(0,-0.5); \end{tikzpicture} &
\begin{tikzpicture} \node at (0,-1) {}; \node at (0,1) {}; \draw[line width = 1pt] (0,0)--(-0.5,0); \draw[line width = 1pt] (0,0)--(0.5,0); \draw[line width = 1pt, dashed] (0,0)--(0,0.5); \draw[line width = 1pt, dashed] (0,0)--(0,-0.5); \end{tikzpicture} &
\begin{tikzpicture} \node at (0,-1) {}; \node at (0,1) {}; \draw[line width = 1pt] (0,0)--(-0.5,0); \draw[line width = 1pt, dashed] (0,0)--(0.5,0); \draw[line width = 1pt] (0,0)--(0,0.5); \draw[line width = 1pt, dashed] (0,0)--(0,-0.5); \end{tikzpicture} &
\begin{tikzpicture} \node at (0,-1) {}; \node at (0,1) {}; \draw[line width = 1pt, dashed] (0,0)--(-0.5,0); \draw[line width = 1pt] (0,0)--(0.5,0); \draw[line width = 1pt, dashed] (0,0)--(0,0.5); \draw[line width = 1pt] (0,0)--(0,-0.5); \end{tikzpicture} &
\begin{tikzpicture} \node at (0,-1) {}; \node at (0,1) {}; \draw[line width = 1pt, dashed] (0,0)--(-0.5,0); \draw[line width = 1pt] (0,0)--(0.5,0); \draw[line width = 1pt] (0,0)--(0,0.5); \draw[line width = 1pt, dashed] (0,0)--(0,-0.5); \end{tikzpicture} &
\begin{tikzpicture} \node at (0,-1) {}; \node at (0,1) {}; \draw[line width = 1pt] (0,0)--(-0.5,0); \draw[line width = 1pt, dashed] (0,0)--(0.5,0); \draw[line width = 1pt, dashed] (0,0)--(0,0.5); \draw[line width = 1pt] (0,0)--(0,-0.5); \end{tikzpicture} &
\begin{tikzpicture} \node at (0,-1) {}; \node at (0,1) {}; \node at (0,0) {\cite{fan1969w,fan1970w},\cite{hsue1975lw}\footnotemark[1],\cite{lin1977w,tanaka1992m,wu2004k}\footnotemark[1]}; \end{tikzpicture}  \\\hline
\begin{tikzpicture} \node at (0,-1) {}; \node at (0,1) {}; \draw[line width = 1pt] (0,0)--(-0.5,0);
\draw[line width = 1pt] (0,0)--(0.5,0); \draw[line width = 1pt] (0,0)--(0,0.5); \draw[line width = 1pt] (0,0)--(0,-0.5); \end{tikzpicture} &
\begin{tikzpicture} \node at (0,-1) {}; \node at (0,1) {}; \draw[line width = 1pt, dashed] (0,0)--(-0.5,0); \draw[line width = 1pt, dashed] (0,0)--(0.5,0); \draw[line width = 1pt, dashed] (0,0)--(0,0.5); \draw[line width = 1pt, dashed] (0,0)--(0,-0.5); \end{tikzpicture} &
\begin{tikzpicture} \node at (0,-1) {}; \node at (0,1) {}; \draw[line width = 1pt, dashed] (0,0)--(-0.5,0); \draw[line width = 1pt, dashed] (0,0)--(0.5,0); \draw[line width = 1pt] (0,0)--(0,0.5); \draw[line width = 1pt] (0,0)--(0,-0.5); \end{tikzpicture} &
\begin{tikzpicture} \node at (0,-1) {}; \node at (0,1) {}; \draw[line width = 1pt] (0,0)--(-0.5,0); \draw[line width = 1pt] (0,0)--(0.5,0); \draw[line width = 1pt, dashed] (0,0)--(0,0.5); \draw[line width = 1pt, dashed] (0,0)--(0,-0.5); \end{tikzpicture} &
\begin{tikzpicture} \node at (0,-1) {}; \node at (0,1) {}; \draw[line width = 1pt, dashed] (0,0)--(-0.5,0); \draw[line width = 1pt] (0,0)--(0.5,0); \draw[line width = 1pt, dashed] (0,0)--(0,0.5); \draw[line width = 1pt] (0,0)--(0,-0.5); \end{tikzpicture} &
\begin{tikzpicture} \node at (0,-1) {}; \node at (0,1) {}; \draw[line width = 1pt] (0,0)--(-0.5,0); \draw[line width = 1pt, dashed] (0,0)--(0.5,0); \draw[line width = 1pt] (0,0)--(0,0.5); \draw[line width = 1pt, dashed] (0,0)--(0,-0.5); \end{tikzpicture} &
\begin{tikzpicture} \node at (0,-1) {}; \node at (0,1) {}; \draw[line width = 1pt] (0,0)--(-0.5,0); \draw[line width = 1pt, dashed] (0,0)--(0.5,0); \draw[line width = 1pt, dashed] (0,0)--(0,0.5); \draw[line width = 1pt] (0,0)--(0,-0.5); \end{tikzpicture} &
\begin{tikzpicture} \node at (0,-1) {}; \node at (0,1) {}; \draw[line width = 1pt, dashed] (0,0)--(-0.5,0); \draw[line width = 1pt] (0,0)--(0.5,0); \draw[line width = 1pt] (0,0)--(0,0.5); \draw[line width = 1pt, dashed] (0,0)--(0,-0.5); \end{tikzpicture} &
\begin{tikzpicture} \node at (0,-1) {}; \node at (0,1) {}; \node at (0,0) {\cite{bazhanov1985s,bazhanov1985s2,bazhanov1985s3}}; \end{tikzpicture}  \\\hline

\begin{tikzpicture} \node at (0,-1) {}; \node at (0,1) {}; \draw[line width = 1pt, dashed] (0,0)--(-0.5,0);
\draw[line width = 1pt, dashed] (0,0)--(0.5,0); \draw[line width = 1pt, dashed] (0,0)--(0,0.5); \draw[line width = 1pt, dashed] (0,0)--(0,-0.5); \end{tikzpicture} &
\begin{tikzpicture} \node at (0,-1) {}; \node at (0,1) {}; \draw[line width = 1pt] (0,0)--(-0.5,0); \draw[line width = 1pt] (0,0)--(0.5,0); \draw[line width = 1pt] (0,0)--(0,0.5); \draw[line width = 1pt] (0,0)--(0,-0.5); \end{tikzpicture} &
\begin{tikzpicture} \node at (0,-1) {}; \node at (0,1) {}; \draw[line width = 1pt] (0,0)--(-0.5,0); \draw[line width = 1pt] (0,0)--(0.5,0); \draw[line width = 1pt, dashed] (0,0)--(0,0.5); \draw[line width = 1pt, dashed] (0,0)--(0,-0.5); \end{tikzpicture} &
\begin{tikzpicture} \node at (0,-1) {}; \node at (0,1) {}; \draw[line width = 1pt, dashed] (0,0)--(-0.5,0); \draw[line width = 1pt, dashed] (0,0)--(0.5,0); \draw[line width = 1pt] (0,0)--(0,0.5); \draw[line width = 1pt] (0,0)--(0,-0.5); \end{tikzpicture} &
\begin{tikzpicture} \node at (0,-1) {}; \node at (0,1) {}; \draw[line width = 1pt] (0,0)--(-0.5,0); \draw[line width = 1pt, dashed] (0,0)--(0.5,0); \draw[line width = 1pt] (0,0)--(0,0.5); \draw[line width = 1pt, dashed] (0,0)--(0,-0.5); \end{tikzpicture} &
\begin{tikzpicture} \node at (0,-1) {}; \node at (0,1) {}; \draw[line width = 1pt, dashed] (0,0)--(-0.5,0); \draw[line width = 1pt] (0,0)--(0.5,0); \draw[line width = 1pt, dashed] (0,0)--(0,0.5); \draw[line width = 1pt] (0,0)--(0,-0.5); \end{tikzpicture} &
\begin{tikzpicture} \node at (0,-1) {}; \node at (0,1) {}; \draw[line width = 1pt, dashed] (0,0)--(-0.5,0); \draw[line width = 1pt] (0,0)--(0.5,0); \draw[line width = 1pt] (0,0)--(0,0.5); \draw[line width = 1pt, dashed] (0,0)--(0,-0.5); \end{tikzpicture} &
\begin{tikzpicture} \node at (0,-1) {}; \node at (0,1) {}; \draw[line width = 1pt] (0,0)--(-0.5,0); \draw[line width = 1pt, dashed] (0,0)--(0.5,0); \draw[line width = 1pt, dashed] (0,0)--(0,0.5); \draw[line width = 1pt] (0,0)--(0,-0.5); \end{tikzpicture} &
\begin{tikzpicture} \node at (0,-1) {}; \node at (0,1) {}; \node at (0,0) {\cite{jaekel1983m}}; \end{tikzpicture}  \\\hline
\begin{tikzpicture} \node at (0,-1) {}; \node at (0,1) {}; \draw[line width = 1pt, dashed] (0,0)--(-0.5,0);
\draw[line width = 1pt, dashed] (0,0)--(0.5,0); \draw[line width = 1pt, dashed] (0,0)--(0,0.5); \draw[line width = 1pt, dashed] (0,0)--(0,-0.5); \end{tikzpicture} &
\begin{tikzpicture} \node at (0,-1) {}; \node at (0,1) {}; \draw[line width = 1pt] (0,0)--(-0.5,0); \draw[line width = 1pt] (0,0)--(0.5,0); \draw[line width = 1pt] (0,0)--(0,0.5); \draw[line width = 1pt] (0,0)--(0,-0.5); \end{tikzpicture} &
\begin{tikzpicture} \node at (0,-1) {}; \node at (0,1) {}; \draw[line width = 1pt] (0,0)--(-0.5,0); \draw[line width = 1pt] (0,0)--(0.5,0); \draw[line width = 1pt, dashed] (0,0)--(0,0.5); \draw[line width = 1pt, dashed] (0,0)--(0,-0.5); \end{tikzpicture} &
\begin{tikzpicture} \node at (0,-1) {}; \node at (0,1) {}; \draw[line width = 1pt, dashed] (0,0)--(-0.5,0); \draw[line width = 1pt, dashed] (0,0)--(0.5,0); \draw[line width = 1pt] (0,0)--(0,0.5); \draw[line width = 1pt] (0,0)--(0,-0.5); \end{tikzpicture} &
\begin{tikzpicture} \node at (0,-1) {}; \node at (0,1) {}; \draw[line width = 1pt] (0,0)--(-0.5,0); \draw[line width = 1pt, dashed] (0,0)--(0.5,0); \draw[line width = 1pt] (0,0)--(0,0.5); \draw[line width = 1pt, dashed] (0,0)--(0,-0.5); \end{tikzpicture} &
\begin{tikzpicture} \node at (0,-1) {}; \node at (0,1) {}; \draw[line width = 1pt, dashed] (0,0)--(-0.5,0); \draw[line width = 1pt] (0,0)--(0.5,0); \draw[line width = 1pt, dashed] (0,0)--(0,0.5); \draw[line width = 1pt] (0,0)--(0,-0.5); \end{tikzpicture} &
\begin{tikzpicture} \node at (0,-1) {}; \node at (0,1) {}; \draw[line width = 1pt] (0,0)--(-0.5,0); \draw[line width = 1pt, dashed] (0,0)--(0.5,0); \draw[line width = 1pt, dashed] (0,0)--(0,0.5); \draw[line width = 1pt] (0,0)--(0,-0.5); \end{tikzpicture} &
\begin{tikzpicture} \node at (0,-1) {}; \node at (0,1) {}; \draw[line width = 1pt, dashed] (0,0)--(-0.5,0); \draw[line width = 1pt] (0,0)--(0.5,0); \draw[line width = 1pt] (0,0)--(0,0.5); \draw[line width = 1pt, dashed] (0,0)--(0,-0.5); \end{tikzpicture} &
\begin{tikzpicture} \node at (0,-1) {}; \node at (0,1) {}; \node at (0,0) {\cite{hsue1975lw}\footnotemark[1],\cite{wu2004k}\footnotemark[1]}; \end{tikzpicture}  \\\hline
\end{tabular}
\end{center}
\caption{Notation conventions for the even 8-vertex model weights. We use the first convention. \label{tableeven}}
\end{table}
\footnotetext[1]{In the derivation in the appendix of~\cite{hsue1975lw}, it is clear from their Figure 4 that they are using the previous convention of~\cite{fan1970w} and not the convention shown in Figure 1 of~\cite{hsue1975lw}. Also, in carrying out their derivation we see that in the quoted free-energy an angle transformation $\theta\to-\theta$ has been performed, which is equivalent to a further change of notation, swapping $w_3$ and $w_4$ plus swapping $w_7$ and $w_8$. Therefore, in practice the notation of~\cite{hsue1975lw} corresponds to the last entry of table~\ref{tableeven}. The results of~\cite{wu2004k} use the results of~\cite{hsue1975lw} in contrast to Figure 2 of~\cite{wu2004k}.\label{evenfootnote}}

\begin{savenotes}
\begin{table}[htpb]
\begin{center}
\begin{tabular}{|cccccccc|c|}
\hline
$v_1$ & $v_2$ & $v_3$ & $v_4$ & $v_5$ & $v_6$ & $v_7$ & $v_8$ & References \\\hline
\begin{tikzpicture} \node at (0,-1) {}; \node at (0,1) {}; \draw[line width = 1pt, dashed] (0,0)--(-0.5,0); \draw[line width = 1pt, dashed] (0,0)--(0.5,0); \draw[line width = 1pt, dashed] (0,0)--(0,0.5); \draw[line width = 1pt] (0,0)--(0,-0.5); \end{tikzpicture} &
\begin{tikzpicture} \node at (0,-1) {}; \node at (0,1) {}; \draw[line width = 1pt] (0,0)--(-0.5,0); \draw[line width = 1pt] (0,0)--(0.5,0); \draw[line width = 1pt] (0,0)--(0,0.5); \draw[line width = 1pt, dashed] (0,0)--(0,-0.5); \end{tikzpicture} &
\begin{tikzpicture} \node at (0,-1) {}; \node at (0,1) {}; \draw[line width = 1pt, dashed] (0,0)--(-0.5,0); \draw[line width = 1pt, dashed] (0,0)--(0.5,0); \draw[line width = 1pt] (0,0)--(0,0.5); \draw[line width = 1pt, dashed] (0,0)--(0,-0.5); \end{tikzpicture} &
\begin{tikzpicture} \node at (0,-1) {}; \node at (0,1) {}; \draw[line width = 1pt] (0,0)--(-0.5,0); \draw[line width = 1pt] (0,0)--(0.5,0); \draw[line width = 1pt, dashed] (0,0)--(0,0.5); \draw[line width = 1pt] (0,0)--(0,-0.5); \end{tikzpicture} &
\begin{tikzpicture} \node at (0,-1) {}; \node at (0,1) {}; \draw[line width = 1pt, dashed] (0,0)--(-0.5,0); \draw[line width = 1pt] (0,0)--(0.5,0); \draw[line width = 1pt, dashed] (0,0)--(0,0.5); \draw[line width = 1pt, dashed] (0,0)--(0,-0.5); \end{tikzpicture} &
\begin{tikzpicture} \node at (0,-1) {}; \node at (0,1) {}; \draw[line width = 1pt] (0,0)--(-0.5,0); \draw[line width = 1pt, dashed] (0,0)--(0.5,0); \draw[line width = 1pt] (0,0)--(0,0.5); \draw[line width = 1pt] (0,0)--(0,-0.5); \end{tikzpicture} &
\begin{tikzpicture} \node at (0,-1) {}; \node at (0,1) {}; \draw[line width = 1pt] (0,0)--(-0.5,0); \draw[line width = 1pt, dashed] (0,0)--(0.5,0); \draw[line width = 1pt, dashed] (0,0)--(0,0.5); \draw[line width = 1pt, dashed] (0,0)--(0,-0.5); \end{tikzpicture} &
\begin{tikzpicture} \node at (0,-1) {}; \node at (0,1) {}; \draw[line width = 1pt, dashed] (0,0)--(-0.5,0); \draw[line width = 1pt] (0,0)--(0.5,0); \draw[line width = 1pt] (0,0)--(0,0.5); \draw[line width = 1pt] (0,0)--(0,-0.5); \end{tikzpicture} &
\begin{tikzpicture} \node at (0,-1) {}; \node at (0,1) {}; \node at (0,0) {\cite{suzuki1971f,wu2004k}${}^1$}; \end{tikzpicture}  \\\hline
\begin{tikzpicture} \node at (0,-1) {}; \node at (0,1) {}; \draw[line width = 1pt, dashed] (0,0)--(-0.5,0); \draw[line width = 1pt, dashed] (0,0)--(0.5,0); \draw[line width = 1pt, dashed] (0,0)--(0,0.5); \draw[line width = 1pt] (0,0)--(0,-0.5); \end{tikzpicture} &
\begin{tikzpicture} \node at (0,-1) {}; \node at (0,1) {}; \draw[line width = 1pt] (0,0)--(-0.5,0); \draw[line width = 1pt, dashed] (0,0)--(0.5,0); \draw[line width = 1pt, dashed] (0,0)--(0,0.5); \draw[line width = 1pt, dashed] (0,0)--(0,-0.5); \end{tikzpicture} &
\begin{tikzpicture} \node at (0,-1) {}; \node at (0,1) {}; \draw[line width = 1pt, dashed] (0,0)--(-0.5,0); \draw[line width = 1pt, dashed] (0,0)--(0.5,0); \draw[line width = 1pt] (0,0)--(0,0.5); \draw[line width = 1pt, dashed] (0,0)--(0,-0.5); \end{tikzpicture} &
\begin{tikzpicture} \node at (0,-1) {}; \node at (0,1) {}; \draw[line width = 1pt, dashed] (0,0)--(-0.5,0); \draw[line width = 1pt] (0,0)--(0.5,0); \draw[line width = 1pt, dashed] (0,0)--(0,0.5); \draw[line width = 1pt, dashed] (0,0)--(0,-0.5); \end{tikzpicture} &
\begin{tikzpicture} \node at (0,-1) {}; \node at (0,1) {}; \draw[line width = 1pt] (0,0)--(-0.5,0); \draw[line width = 1pt] (0,0)--(0.5,0); \draw[line width = 1pt] (0,0)--(0,0.5); \draw[line width = 1pt, dashed] (0,0)--(0,-0.5); \end{tikzpicture} &
\begin{tikzpicture} \node at (0,-1) {}; \node at (0,1) {}; \draw[line width = 1pt, dashed] (0,0)--(-0.5,0); \draw[line width = 1pt] (0,0)--(0.5,0); \draw[line width = 1pt] (0,0)--(0,0.5); \draw[line width = 1pt] (0,0)--(0,-0.5); \end{tikzpicture} &
\begin{tikzpicture} \node at (0,-1) {}; \node at (0,1) {}; \draw[line width = 1pt] (0,0)--(-0.5,0); \draw[line width = 1pt] (0,0)--(0.5,0); \draw[line width = 1pt, dashed] (0,0)--(0,0.5); \draw[line width = 1pt] (0,0)--(0,-0.5); \end{tikzpicture} &
\begin{tikzpicture} \node at (0,-1) {}; \node at (0,1) {}; \draw[line width = 1pt] (0,0)--(-0.5,0); \draw[line width = 1pt, dashed] (0,0)--(0.5,0); \draw[line width = 1pt] (0,0)--(0,0.5); \draw[line width = 1pt] (0,0)--(0,-0.5); \end{tikzpicture} &
\begin{tikzpicture} \node at (0,-1) {}; \node at (0,1) {}; \node at (0,0) {\cite{lieb1972w,rae1973,stilck1983}}; \end{tikzpicture}  \\\hline
\begin{tikzpicture} \node at (0,-1) {}; \node at (0,1) {}; \draw[line width = 1pt, dashed] (0,0)--(-0.5,0); \draw[line width = 1pt, dashed] (0,0)--(0.5,0); \draw[line width = 1pt, dashed] (0,0)--(0,0.5); \draw[line width = 1pt] (0,0)--(0,-0.5); \end{tikzpicture} &
\begin{tikzpicture} \node at (0,-1) {}; \node at (0,1) {}; \draw[line width = 1pt] (0,0)--(-0.5,0); \draw[line width = 1pt] (0,0)--(0.5,0); \draw[line width = 1pt] (0,0)--(0,0.5); \draw[line width = 1pt, dashed] (0,0)--(0,-0.5); \end{tikzpicture} &
\begin{tikzpicture} \node at (0,-1) {}; \node at (0,1) {}; \draw[line width = 1pt] (0,0)--(-0.5,0); \draw[line width = 1pt] (0,0)--(0.5,0); \draw[line width = 1pt, dashed] (0,0)--(0,0.5); \draw[line width = 1pt] (0,0)--(0,-0.5); \end{tikzpicture} &
\begin{tikzpicture} \node at (0,-1) {}; \node at (0,1) {}; \draw[line width = 1pt, dashed] (0,0)--(-0.5,0); \draw[line width = 1pt, dashed] (0,0)--(0.5,0); \draw[line width = 1pt] (0,0)--(0,0.5); \draw[line width = 1pt, dashed] (0,0)--(0,-0.5); \end{tikzpicture} &
\begin{tikzpicture} \node at (0,-1) {}; \node at (0,1) {}; \draw[line width = 1pt] (0,0)--(-0.5,0); \draw[line width = 1pt, dashed] (0,0)--(0.5,0); \draw[line width = 1pt, dashed] (0,0)--(0,0.5); \draw[line width = 1pt, dashed] (0,0)--(0,-0.5); \end{tikzpicture} &
\begin{tikzpicture} \node at (0,-1) {}; \node at (0,1) {}; \draw[line width = 1pt, dashed] (0,0)--(-0.5,0); \draw[line width = 1pt] (0,0)--(0.5,0); \draw[line width = 1pt] (0,0)--(0,0.5); \draw[line width = 1pt] (0,0)--(0,-0.5); \end{tikzpicture} &
\begin{tikzpicture} \node at (0,-1) {}; \node at (0,1) {}; \draw[line width = 1pt] (0,0)--(-0.5,0); \draw[line width = 1pt, dashed] (0,0)--(0.5,0); \draw[line width = 1pt] (0,0)--(0,0.5); \draw[line width = 1pt] (0,0)--(0,-0.5); \end{tikzpicture} &
\begin{tikzpicture} \node at (0,-1) {}; \node at (0,1) {}; \draw[line width = 1pt, dashed] (0,0)--(-0.5,0); \draw[line width = 1pt] (0,0)--(0.5,0); \draw[line width = 1pt, dashed] (0,0)--(0,0.5); \draw[line width = 1pt, dashed] (0,0)--(0,-0.5); \end{tikzpicture} &
\begin{tikzpicture} \node at (0,-1) {}; \node at (0,1) {}; \node at (0,0) {\cite{wu1972,gaaff1975h}}; \end{tikzpicture}  \\\hline
\begin{tikzpicture} \node at (0,-1) {}; \node at (0,1) {}; \draw[line width = 1pt, dashed] (0,0)--(-0.5,0); \draw[line width = 1pt, dashed] (0,0)--(0.5,0); \draw[line width = 1pt, dashed] (0,0)--(0,0.5); \draw[line width = 1pt] (0,0)--(0,-0.5); \end{tikzpicture} &
\begin{tikzpicture} \node at (0,-1) {}; \node at (0,1) {}; \draw[line width = 1pt] (0,0)--(-0.5,0); \draw[line width = 1pt, dashed] (0,0)--(0.5,0); \draw[line width = 1pt] (0,0)--(0,0.5); \draw[line width = 1pt] (0,0)--(0,-0.5); \end{tikzpicture} &
\begin{tikzpicture} \node at (0,-1) {}; \node at (0,1) {}; \draw[line width = 1pt] (0,0)--(-0.5,0); \draw[line width = 1pt] (0,0)--(0.5,0); \draw[line width = 1pt, dashed] (0,0)--(0,0.5); \draw[line width = 1pt] (0,0)--(0,-0.5); \end{tikzpicture} &
\begin{tikzpicture} \node at (0,-1) {}; \node at (0,1) {}; \draw[line width = 1pt] (0,0)--(-0.5,0); \draw[line width = 1pt, dashed] (0,0)--(0.5,0); \draw[line width = 1pt, dashed] (0,0)--(0,0.5); \draw[line width = 1pt, dashed] (0,0)--(0,-0.5); \end{tikzpicture} &
\begin{tikzpicture} \node at (0,-1) {}; \node at (0,1) {}; \draw[line width = 1pt] (0,0)--(-0.5,0); \draw[line width = 1pt] (0,0)--(0.5,0); \draw[line width = 1pt] (0,0)--(0,0.5); \draw[line width = 1pt, dashed] (0,0)--(0,-0.5); \end{tikzpicture} &
\begin{tikzpicture} \node at (0,-1) {}; \node at (0,1) {}; \draw[line width = 1pt, dashed] (0,0)--(-0.5,0); \draw[line width = 1pt] (0,0)--(0.5,0); \draw[line width = 1pt, dashed] (0,0)--(0,0.5); \draw[line width = 1pt, dashed] (0,0)--(0,-0.5); \end{tikzpicture} &
\begin{tikzpicture} \node at (0,-1) {}; \node at (0,1) {}; \draw[line width = 1pt, dashed] (0,0)--(-0.5,0); \draw[line width = 1pt, dashed] (0,0)--(0.5,0); \draw[line width = 1pt] (0,0)--(0,0.5); \draw[line width = 1pt, dashed] (0,0)--(0,-0.5); \end{tikzpicture} &
\begin{tikzpicture} \node at (0,-1) {}; \node at (0,1) {}; \draw[line width = 1pt, dashed] (0,0)--(-0.5,0); \draw[line width = 1pt] (0,0)--(0.5,0); \draw[line width = 1pt] (0,0)--(0,0.5); \draw[line width = 1pt] (0,0)--(0,-0.5); \end{tikzpicture} &
\begin{tikzpicture} \node at (0,-1) {}; \node at (0,1) {}; \node at (0,0) {\cite{wu1969}}; \end{tikzpicture}  \\\hline
\begin{tikzpicture} \node at (0,-1) {}; \node at (0,1) {}; \draw[line width = 1pt] (0,0)--(-0.5,0); \draw[line width = 1pt, dashed] (0,0)--(0.5,0); \draw[line width = 1pt, dashed] (0,0)--(0,0.5); \draw[line width = 1pt, dashed] (0,0)--(0,-0.5); \end{tikzpicture} &
\begin{tikzpicture} \node at (0,-1) {}; \node at (0,1) {}; \draw[line width = 1pt, dashed] (0,0)--(-0.5,0); \draw[line width = 1pt] (0,0)--(0.5,0); \draw[line width = 1pt] (0,0)--(0,0.5); \draw[line width = 1pt] (0,0)--(0,-0.5); \end{tikzpicture} &
\begin{tikzpicture} \node at (0,-1) {}; \node at (0,1) {}; \draw[line width = 1pt, dashed] (0,0)--(-0.5,0); \draw[line width = 1pt] (0,0)--(0.5,0); \draw[line width = 1pt, dashed] (0,0)--(0,0.5); \draw[line width = 1pt, dashed] (0,0)--(0,-0.5); \end{tikzpicture} &
\begin{tikzpicture} \node at (0,-1) {}; \node at (0,1) {}; \draw[line width = 1pt] (0,0)--(-0.5,0); \draw[line width = 1pt, dashed] (0,0)--(0.5,0); \draw[line width = 1pt] (0,0)--(0,0.5); \draw[line width = 1pt] (0,0)--(0,-0.5); \end{tikzpicture} &
\begin{tikzpicture} \node at (0,-1) {}; \node at (0,1) {}; \draw[line width = 1pt, dashed] (0,0)--(-0.5,0); \draw[line width = 1pt, dashed] (0,0)--(0.5,0); \draw[line width = 1pt] (0,0)--(0,0.5); \draw[line width = 1pt, dashed] (0,0)--(0,-0.5); \end{tikzpicture} &
\begin{tikzpicture} \node at (0,-1) {}; \node at (0,1) {}; \draw[line width = 1pt] (0,0)--(-0.5,0); \draw[line width = 1pt] (0,0)--(0.5,0); \draw[line width = 1pt, dashed] (0,0)--(0,0.5); \draw[line width = 1pt] (0,0)--(0,-0.5); \end{tikzpicture} &
\begin{tikzpicture} \node at (0,-1) {}; \node at (0,1) {}; \draw[line width = 1pt, dashed] (0,0)--(-0.5,0); \draw[line width = 1pt, dashed] (0,0)--(0.5,0); \draw[line width = 1pt, dashed] (0,0)--(0,0.5); \draw[line width = 1pt] (0,0)--(0,-0.5); \end{tikzpicture} &
\begin{tikzpicture} \node at (0,-1) {}; \node at (0,1) {}; \draw[line width = 1pt] (0,0)--(-0.5,0); \draw[line width = 1pt] (0,0)--(0.5,0); \draw[line width = 1pt] (0,0)--(0,0.5); \draw[line width = 1pt, dashed] (0,0)--(0,-0.5); \end{tikzpicture} &
\begin{tikzpicture} \node at (0,-1) {}; \node at (0,1) {}; \node at (0,0) {\cite{wu2004k}${}^1$}; \end{tikzpicture} \\\hline
\end{tabular}
\end{center}
\caption{Notation conventions for the odd 8-vertex model weights. We use the first convention. \label{tableodd}}
\end{table}
\end{savenotes}

\vspace{1in}\phantom{a}

\section{Partition function symmetries}
\label{app:parttrans}
\footnotetext[1]{Because of the issues noted in footnote~\ref{evenfootnote} on page~\pageref{evenfootnote}, the notation used in practice in~\cite{wu2004k} is not the one shown in their Figure 1, but the last entry of table~\ref{tableodd}.}
The partition function of the full 16-vertex model must be invariant under various lattice and bond-reversal transformations. There are four generators for the symmetries, two for the dihedral group of order 8 of lattice rotations and reflections, and two due to horizontal and vertical bond reversal symmetries. We call the counter-clockwise rotation generator $c$, the horizontal reflection generator $r$, and the horizontal and vertical bond reversal operations $h$ and $v$ respectively. Together, they generate the group $C_2\times C_2\times D_8$ of order 32, where $C_2$ is the cyclic group of order 2 and $D_8$ is the dihedral group of order 8. We list each of the transformations below, from which the even and odd 8-vertex model transformations can be found by specialization. We note that this list is larger than those found in~\cite{fan1969w,fan1970w,lieb1972w,wu2004k}.
\beqr
I: (w_1,w_2,w_3,w_4,w_5,w_6,w_7,w_8;v_1,v_2,v_3,v_4,v_5,v_6,v_7,v_8) \\
c^2: (w_1,w_2,w_3,w_4,w_6,w_5,w_8,w_7;v_3,v_4,v_1,v_2,v_7,v_8,v_5,v_6) \\
r: (w_1,w_2,w_3,w_4,w_7,w_8,w_5,w_6;v_1,v_2,v_3,v_4,v_7,v_8,v_5,v_6) \\
c^2r: (w_1,w_2,w_3,w_4,w_8,w_7,w_6,w_5;v_3,v_4,v_1,v_2,v_5,v_6,v_7,v_8) \\
c^3r: (w_1,w_2,w_4,w_3,w_5,w_6,w_8,w_7;v_5,v_6,v_7,v_8,v_1,v_2,v_3,v_4) \\
cr: (w_1,w_2,w_4,w_3,w_6,w_5,w_7,w_8;v_7,v_8,v_5,v_6,v_3,v_4,v_1,v_2) \\
c^3: (w_1,w_2,w_4,w_3,w_7,w_8,w_6,w_5;v_7,v_8,v_5,v_6,v_1,v_2,v_3,v_4) \\
c: (w_1,w_2,w_4,w_3,w_8,w_7,w_5,w_6;v_5,v_6,v_7,v_8,v_3,v_4,v_1,v_2) \\
crhv: (w_2,w_1,w_3,w_4,w_5,w_6,w_8,w_7;v_8,v_7,v_6,v_5,v_4,v_3,v_2,v_1) \\
c^3rhv: (w_2,w_1,w_3,w_4,w_6,w_5,w_7,w_8;v_6,v_5,v_8,v_7,v_2,v_1,v_4,v_3) \\
chv: (w_2,w_1,w_3,w_4,w_7,w_8,w_6,w_5;v_6,v_5,v_8,v_7,v_4,v_3,v_2,v_1) \\
c^3hv: (w_2,w_1,w_3,w_4,w_8,w_7,w_5,w_6;v_8,v_7,v_6,v_5,v_2,v_1,v_4,v_3) \\
c^2hv: (w_2,w_1,w_4,w_3,w_5,w_6,w_7,w_8;v_4,v_3,v_2,v_1,v_8,v_7,v_6,v_5) \\
hv: (w_2,w_1,w_4,w_3,w_6,w_5,w_8,w_7;v_2,v_1,v_4,v_3,v_6,v_5,v_8,v_7) \\
c^2rhv: (w_2,w_1,w_4,w_3,w_7,w_8,w_5,w_6;v_4,v_3,v_2,v_1,v_6,v_5,v_8,v_7) \\
rhv: (w_2,w_1,w_4,w_3,w_8,w_7,w_6,w_5;v_2,v_1,v_4,v_3,v_8,v_7,v_6,v_5) \\
c^2rv: (w_3,w_4,w_1,w_2,w_5,w_6,w_7,w_8;v_1,v_2,v_3,v_4,v_8,v_7,v_6,v_5) \\
rv: (w_3,w_4,w_1,w_2,w_6,w_5,w_8,w_7;v_3,v_4,v_1,v_2,v_6,v_5,v_8,v_7) \\
c^2v: (w_3,w_4,w_1,w_2,w_7,w_8,w_5,w_6;v_1,v_2,v_3,v_4,v_6,v_5,v_8,v_7) \\
v: (w_3,w_4,w_1,w_2,w_8,w_7,w_6,w_5;v_3,v_4,v_1,v_2,v_8,v_7,v_6,v_5) \\
cv: (w_3,w_4,w_2,w_1,w_5,w_6,w_8,w_7;v_8,v_7,v_6,v_5,v_1,v_2,v_3,v_4) \\
c^3v: (w_3,w_4,w_2,w_1,w_6,w_5,w_7,w_8;v_6,v_5,v_8,v_7,v_3,v_4,v_1,v_2) \\
crv: (w_3,w_4,w_2,w_1,w_7,w_8,w_6,w_5;v_6,v_5,v_8,v_7,v_1,v_2,v_3,v_4) \\
c^3rv: (w_3,w_4,w_2,w_1,w_8,w_7,w_5,w_6;v_8,v_7,v_6,v_5,v_3,v_4,v_1,v_2) \\
c^3h: (w_4,w_3,w_1,w_2,w_5,w_6,w_8,w_7;v_5,v_6,v_7,v_8,v_4,v_3,v_2,v_1) \\
ch: (w_4,w_3,w_1,w_2,w_6,w_5,w_7,w_8;v_7,v_8,v_5,v_6,v_2,v_1,v_4,v_3) \\
c^3rh: (w_4,w_3,w_1,w_2,w_7,w_8,w_6,w_5;v_7,v_8,v_5,v_6,v_4,v_3,v_2,v_1) \\
crh: (w_4,w_3,w_1,w_2,w_8,w_7,w_5,w_6;v_5,v_6,v_7,v_8,v_2,v_1,v_4,v_3) \\
rh: (w_4,w_3,w_2,w_1,w_5,w_6,w_7,w_8;v_4,v_3,v_2,v_1,v_5,v_6,v_7,v_8) \\
c^2rh: (w_4,w_3,w_2,w_1,w_6,w_5,w_8,w_7;v_2,v_1,v_4,v_3,v_7,v_8,v_5,v_6) \\
h: (w_4,w_3,w_2,w_1,w_7,w_8,w_5,w_6;v_4,v_3,v_2,v_1,v_7,v_8,v_5,v_6) \\
c^2h: (w_4,w_3,w_2,w_1,w_8,w_7,w_6,w_5;v_2,v_1,v_4,v_3,v_5,v_6,v_7,v_8) 
\eeqr

\section{The \texorpdfstring{$SL(2)\times SL(2)$}{SL(2) x SL(2)} transformation}\label{app:genweakgraph}
Using the $SL(2)\times SL(2)$ transformation~(\ref{genweakgraphtrans}) of Gaaff and Hijmans~\cite{gaaff1975h}, we can convert it into a $16\times 16$ matrix acting on the weight vector $\mathbf{w}$ defined in section~\ref{sec:weakgraph}. In the matrix below, we assume without loss of generality that 
\beq
s_1s_3-s_2s_4=1,\qquad t_1t_3-t_2t_4=1
\eeq
so that the matrices $S$ and $T$ both have unit determinant. The matrix below then has determinant equal to -1, so that it corresponds to an involution. A comparison with the weak-graph transformation in (\ref{weakgraphtrans}) reveals that this transformation does not specialize to the weak-graph transformation, although the the partition function is invariant under both sets of transformations.
\begin{multline}
\left(\begin{matrix}
s_1t_3s_4t_4 & s_2t_1s_3t_2 & s_3t_1s_4t_4 & s_1t_2s_2t_3 & -s_3t_2s_4t_3 & -s_1t_1s_2t_4 & s_2t_2s_3t_3 & s_1t_1s_2t_2 \\
s_2t_3s_3t_4 & s_1t_1s_4t_2 & s_3t_2s_4t_3 & s_1t_1s_2t_4 & -s_3t_1s_4t_4 & -s_1t_2s_2t_3 & s_1t_1s_4t_4 & s_1t_1s_2t_2 \\
-s_2t_3s_3t_4 & -s_1t_1s_4t_2 & -s_3t_1s_4t_4 & -s_1t_2s_2t_3 & s_3t_2s_4t_3 & s_1t_1s_2t_4 & -s_1t_2s_4t_3 & -s_1t_1s_2t_2 \\
-s_1t_3s_4t_4 & -s_2t_1s_3t_2 & -s_3t_2s_4t_3 & -s_1t_1s_2t_4 & s_3t_1s_4t_4 & s_1t_2s_2t_3 & -s_2t_1s_3t_4 & -s_1t_1s_2t_2 \\
-s_1s_3t_4^2 & -s_1s_3t_2^2 & -s_3^2t_2t_4 & -s_1^2t_2t_4 & s_3^2t_2t_4 & s_1^2t_2t_4 & -s_1s_3t_2t_4 & -s_1^2t_2^2 \\
-s_2s_4t_3^2 & -s_2s_4t_1^2 & -s_4^2t_1t_3 & -s_2^2t_1t_3 & s_4^2t_1t_3 & s_2^2t_1t_3 & -s_2s_4t_1t_3 & -s_2^2t_1^2 \\
s_1s_3t_3^2 & s_1s_3t_1^2 & s_3^2t_1t_3 & s_1^2t_1t_3 & -s_3^2t_1t_3 & -s_1^2t_1t_3 & s_1s_3t_1t_3 & s_1^2t_1^2 \\
s_2s_4t_4^2 & s_2s_4t_2^2 & s_4^2t_2t_4 & s_2^2t_2t_4 & -s_4^2t_2t_4 & -s_2^2t_2t_4 & s_2s_4t_2t_4 & s_2^2t_2^2 \\
-s_1s_4t_3^2 & -s_2s_3t_1^2 & -s_3s_4t_1t_3 & -s_1s_2t_1t_3 & s_3s_4t_1t_3 & s_1s_2t_1t_3 & -s_2s_3t_1t_3 & -s_1s_2t_1^2 \\
-s_2s_3t_4^2 & -s_1s_4t_2^2 & -s_3s_4t_2t_4 & -s_1s_2t_2t_4 & s_3s_4t_2t_4 & s_1s_2t_2t_4 & -s_1s_4t_2t_4 & -s_1s_2t_2^2 \\
s_2s_3t_3^2 & s_1s_4t_1^2 & s_3s_4t_1t_3 & s_1s_2t_1t_3 & -s_3s_4t_1t_3 & -s_1s_2t_1t_3 & s_1s_4t_1t_3 & s_1s_2t_1^2 \\
s_1s_4t_4^2 & s_2s_3t_2^2 & s_3s_4t_2t_4 & s_1s_2t_2t_4 & -s_3s_4t_2t_4 & -s_1s_2t_2t_4 & s_2s_3t_2t_4 & s_1s_2t_2^2 \\
s_1s_3t_3t_4 & s_1s_3t_1t_2 & s_3^2t_2t_3 & s_1^2t_1t_4 & -s_3^2t_1t_4 & -s_1^2t_2t_3 & s_1s_3t_1t_4 & s_1^2t_1t_2 \\
s_2s_4t_3t_4 & s_2s_4t_1t_2 & s_4^2t_1t_4 & s_2^2t_2t_3 & -s_4^2t_2t_3 & -s_2^2t_1t_4 & s_2s_4t_2t_3 & s_2^2t_1t_2 \\
-s_2s_4t_3t_4 & -s_2s_4t_1t_2 & -s_4^2t_2t_3 & -s_2^2t_1t_4 & s_4^2t_1t_4 & s_2^2t_2t_3 & -s_2s_4t_1t_4 & -s_2^2t_1t_2 \\
-s_1s_3t_3t_4 & -s_1s_3t_1t_2 & -s_3^2t_1t_4 & -s_1^2t_2t_3 & s_3^2t_2t_3 & s_1^2t_1t_4 & -s_1s_3t_2t_3 & -s_1^2t_1t_2 \\
\end{matrix}\right.
\\
\left.\begin{matrix}
-s_1t_3s_2t_4 & -s_2t_1s_3t_4 & -s_1t_2s_4t_3 & -s_3t_1s_4t_2 & s_3t_3s_4t_4 & s_1t_1s_4t_4 & -s_1t_1s_4t_2 & -s_2t_3s_3t_4 \\
-s_1t_3s_2t_4 & -s_1t_2s_4t_3 & -s_2t_1s_3t_4 & -s_3t_1s_4t_2 & s_3t_3s_4t_4 & s_2t_2s_3t_3 & -s_2t_1s_3t_2 & -s_1t_3s_4t_4 \\
s_1t_3s_2t_4 & s_1t_1s_4t_4 & s_2t_2s_3t_3 & s_3t_1s_4t_2 & -s_3t_3s_4t_4 & -s_2t_1s_3t_4 & s_2t_1s_3t_2 & s_1t_3s_4t_4 \\
s_1t_3s_2t_4 & s_2t_2s_3t_3 & s_1t_1s_4t_4 & s_3t_1s_4t_2 & -s_3t_3s_4t_4 & -s_1t_2s_4t_3 & s_1t_1s_4t_2 & s_2t_3s_3t_4 \\
s_1^2t_4^2 & s_1s_3t_2t_4 & s_1s_3t_2t_4 & s_3^2t_2^2 & -s_3^2t_4^2 & -s_1s_3t_2t_4 & s_1s_3t_2^2 & s_1s_3t_4^2 \\
s_2^2t_3^2 & s_2s_4t_1t_3 & s_2s_4t_1t_3 & s_4^2t_1^2 & -s_4^2t_3^2 & -s_2s_4t_1t_3 & s_2s_4t_1^2 & s_2s_4t_3^2 \\
-s_1^2t_3^2 & -s_1s_3t_1t_3 & -s_1s_3t_1t_3 & -s_3^2t_1^2 & s_3^2t_3^2 & s_1s_3t_1t_3 & -s_1s_3t_1^2 & -s_1s_3t_3^2 \\
-s_2^2t_4^2 & -s_2s_4t_2t_4 & -s_2s_4t_2t_4 & -s_4^2t_2^2 & s_4^2t_4^2 & s_2s_4t_2t_4 & -s_2s_4t_2^2 & -s_2s_4t_4^2 \\
s_1s_2t_3^2 & s_2s_3t_1t_3 & s_1s_4t_1t_3 & s_3s_4t_1^2 & -s_3s_4t_3^2 & -s_1s_4t_1t_3 & s_1s_4t_1^2 & s_2s_3t_3^2 \\
s_1s_2t_4^2 & s_1s_4t_2t_4 & s_2s_3t_2t_4 & s_3s_4t_2^2 & -s_3s_4t_4^2 & -s_2s_3t_2t_4 & s_2s_3t_2^2 & s_1s_4t_4^2 \\
-s_1s_2t_3^2 & -s_1s_4t_1t_3 & -s_2s_3t_1t_3 & -s_3s_4t_1^2 & s_3s_4t_3^2 & s_2s_3t_1t_3 & -s_2s_3t_1^2 & -s_1s_4t_3^2 \\
-s_1s_2t_4^2 & -s_2s_3t_2t_4 & -s_1s_4t_2t_4 & -s_3s_4t_2^2 & s_3s_4t_4^2 & s_1s_4t_2t_4 & -s_1s_4t_2^2 & -s_2s_3t_4^2 \\
-s_1^2t_3t_4 & -s_1s_3t_2t_3 & -s_1s_3t_1t_4 & -s_3^2t_1t_2 & s_3^2t_3t_4 & s_1s_3t_2t_3 & -s_1s_3t_1t_2 & -s_1s_3t_3t_4 \\
-s_2^2t_3t_4 & -s_2s_4t_1t_4 & -s_2s_4t_2t_3 & -s_4^2t_1t_2 & s_4^2t_3t_4 & s_2s_4t_1t_4 & -s_2s_4t_1t_2 & -s_2s_4t_3t_4 \\
s_2^2t_3t_4 & s_2s_4t_2t_3 & s_2s_4t_1t_4 & s_4^2t_1t_2 & -s_4^2t_3t_4 & -s_2s_4t_2t_3 & s_2s_4t_1t_2 & s_2s_4t_3t_4 \\
s_1^2t_3t_4 & s_1s_3t_1t_4 & s_1s_3t_2t_3 & s_3^2t_1t_2 & -s_3^2t_3t_4 & -s_1s_3t_1t_4 & s_1s_3t_1t_2 & s_1s_3t_3t_4 \\
\end{matrix}\right)
\end{multline}

\section{Dimer constructions for even and odd free-fermion 8-vertex models}\label{app:dimerconst}
In the following subsections we outline the dimer solution method of~\cite{mccoy1973w,mccoy2014w} to construct the free energies of the even and odd 8-vertex staggered free-fermion models, appropriate for finite lattices. 

\subsection{The odd 8-vertex models}
\begin{figure}[htpb]
\begin{center}
\scalebox{.6}{
\begin{tikzpicture}
\begin{scope}[shift={(10,-3)}]
\node at (-1.5,9.5) {{\Large $v_4$}};
\draw[line width = 2pt] (0,8) -- (0,9);
\draw[line width = 2pt,dashed] (0,10) -- (0,9);
\draw[line width = 2pt] (-1,9) -- (0,9);
\draw[line width = 2pt] (1,9) -- (0,9);
\node at (1.5,9) {\large $\bm{=}$};
\draw[line width = 2pt] (3,8) -- (3,9);
\draw[line width = 2pt] (3,10) -- (3,9);
\draw[line width = 2pt] (2,9) -- (3,9);
\draw[line width = 2pt] (4,9) -- (3,9);
\draw[line width = 2pt] (2.4,9) -- (3,9.6) -- (3.6,9) -- (3,8.4) -- (2.4,9);
\draw[line width = 1.3pt] (2.2,9) ellipse (0.4 and 0.1);
\draw[line width = 1.3pt] (3.8,9) ellipse (0.4 and 0.1);
\draw[line width = 1.3pt,rotate around={90:(3,8.2)}] (3,8.2) ellipse (0.4 and 0.1);
\draw[line width = 1.3pt,rotate around={90:(3,9.3)}] (3,9.3) ellipse (0.4 and 0.1);
\end{scope}
\begin{scope}[shift={(-7,-3)}]
\node at (5.5,9.5) {{\Large $v_3$}};
\draw[line width = 2pt,dashed] (7,8) -- (7,9);
\draw[line width = 2pt] (7,10) -- (7,9);
\draw[line width = 2pt,dashed] (6,9) -- (7,9);
\draw[line width = 2pt,dashed] (8,9) -- (7,9);
\node at (8.5,9) {\large $\bm{=}$};
\draw[line width = 2pt] (10,8) -- (10,9);
\draw[line width = 2pt] (10,10) -- (10,9);
\draw[line width = 2pt] (9,9) -- (10,9);
\draw[line width = 2pt] (11,9) -- (10,9);
\draw[line width = 2pt] (9.4,9) -- (10,9.6) -- (10.6,9) -- (10,8.4) -- (9.4,9);
\draw[line width = 1.3pt] (10.3,9) ellipse (0.4 and 0.1);
\draw[line width = 1.3pt,rotate around={135:(9.7,8.7)}] (9.7,8.7) ellipse (0.4 and 0.1);
\draw[line width = 1.3pt,rotate around={90:(10,9.8)}] (10,9.8) ellipse (0.4 and 0.1);
\node at (11.5,9) {\large $\bm{+}$};
\draw[line width = 2pt] (13,8) -- (13,9);
\draw[line width = 2pt] (13,10) -- (13,9);
\draw[line width = 2pt] (12,9) -- (13,9);
\draw[line width = 2pt] (14,9) -- (13,9);
\draw[line width = 2pt] (12.4,9) -- (13,9.6) -- (13.6,9) -- (13,8.4) -- (12.4,9);
\draw[line width = 1.3pt] (12.7,9) ellipse (0.4 and 0.1);
\draw[line width = 1.3pt,rotate around={45:(13.3,8.7)}] (13.3,8.7) ellipse (0.4 and 0.1);
\draw[line width = 1.3pt,rotate around={90:(13,9.8)}] (13,9.8) ellipse (0.4 and 0.1);
\end{scope}

\begin{scope}[shift={(10,3)}]
\node at (-1.5,6.5) {{\Large $v_2$}};
\draw[line width = 2pt,dashed] (0,5) -- (0,6);
\draw[line width = 2pt] (0,7) -- (0,6);
\draw[line width = 2pt] (-1,6) -- (0,6);
\draw[line width = 2pt] (1,6) -- (0,6);
\node at (1.5,6) {\large $\bm{=}$};
\draw[line width = 2pt] (3,5) -- (3,6);
\draw[line width = 2pt] (3,7) -- (3,6);
\draw[line width = 2pt] (2,6) -- (3,6);
\draw[line width = 2pt] (4,6) -- (3,6);
\draw[line width = 2pt] (2.4,6) -- (3,6.6) -- (3.6,6) -- (3,5.4) -- (2.4,6);
\draw[line width = 1.3pt] (2.2,6) ellipse (0.4 and 0.1);
\draw[line width = 1.3pt] (3.8,6) ellipse (0.4 and 0.1);
\draw[line width = 1.3pt,rotate around={90:(3,5.7)}] (3,5.7) ellipse (0.4 and 0.1);
\draw[line width = 1.3pt,rotate around={90:(3,6.8)}] (3,6.8) ellipse (0.4 and 0.1);
\end{scope}
\begin{scope}[shift={(-7,3)}]
\node at (5.5,6.5) {{\Large $v_1$}};
\draw[line width = 2pt] (7,5) -- (7,6);
\draw[line width = 2pt,dashed] (7,7) -- (7,6);
\draw[line width = 2pt,dashed] (6,6) -- (7,6);
\draw[line width = 2pt,dashed] (8,6) -- (7,6);
\node at (8.5,6) {\large $\bm{=}$};
\draw[line width = 2pt] (10,5) -- (10,6);
\draw[line width = 2pt] (10,7) -- (10,6);
\draw[line width = 2pt] (9,6) -- (10,6);
\draw[line width = 2pt] (11,6) -- (10,6);
\draw[line width = 2pt] (9.4,6) -- (10,6.6) -- (10.6,6) -- (10,5.4) -- (9.4,6);
\draw[line width = 1.3pt] (9.7,6) ellipse (0.4 and 0.1);
\draw[line width = 1.3pt,rotate around={90:(10,5.2)}] (10,5.2) ellipse (0.4 and 0.1);
\draw[line width = 1.3pt,rotate around={135:(10.3,6.3)}] (10.3,6.3) ellipse (0.4 and 0.1);
\node at (11.5,6) {\large $\bm{+}$};
\draw[line width = 2pt] (13,5) -- (13,6);
\draw[line width = 2pt] (13,7) -- (13,6);
\draw[line width = 2pt] (12,6) -- (13,6);
\draw[line width = 2pt] (14,6) -- (13,6);
\draw[line width = 2pt] (12.4,6) -- (13,6.6) -- (13.6,6) -- (13,5.4) -- (12.4,6);
\draw[line width = 1.3pt] (13.3,6) ellipse (0.4 and 0.1);
\draw[line width = 1.3pt,rotate around={90:(13,5.2)}] (13,5.2) ellipse (0.4 and 0.1);
\draw[line width = 1.3pt,rotate around={45:(12.7,6.3)}] (12.7,6.3) ellipse (0.4 and 0.1);
\end{scope}

\begin{scope}[shift={(10,-3)}]
\node at (-1.5,3.5) {{\Large $v_8$}};
\draw[line width = 2pt] (0,2) -- (0,3);
\draw[line width = 2pt] (0,4) -- (0,3);
\draw[line width = 2pt,dashed] (-1,3) -- (0,3);
\draw[line width = 2pt] (1,3) -- (0,3);
\node at (1.5,3) {\large $\bm{=}$};
\draw[line width = 2pt] (3,2) -- (3,3);
\draw[line width = 2pt] (3,4) -- (3,3);
\draw[line width = 2pt] (2,3) -- (3,3);
\draw[line width = 2pt] (4,3) -- (3,3);
\draw[line width = 2pt] (2.4,3) -- (3,3.6) -- (3.6,3) -- (3,2.4) -- (2.4,3);
\draw[line width = 1.3pt] (2.7,3) ellipse (0.4 and 0.1);
\draw[line width = 1.3pt] (3.8,3) ellipse (0.4 and 0.1);
\draw[line width = 1.3pt,rotate around={90:(3,2.2)}] (3,2.2) ellipse (0.4 and 0.1);
\draw[line width = 1.3pt,rotate around={90:(3,3.8)}] (3,3.8) ellipse (0.4 and 0.1);
\end{scope}
\begin{scope}[shift={(-7,-3)}]
\node at (5.5,3.5) {{\Large $v_7$}};
\draw[line width = 2pt,dashed] (7,2) -- (7,3);
\draw[line width = 2pt,dashed] (7,4) -- (7,3);
\draw[line width = 2pt] (6,3) -- (7,3);
\draw[line width = 2pt,dashed] (8,3) -- (7,3);
\node at (8.5,3) {\large $\bm{=}$};
\draw[line width = 2pt] (10,2) -- (10,3);
\draw[line width = 2pt] (10,4) -- (10,3);
\draw[line width = 2pt] (9,3) -- (10,3);
\draw[line width = 2pt] (11,3) -- (10,3);
\draw[line width = 2pt] (9.4,3) -- (10,3.6) -- (10.6,3) -- (10,2.4) -- (9.4,3);
\draw[line width = 1.3pt] (9.2,3) ellipse (0.4 and 0.1);
\draw[line width = 1.3pt,rotate around={135:(10.3,3.3)}] (10.3,3.3) ellipse (0.4 and 0.1);
\draw[line width = 1.3pt,rotate around={90:(10,2.7)}] (10,2.7) ellipse (0.4 and 0.1);
\node at (11.5,3) {\large $\bm{+}$};
\draw[line width = 2pt] (13,2) -- (13,3);
\draw[line width = 2pt] (13,4) -- (13,3);
\draw[line width = 2pt] (12,3) -- (13,3);
\draw[line width = 2pt] (14,3) -- (13,3);
\draw[line width = 2pt] (12.4,3) -- (13,3.6) -- (13.6,3) -- (13,2.4) -- (12.4,3);
\draw[line width = 1.3pt] (12.2,3) ellipse (0.4 and 0.1);
\draw[line width = 1.3pt,rotate around={45:(13.3,2.7)}] (13.3,2.7) ellipse (0.4 and 0.1);
\draw[line width = 1.3pt,rotate around={90:(13,3.3)}] (13,3.3) ellipse (0.4 and 0.1);
\end{scope}

\begin{scope}[shift={(10,3)}]
\node at (-1.5,0.5) {{\Large $v_6$}};
\draw[line width = 2pt] (0,-1) -- (0,0);
\draw[line width = 2pt] (0,1) -- (0,0);
\draw[line width = 2pt] (-1,0) -- (0,0);
\draw[line width = 2pt,dashed] (1,0) -- (0,0);
\node at (1.5,0) {\large $\bm{=}$};
\draw[line width = 2pt] (3,-1) -- (3,0);
\draw[line width = 2pt] (3,1) -- (3,0);
\draw[line width = 2pt] (2,0) -- (3,0);
\draw[line width = 2pt] (4,0) -- (3,0);
\draw[line width = 2pt] (2.4,0) -- (3,0.6) -- (3.6,0) -- (3,-0.6) -- (2.4,0);
\draw[line width = 1.3pt] (2.2,0) ellipse (0.4 and 0.1);
\draw[line width = 1.3pt] (3.3,0) ellipse (0.4 and 0.1);
\draw[line width = 1.3pt,rotate around={90:(3,-0.8)}] (3,-0.8) ellipse (0.4 and 0.1);
\draw[line width = 1.3pt,rotate around={90:(3,0.8)}] (3,0.8) ellipse (0.4 and 0.1);
\end{scope}
\begin{scope}[shift={(-7,3)}]
\node at (5.5,0.5) {{\Large $v_5$}};
\draw[line width = 2pt,dashed] (7,-1) -- (7,0);
\draw[line width = 2pt,dashed] (7,1) -- (7,0);
\draw[line width = 2pt,dashed] (6,0) -- (7,0);
\draw[line width = 2pt] (8,0) -- (7,0);
\node at (8.5,0) {\large $\bm{=}$};
\draw[line width = 2pt] (10,-1) -- (10,0);
\draw[line width = 2pt] (10,1) -- (10,0);
\draw[line width = 2pt] (9,0) -- (10,0);
\draw[line width = 2pt] (11,0) -- (10,0);
\draw[line width = 2pt] (9.4,0) -- (10,0.6) -- (10.6,0) -- (10,-0.6) -- (9.4,0);
\draw[line width = 1.3pt,rotate around={135:(9.7,-0.3)}] (9.7,-0.3) ellipse (0.4 and 0.1);
\draw[line width = 1.3pt] (10.8,0) ellipse (0.4 and 0.1);
\draw[line width = 1.3pt,rotate around={90:(10,0.3)}] (10,0.3) ellipse (0.4 and 0.1);
\node at (11.5,0) {\large $\bm{+}$};
\draw[line width = 2pt] (13,-1) -- (13,0);
\draw[line width = 2pt] (13,1) -- (13,0);
\draw[line width = 2pt] (12,0) -- (13,0);
\draw[line width = 2pt] (14,0) -- (13,0);
\draw[line width = 2pt] (12.4,0) -- (13,0.6) -- (13.6,0) -- (13,-0.6) -- (12.4,0);
\draw[line width = 1.3pt,rotate around={45:(12.7,0.3)}] (12.7,0.3) ellipse (0.4 and 0.1);
\draw[line width = 1.3pt] (13.8,0) ellipse (0.4 and 0.1);
\draw[line width = 1.3pt,rotate around={90:(13,-0.3)}] (13,-0.3) ellipse (0.4 and 0.1);
\end{scope}
\end{tikzpicture}
}
\caption{The correspondence between the odd 8-vertex model weights and the dimer weights.\label{fig:weights}}
\end{center}
\end{figure}
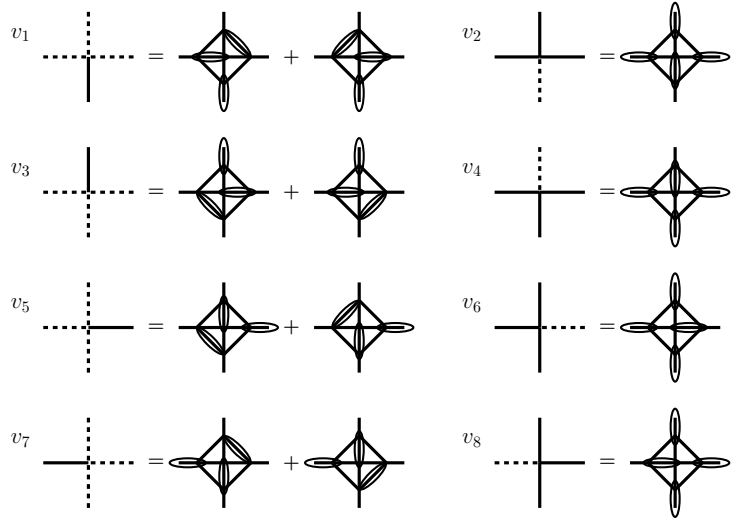

\begin{figure}[htpb]
\begin{center}
\begin{subfigure}{0.45\textwidth}
\centering
\scalebox{0.85}{
\begin{tikzpicture}
\draw[line width = 1pt] (-2.25,0) -- (2.25,0);
\draw[line width = 1pt] (0,-2.25) -- (0,2.25); 
\draw[line width = 1pt] (0,-1) -- (1,0) -- (0,1) -- (-1,0) -- (0,-1);
\node[above=4.5pt, right=-2pt] at (0,1) {\large\textbf{U}};
\node[below=5.5pt, right=-2pt] at (0,-1) {\large\textbf{D}};
\node[above=5.5pt, left=-2pt] at (-1,0) {\large\textbf{L}};
\node[above=5.5pt, right=-2.5pt] at (1,0) {\large\textbf{R}};
\node[above=5.5pt, right=-2.5pt] at (0,0) {\large\textbf{C}};
\node[above=3.5pt, left=-3.5pt] at (-0.5,0.5) {$z_1$};
\node[below=3.5pt, right=-3.5pt] at (0.5,-0.5) {$z_2$};
\node[above=3.5pt, right=-3.5pt] at (0.5,0.5) {$z_3$};
\node[below=3.5pt, left=-3.5pt] at (-0.5,-0.5) {$z_4$};
\node[below=3.5pt, left=-3.5pt] at (-0.5,-0.5) {$z_4$};
\node[below=1pt,left=-3.5pt] at (0,0.5) {$z_5$};
\node[above=1pt,left=-3pt] at (0,-0.5) {$z_6$};
\node[above=-2.5pt] at (-0.4,0) {$z_7$};
\node[below=-1.5pt] at (0.4,0) {$z_8$};
\node[above=4.5pt, right=-2pt] at (0,1.6) {1};
\node[below=5.5pt, right=-2pt] at (0,-1.6) {1};
\node[above=5.5pt, left=-2pt] at (-1.6,0) {1};
\node[above=5.5pt, right=-2.5pt] at (1.6,0) {1};
\end{tikzpicture}
}
\subcaption{Site and bond weight definitions around a cluster for the odd 8-vertex model.\label{fig:bonddefs1}}
\end{subfigure}
\begin{subfigure}{0.45\textwidth}
\centering
\scalebox{.55}{
\includegraphics{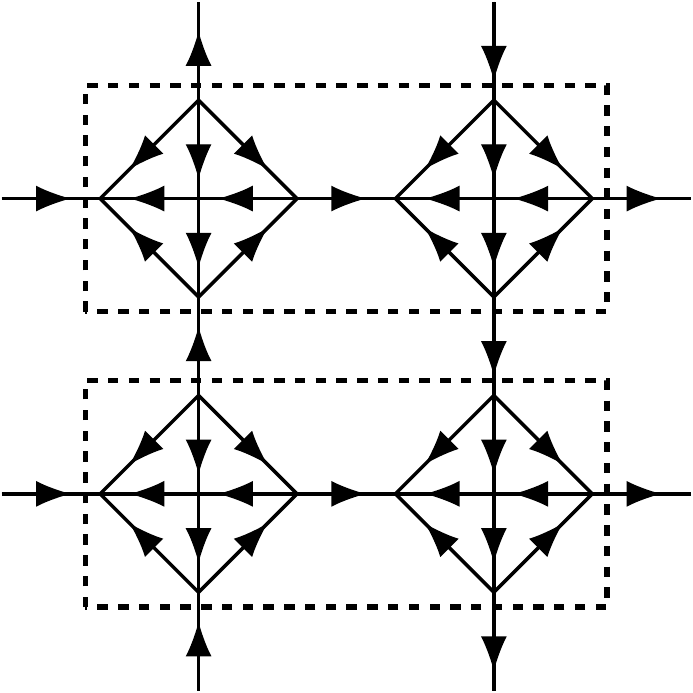}
}
\subcaption{Orientation graph convention on the column staggered lattice for the odd 8-vertex model.\label{fig:orientgraph1}}
\end{subfigure}
\caption{\label{fig:bondorient1}}
\end{center}
\end{figure}

We consider a square lattice with toroidal boundary conditions, of size $M\times N$. Using the dimer construction of Fan and Wu~\cite{fan1970w}, we will eliminate the central bond ``$w_2$" in their notation in order for the dimer coverings to match the odd 8-vertex weights, as shown in figure~\ref{fig:weights}. Each of the weights can be written in terms of the lattice bond weights $z_i$, defined in figure~\ref{fig:bonddefs1}. 

\begin{figure}[htpb]
\begin{center}
\scalebox{.55}{
\includegraphics{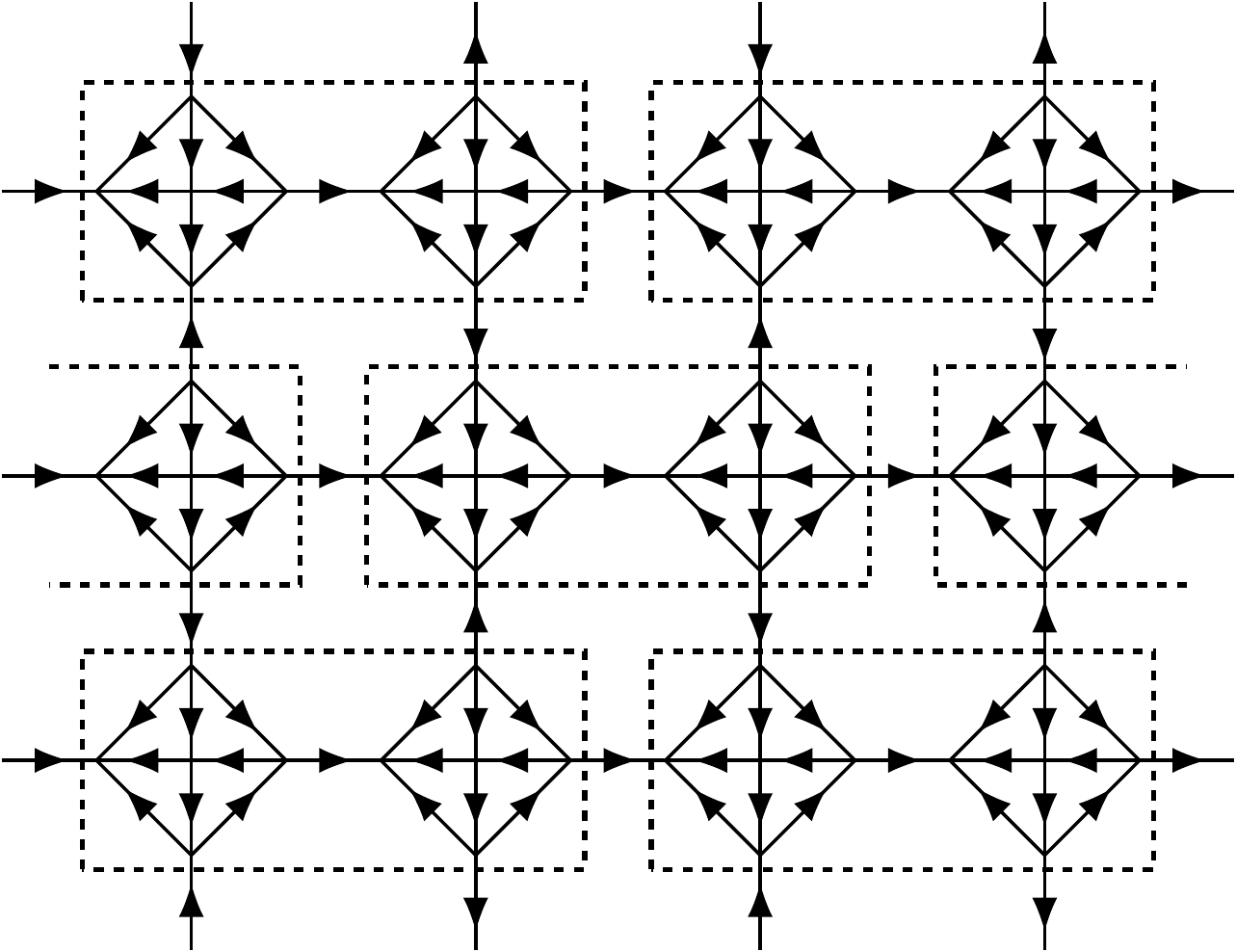}
}
\caption{Orientation graph convention on the bi-partite staggered lattice for the odd 8-vertex model.\label{fig:bondorient2}}
\end{center}
\end{figure}

From figures~\ref{fig:weights} and~\ref{fig:bonddefs1}, the odd 8-vertex weights have the following expressions in terms of bond weights
\bat{4}
v_1 &= z_1z_8+z_3z_7,\quad & v_2 &= z_6,\quad & v_3  &= z_2z_7+z_4z_8,\quad & v_4 &= z_5, \\
v_5 &= z_1z_6+z_4z_5,\quad & v_6 &= z_8,\quad & v_7  &= z_2z_5+z_3z_6,\quad & v_8 &= z_7      
\eat
It can be seen from these relations that the model weights $v_i$ follow the free-fermion condition
\beq
v_1v_2+v_3v_4 = v_5v_6 + v_7v_8 
\eeq
One of the weights $z_i$ is superfluous and can be made arbitrary. We here take $z_2=1$.

We can now solve for the bond weights $z_i$ in terms of the vertex weights $v_i$,
\beqr
z_1 &=& \frac{v_4v_8 +v_5v_6 -v_3v_4}{v_2v_6} = \frac{v_1v_2+v_4v_8-v_7v_8}{v_2v_6}\\
z_2 &=& 1,\qquad z_3 = \frac{v_7-v_4}{v_2},\qquad z_4 = \frac{v_3-v_8}{v_6} \\
z_5 &=& v_4,\qquad z_6 = v_2,\qquad z_7 = v_8,\qquad z_8 = v_6 
\eeqr

Because there are an odd number of sites in each cluster surrounding the vertex of the original lattice, the basic unit must consist of two clusters in order to allow a full dimer covering of the sites. As a result, ensuring that every transition cycle has a negative sign requires an alternation of bond orientations of one bond across rows or columns, or else across sub-lattices. As opposed to the even 8-vertex model, the odd 8-vertex model is naturally a staggered model when solved as a dimer model. We choose the convention shown in the orientation graph of figure~\ref{fig:bondorient1} for the column staggered model, and the convention shown in the orientation graph of figure~\ref{fig:bondorient2} for the bi-partite staggered model. For the staggered odd 8-vertex free-fermion models, we define new bond weights on the second cluster, denoted as $\bar{z}_i$, as well as corresponding vertex weights $\bar{v}_i$. Again we have the free-fermion constraint
\beq
\bar{v}_1\bar{v}_2 + \bar{v}_3\bar{v}_4 = \bar{v}_5\bar{v}_6+\bar{v}_7\bar{v}_8 \label{ffcond2}
\eeq
on the staggered weights.

We consider first the column staggered case. Its partition function will be given by a Pfaffian whose square is given by the determinant of the following matrix
\beq
M_{8OC} = T\otimes I_N \otimes I_M + A_1\otimes H_N^\mathrm{T}\otimes I_M + A_2\otimes H_N\otimes I_M + B_1\otimes I_N\otimes H_M^\mathrm{T}+ B_2\otimes I_N\otimes H_M
\eeq
where the $I_n$ are $n\times n$ identity matrices, where $T$ is defined as
\beq
T = \bordermatrix{
~ & U_1 & D_1 & L_1 & R_1 & C_1 & U_2 & D_2 & L_2 & R_2 & C_2 \cr
U_1 & 0 & 0 & z_1 & z_3 & z_5 & 0 & 0 & 0 & 0 & 0 \cr
D_1 & 0 & 0 & z_4 & z_2 & -z_6 & 0 & 0 & 0 & 0 & 0 \cr
L_1 & -z_1 & -z_4 & 0 & 0 & -z_7 & 0 & 0 & 0 & 0 & 0 \cr
R_1 & -z_3 & -z_2 & 0 & 0 & z_8 & 0 & 0 & 1 & 0 & 0 \cr
C_1 & -z_5 & z_6 & z_7 & -z_8 & 0 & 0 & 0 & 0 & 0 & 0 \cr
U_2 & 0 & 0 & 0 & 0 & 0 & 0 & 0 & \bar{z}_1 & \bar{z}_3 & \bar{z}_5 \cr
D_2 & 0 & 0 & 0 & 0 & 0 & 0 & 0 & \bar{z}_4 & \bar{z}_2 & -\bar{z}_6 \cr
L_2 & 0 & 0 & 0 & -1 & 0 & -\bar{z}_1 & -\bar{z}_4 & 0 & 0 & -\bar{z}_7 \cr
R_2 & 0 & 0 & 0 & 0 & 0 & -\bar{z}_3 & -\bar{z}_2 & 0 & 0 & \bar{z}_8 \cr
C_2 & 0 & 0 & 0 & 0 & 0 & -\bar{z}_5 & \bar{z}_6 & \bar{z}_7 & -\bar{z}_8 & 0 \cr
}
\eeq
with the left cluster having index 1 and the right cluster index 2, where the $n\times n$ matrix $H_n$ is defined as
\beq
H_n =
 \begin{pmatrix}
  0 & 1 & 0 & \cdots & 0 \\
  0 & 0 & 1 & \cdots & 0 \\
  \vdots  & \vdots & \vdots  & \ddots & \vdots  \\
  0 & 0 & 0 & \cdots & 1 \\
  1 & 0 & 0 & \cdots & 0
 \end{pmatrix}
\eeq
and where the non-zero elements of the $10\times10$ matrices $A_1,A_2,B_1,B_2$ are given by 
\beqr
(A_1)_{L_1,R_2} &=& -(A_2)_{R_2,L_1} = -1 \\
(B_1)_{U_1,D_1} &=& -(B_1)_{U_2,D_2} = 1 \\
(B_2)_{D_1,U_1} &=& -(B_2)_{D_2,U_2} = -1 
\eeqr

The determinant of $M_{8OC}$ is then given by 
\beqr
\mathrm{Det}(M_{8OC}) &=& \prod_{\theta_1}\prod_{\theta_2}  D(\theta_1,\theta_2) \\
 &=& \prod_{\theta_1}\prod_{\theta_2} 
\begin{vmatrix}
0 & e^{-i\theta_2} & z_1 & z_3 & z_5 & 0 & 0 & 0 & 0 & 0 \\
-e^{i\theta_2} & 0 & z_4 & z_2 & -z_6 & 0 & 0 & 0 & 0 & 0 \\
-z_1 & -z_4 & 0 & 0 & -z_7 & 0 & 0 & 0 & -e^{-i\theta_1} & 0 \\
-z_3 & -z_2 & 0 & 0 & z_8 & 0 & 0 & 1 & 0 & 0 \\
-z_5 & z_6 & z_7 & -z_8 & 0 & 0 & 0 & 0 & 0 & 0 \\
0 & 0 & 0 & 0 & 0 & 0 & -e^{-i\theta_2} & \bar{z}_1 & \bar{z}_3 & \bar{z}_5 \\
0 & 0 & 0 & 0 & 0 & e^{i\theta_2} & 0 & \bar{z}_4 & \bar{z}_2 & -\bar{z}_6 \\
0 & 0 & 0 & -1 & 0 & -\bar{z}_1 & -\bar{z}_4 & 0 & 0 & -\bar{z}_7 \\
0 & 0 & e^{i\theta_1} & 0 & 0 & -\bar{z}_3 & -\bar{z}_2 & 0 & 0 & \bar{z}_8 \\
0 & 0 & 0 & 0 & 0 & -\bar{z}_5 & \bar{z}_6 & \bar{z}_7 & -\bar{z}_8 & 0 \\
 \end{vmatrix} 
\eeqr
where
\beq
\theta_1 = \frac{2\pi n}{N},\qquad\theta_2 = \frac{2\pi m}{M}
\eeq
and where $n=1,\ldots,N$ and $m=1,\ldots,M$.

The partition function $Z_{8OC}$ is the square root of the determinant of $M_{8OC}$\footnote{On the finite lattice, for toroidal boundary conditions, four Pfaffians are actually needed, but they become degenerate in the thermodynamic limit~\cite{mccoy1973w,mccoy2014w}.}. In the thermodynamic limit, the free energy $f_{8OC}$ is given by 
\beq
f_{8OC} = -\frac{1}{2\beta}\lim_{M\to\infty}\lim_{N\to\infty}\frac{1}{MN}\ln Z_{8OC} = -\frac{1}{2\beta}\lim_{M\to\infty}\lim_{N\to\infty}\frac{1}{MN}\ln \mathrm{Det}(M_{8OC})^{1/2}
\eeq
where the leading factor of 2 is due to the fact that there are two clusters in each unit. 

The logarithm of the products in the determinant can be expanded and written as integrals in the thermodynamic limit, giving the following free energy expression
\beq
-\beta f_{8OC} =  \frac{1}{16\pi^2}\int_0^{2\pi}\int_0^{2\pi} \ln\big[D(\theta_1,\theta_2)\big] d\theta_1d\theta_2
\eeq

The bi-partite staggered lattice can be solved analogously, for which we have the following altered expressions
\beq
M_{8OB} = \hspace{6in}\vspace{-0.1in}\nonumber
\eeq
\beq
T\otimes I_N \otimes I_M + A_1\otimes H_N^\mathrm{T}\otimes I_M + A_2\otimes H_N\otimes I_M + B_1\otimes I_N\otimes H_M^\mathrm{T}+ B_2\otimes I_N\otimes H_M + C_1\otimes H_N^\mathrm{T}\otimes H_M + C_2\otimes H_N\otimes H_M^\mathrm{T}
\eeq
with
\beqr
(A_1)_{L_1,R_2} &=& -(A_2)_{R_2,L_1} = -1 \\
(B_1)_{U_2,D_1} &=& -(B_2)_{D_1,U_2} = 1 \\
(C_1)_{U_1,D_2} &=& -(C_2)_{D_2,U_1} = -1 
\eeqr
and now
\beq
D(\theta_1,\theta_2) = \begin{vmatrix}
0 & 0 & z_1 & z_3 & z_5 & 0 & -e^{i(\theta_2-\theta_1)} & 0 & 0 & 0 \\
0 & 0 & z_4 & z_2 & -z_6 & -e^{-i\theta_2} & 0 & 0 & 0 & 0 \\
-z_1 & -z_4 & 0 & 0 & -z_7 & 0 & 0 & 0 & -e^{-i\theta_1} & 0 \\
-z_3 & -z_2 & 0 & 0 & z_8 & 0 & 0 & 1 & 0 & 0 \\
-z_5 & z_6 & z_7 & -z_8 & 0 & 0 & 0 & 0 & 0 & 0 \\
0 & e^{i\theta_2} & 0 & 0 & 0 & 0 & 0 & \bar{z}_1 & \bar{z}_3 & \bar{z}_5 \\
e^{i(\theta_1-\theta_1)} & 0 & 0 & 0 & 0 & 0 & 0 & \bar{z}_4 & \bar{z}_2 & -\bar{z}_6 \\
0 & 0 & 0 & -1 & 0 & -\bar{z}_1 & -\bar{z}_4 & 0 & 0 & -\bar{z}_7 \\
0 & 0 & e^{i\theta_1} & 0 & 0 & -\bar{z}_3 & -\bar{z}_2 & 0 & 0 & \bar{z}_8 \\
0 & 0 & 0 & 0 & 0 & -\bar{z}_5 & \bar{z}_6 & \bar{z}_7 & -\bar{z}_8 & 0 \\
 \end{vmatrix} 
\eeq

\subsection{The even free-fermion 8-vertex models}
\begin{figure}[htpb]
\begin{center}
\scalebox{0.5}{
\includegraphics{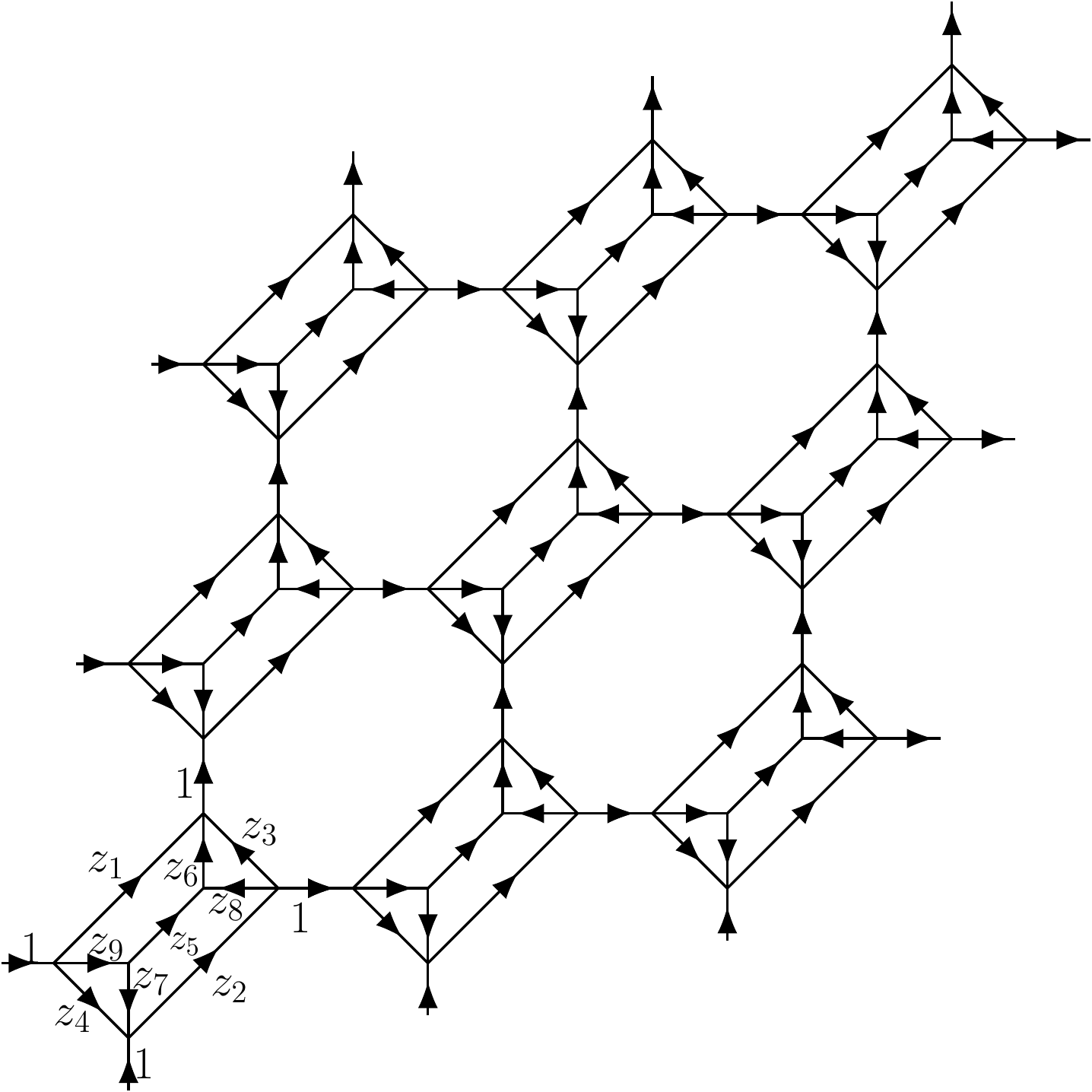}
}
\end{center}
\caption{Cluster definition, and bond orientation convention for the even 8-vertex models.\label{fig:evenstaggered}}
\end{figure}

For the even free-fermion 8-vertex model, we follow a similar procedure as for the odd 8-vertex model. In the homogeneous case it is not necessary to stagger the lattice in order to solve it, as opposed to the odd 8-vertex model. In figure~\ref{fig:evenstaggered}, we show the dimer cluster we use and the bond orientation convention.

From dimer coverings, we find the following correspondence between the even 8-vertex weights $w_i$ and the internal bonds $z_i$, as well as the inverse mapping below between $z_i$ and $w_i$ after choosing to eliminate the extra degrees of freedom in the $z_i$ by choosing $z_7=z_9=1$
\bat{3}
w_1 &= z_1z_7z_8 + z_2z_6z_9 + z_5(z_1z_2+z_3z_4) \quad\quad&  &  \quad\quad& z_1 &= \frac{w_5-w_4}{w_2}  \label{bondvertex1}\\
w_2 &= z_5 \quad\quad&  &  \quad\quad& z_2 &= \frac{w_6-w_3}{w_2}  \\
w_3 &= z_8z_9 \quad\quad&  &  \quad\quad& z_3 &=  \frac{w_7}{w_2} \\
w_4 &= z_6z_7 \quad\quad&  \Rightarrow&  \quad\quad& z_4 &=  \frac{w_8}{w_2} \\
w_5 &= z_1z_5+z_6z_9 \quad\quad&  &  \quad\quad& z_5 &=  w_2 \\
w_6 &= z_2z_5+z_7z_8 \quad\quad&  &  \quad\quad& z_6 &=  w_4 \\
w_7 &= z_3z_5 \quad\quad&  &  \quad\quad& z_7 &=  1 \\
w_8 &= z_4z_5 \quad\quad&  &  \quad\quad& z_8 &=  w_3 \\
&  \quad\quad&  &  \quad\quad& z_9 &=  1 \label{bondvertex2}
\eat

Again, the relations~(\ref{bondvertex1})--(\ref{bondvertex2}) enforce the free-fermion condition 
\beq
w_2w_1 + w_4w_3 = w_6w_5+w_7w_8
\eeq
as was the case for the odd weights. For the staggered lattices, staggered bond weights $\bar{z}_i$ and staggered vertex weights $\bar{w}_i$ are defined in the same manner, which also obey an analogous free-fermion condition. 

We define a matrix $T$ as above, valid both for the column and bi-partite staggered even 8-vertex models
\beq
T = \bordermatrix{
~ & U_1 & D_1 & L_1 & R_1 & C_1 & E_1 & U_2 & D_2 & L_2 & R_2 & C_2 & E_2 \cr
U_1 & 0 & 0 & -z_1 & -z_3 & -z_6 & 0 & 0 & 0 & 0 & 0 & 0 & 0 \cr
D_1 & 0 & 0 & -z_4 & z_2 & 0 & -z_7 & 0 & 0 & 0 & 0 & 0 & 0 \cr
L_1 & z_1 & z_4 & 0 & 0 & 0 & z_9 & 0 & 0 & 0 & 0 & 0 & 0 \cr
R_1 & z_3 & -z_2 & 0 & 0 & z_8 & 0 & 0 & 0 & 1 & 0 & 0 & 0 \cr
C_1 & z_6 & 0 & 0 & -z_8 & 0 & -z_5 & 0 & 0 & 0 & 0 & 0 & 0 \cr
E_1 & 0 & z_7 & -z_9 & 0 & z_5 & 0 & 0 & 0 & 0 & 0 & 0 & 0 \cr
U_2 & 0 & 0 & 0 & 0 & 0 & 0   & 0 & 0 & -\bar{z}_1 & -\bar{z}_3 & -\bar{z}_6 & 0 \cr
D_2 & 0 & 0 & 0 & 0 & 0 & 0   & 0 & 0 & -\bar{z}_4 & \bar{z}_2 & 0 & -\bar{z}_7 \cr
L_2 & 0 & 0 & 0 & -1 & 0 & 0   & \bar{z}_1 & \bar{z}_4 & 0 & 0 & 0 & \bar{z}_9 \cr
R_2 & 0 & 0 & 0 & 0 & 0 & 0   & \bar{z}_3 & -\bar{z}_2 & 0 & 0 & \bar{z}_8 & 0 \cr
C_2 & 0 & 0 & 0 & 0 & 0 & 0   & \bar{z}_6 & 0 & 0 & -\bar{z}_8 & 0 & -\bar{z}_5 \cr
E_2 & 0 & 0 & 0 & 0 & 0 & 0   & 0 & \bar{z}_7 & -\bar{z}_9 & 0 & \bar{z}_5 & 0 \cr
}
\eeq
with the left cluster having index 1 and the right cluster index 2. The free-energies of the homogeneous, column staggered, and bi-partite staggered can now be found in a straight forward manner by modifying the expressions for the odd 8-vertex model in the previous subsection.

\section{Free-fermion 8-vertex model solutions}\label{app:results}

\subsection{Odd 8-vertex model, bi-partite staggered}
The bi-partite staggered odd free-fermion 8-vertex model free energy is given below
\beqr
-\beta f_{8OB} &=&  \frac{1}{16\pi^2}\int_0^{2\pi}\int_0^{2\pi} \ln\big[A+2B\cos(\theta_1) +2C\cos(\theta_2)+ 2D\cos(\theta_1-\theta_2)   \nonumber\\
&&\qquad\qquad\qquad\qquad + 2E\cos(\theta_1+\theta_2) + 2F\cos(2\theta_1)+ 2G\cos(2\theta_2) \big] d\theta_1d\theta_2\quad
\eeqr
where
\beqr
A &=& \bar{v}_8^2v_6^2+\bar{v}_7^2v_5^2+\bar{v}_5^2v_7^2+\bar{v}_6^2v_8^2+\bar{v}_3^2v_1^2 +\bar{v}_4^2v_2^2 +\bar{v}_1^2v_3^2+\bar{v}_2^2v_4^2+2(v_8v_7+v_5v_6)(\bar{v}_8\bar{v}_7 +\bar{v}_5\bar{v}_6) \\
B &=& (v_1\bar{v}_3+v_2\bar{v}_4)(v_7\bar{v}_5+v_8\bar{v}_6)-(v_3\bar{v}_1+v_4\bar{v}_2)(v_5\bar{v}_7+v_6\bar{v}_8) \\
C &=& (v_3\bar{v}_1+v_4\bar{v}_2)(v_7\bar{v}_5+v_8\bar{v}_6)-(v_1\bar{v}_3+v_2\bar{v}_4)(v_5\bar{v}_7+v_6\bar{v}_8) \\
D &=& v_1v_4\bar{v}_2\bar{v}_3+v_2v_3\bar{v}_1\bar{v}_4-v_6v_8\bar{v}_6\bar{v}_8-v_5v_7\bar{v}_5\bar{v}_7 \\
E &=& v_1v_3\bar{v}_1\bar{v}_3+v_2v_4\bar{v}_2\bar{v}_4-v_6v_7\bar{v}_5\bar{v}_8-v_5v_8\bar{v}_6\bar{v}_7 \\
F &=& -(v_1v_2-v_5v_6)(\bar{v}_1\bar{v}_2-\bar{v}_5\bar{v}_6) \\
G &=& -(v_1v_2-v_7v_8)(\bar{v}_1\bar{v}_2-\bar{v}_7\bar{v}_8) 
\eeqr

Specializing to homogeneous variables and using the transformation $\theta_1\to(-\theta_1+\theta_2)/2$, $\theta_2\to(\theta_1+\theta_2)/2$, we have the following expression
\beq
-\beta f_{8O} 	=  \frac{1}{16\pi^2}\int_0^{2\pi}\int_0^{2\pi} \ln\big[2A + 2D\cos(\theta_1) + 2E\cos(\theta_2)  + 2F\cos(\theta_2-\theta_1) + 2G\cos(\theta_2+\theta_1) \big] d\theta_1d\theta_2\quad
\eeq
where
\beqr
A &=& (v_1v_2+v_3v_4)(v_5v_6+v_7v_8)+(v_1^2v_3^2+v_2^2v_4^2) +(v_5^2v_7^2+v_6^2v_8^2) \\
D &=& 2v_1v_2v_3v_4-(v_5^2v_7^2+v_6^2v_8^2) \\
E &=& (v_1^2v_3^2+v_2^2v_4^2)-2v_5v_6v_7v_8 \\
F &=& (v_1v_2-v_5v_6)(v_3v_4-v_7v_8) \\
G &=& (v_1v_2-v_7v_8)(v_3v_4-v_5v_6)
\eeqr

\subsection{Odd 8-vertex, column staggered}
The column staggered odd 8-vertex free-fermion model's free energy is given by
\beqr
-\beta f_{8OC} &=&  \frac{1}{16\pi^2}\int_0^{2\pi}\int_0^{2\pi} \ln\big[A + 2B\cos(\theta_1) +2C\cos(\theta_2) +2D\cos(\theta_1-\theta_2) +2E\cos(\theta_1 + \theta_2)\nonumber\\
&&\qquad\qquad\qquad\qquad + 2G\cos(2\theta_2)  + 2H\cos(\theta_1-2\theta_2) + 2I\cos(\theta_1+2\theta_2) \big] d\theta_1d\theta_2\quad
\eeqr
where
\beqr
A &=& (v_5^2+v_8^2)(\bar{v}_6^2+\bar{v}_7^2)+(v_6^2+v_7^2)(\bar{v}_5^2 +\bar{v}_8^2)+2v_1v_3\bar{v}_1\bar{v}_3+2v_2v_4\bar{v}_2\bar{v}_4+2v_5v_8\bar{v}_6\bar{v}_7+2v_6v_7\bar{v}_5\bar{v}_8 \nonumber\\&&\\
B &=& v_1v_2\bar{v}_3\bar{v}_4+v_3v_4\bar{v}_1\bar{v}_2-v_5v_6\bar{v}_7\bar{v}_8-v_7v_8\bar{v}_5\bar{v}_6-(v_5v_7-v_6v_8)(\bar{v}_5\bar{v}_7-\bar{v}_6\bar{v}_8) \\
C &=& \bar{v}_5\bar{v}_8(v_6^2+v_7^2)-\bar{v}_6\bar{v}_7(v_5^2+v_8^2)-v_5v_8(\bar{v}_6^2+\bar{v}_7^2)+v_6v_7(\bar{v}_5^2+\bar{v}_8^2) \\
D&=& (v_3v_4-v_5v_6)(\bar{v}_5\bar{v}_7-\bar{v}_6\bar{v}_8)-(v_5v_7-v_6v_8)(\bar{v}_3\bar{v}_4-\bar{v}_5\bar{v}_6) \\
E &=& (v_1v_2-v_5v_6)(\bar{v}_5\bar{v}_7-\bar{v}_6\bar{v}_8)-(v_5v_7-v_6v_8)(\bar{v}_1\bar{v}_2-\bar{v}_5\bar{v}_6) \\
G &=& v_8v_5\bar{v}_7\bar{v}_6+v_7v_6\bar{v}_8\bar{v}_5-v_3v_1\bar{v}_3\bar{v}_1-v_4v_2\bar{v}_4\bar{v}_2 \\
H &=& (v_1v_2-v_7v_8)(\bar{v}_1\bar{v}_2-\bar{v}_7\bar{v}_8) \\
I &=& (v_1v_2-v_5v_6)(\bar{v}_1\bar{v}_2-\bar{v}_5\bar{v}_6) 
\eeqr

Specializing to homogeneous variables and using the transformation $2\theta_2\to\theta_2$ we have the following expression
\beq
-\beta f_{8O} =  \frac{1}{16\pi^2}\int_0^{2\pi}\int_0^{2\pi} \ln\big[A + 2B\cos(\theta_1) + 2D\cos(\theta_2)+ 2H\cos(\theta_1-\theta_2) + 2I\cos(\theta_1+\theta_2) \big] d\theta_1d\theta_2
\eeq
where
\beqr
A &=& 2(v_5^2+v_8^2)(v_6^2+v_7^2)+2v_1^2v_3^2+2v_2^2v_4^2+4v_5v_6v_7v_8 \\
B &=& 2v_1v_2v_3v_4-v_5^2v_7^2-v_6^2v_8^2 \\
G &=& 2v_5v_6v_7v_8-v_1^2v_3^2-v_2^2v_4^2 \\
H &=& (v_1v_2-v_7v_8)^2 \\
I &=& (v_1v_2-v_5v_6)^2 
\eeqr

\subsection{Even 8-vertex model, bi-partite staggered}
The bi-partite staggered even 8-vertex free-fermion model free-energy is given by
\beqr
-\beta f_{8EB} &=&  \frac{1}{16\pi^2}\int_0^{2\pi}\int_0^{2\pi} \ln\big[A +2B\cos(\theta_1)+2C\cos(\theta_2)+ 2D\cos(\theta_1-\theta_2)  \nonumber\\
&&\qquad\qquad\qquad\qquad + 2E\cos(\theta_1+\theta_2) + 2F\cos(2\theta_1)+ 2G\cos(2\theta_2) \big] d\theta_1d\theta_2\quad \label{evenbstagint}
\eeqr
where
\beqr
A &=& w_1^2\bar{w}_1^2+w_2^2\bar{w}_2^2+w_3^2\bar{w}_3^2 +w_4^2\bar{w}_4^2+ w_5^2\bar{w}_6^2+w_6^2\bar{w}_5^2 +w_7^2\bar{w}_8^2+w_8^2\bar{w}_7^2 +2(w_5w_6+w_7w_8)(\bar{w}_5\bar{w}_6+\bar{w}_7\bar{w}_8) \nonumber\\&&\\
B &=& -(w_1\bar{w}_1+w_2\bar{w}_2)(w_5\bar{w}_6+w_6\bar{w}_5)+(w_3\bar{w}_3+w_4\bar{w}_4)(w_7\bar{w}_8+w_8\bar{w}_7) \\
C &=& -(w_1\bar{w}_1+w_2\bar{w}_2)(w_7\bar{w}_8+w_8\bar{w}_7)+(w_3\bar{w}_3+w_4\bar{w}_4)(w_5\bar{w}_6+w_6\bar{w}_5) \\
D &=& -w_1w_4\bar{w}_1\bar{w}_4-w_2w_3\bar{w}_2\bar{w}_3+w_5w_7\bar{w}_6\bar{w}_8+w_6w_8\bar{w}_5\bar{w}_7 \\
E &=& -w_1w_3\bar{w}_1\bar{w}_3-w_2w_4\bar{w}_2\bar{w}_4+w_5w_8\bar{w}_6\bar{w}_7+w_6w_7\bar{w}_5\bar{w}_8 \\
F &=& (w_3w_4-w_5w_6)(\bar{w}_3\bar{w}_4-\bar{w}_5\bar{w}_6) \\
G &=& (w_3w_4-w_7w_8)(\bar{w}_3\bar{w}_4-\bar{w}_7\bar{w}_8)  
\eeqr

Specializing to homogeneous variables, we also have the following free-energy expression
\beqr
-\beta f_{E} &=&  \frac{1}{16\pi^2}\int_0^{2\pi}\int_0^{2\pi} \ln\big[A +2B\cos(\theta_1) +2C\cos(\theta_2) + 2D\cos(\theta_1-\theta_2) \nonumber\\
&&\qquad\qquad\qquad\qquad + 2E\cos(\theta_1+\theta_2) + 2F\cos(2\theta_1)+ 2G\cos(2\theta_2) \big] d\theta_1d\theta_2\quad
\eeqr
where
\beqr
A &=& w_1^4+w_2^4+w_3^4+w_4^4+4w_5^2w_6^2+4w_5w_6w_7w_8+4w_7^2w_8^2 \\
B &=& 2w_7w_8(w_3^2+w_4^2)-2w_5w_6(w_1^2+w_2^2) \\
C &=& 2w_5w_6(w_3^2+w_4^2)-2w_7w_8(w_1^2+w_2^2) \\
D &=& 2w_5w_6w_7w_8-w_1^2w_4^2-w_2^2w_3^2 \\
E &=& 2w_5w_6w_7w_8-w_1^2w_3^2-w_2^2w_4^2 \\
F &=& (w_3w_4-w_5w_6)^2 \\
G &=& (w_3w_4-w_7w_8)^2 
\eeqr

\subsection{Even 8-vertex model, column staggered}
The column staggered even 8-vertex model free-fermion free-energy is given by
\beqr
-\beta f_{8EC} &=&  \frac{1}{16\pi^2}\int_0^{2\pi}\int_0^{2\pi} \ln\big[A + 2B\cos(\theta_1) +2C\cos(\theta_2) +2D\cos(\theta_1 - \theta_2) +2E\cos(\theta_1+\theta_2) \nonumber\\
&&\qquad\qquad\qquad\qquad + 2G\cos(2\theta_2) + 2H\cos(\theta_1-2\theta_2) + 2I\cos(\theta_1+2\theta_2) \big] d\theta_1d\theta_2\quad \label{evencstagint}
\eeqr
where
\beqr
A &=& (w_1^2+w_3^2)(\bar{w}_1^2+\bar{w}_3^2)+(w_2^2+w_4^2)(\bar{w}_2^2+\bar{w}_4^2) +2(w_1w_3\bar{w}_1\bar{w}_3+w_2\bar{w}_2w_4\bar{w}_4+w_5w_8\bar{w}_6\bar{w}_7+w_6w_7\bar{w}_5\bar{w}_8) \nonumber\\&&\\
B &=& -(w_1w_4-w_2w_3)(\bar{w}_1\bar{w}_4-\bar{w}_2\bar{w}_3)+w_1w_2\bar{w}_3\bar{w}_4+w_3w_4\bar{w}_1\bar{w}_2-w_5w_6\bar{w}_7\bar{w}_8-w_7w_8\bar{w}_6\bar{w}_5 \\
C &=& -w_1w_3(\bar{w}_1^2+\bar{w}_3^2)-\bar{w}_1\bar{w}_3(w_1^2+w_3^2)+w_2w_4(\bar{w}_2^2+\bar{w}_4^2)+\bar{w}_2\bar{w}_4(w_2^2+w_4^2) \\
D &=& (w_1w_4-w_2w_3)(\bar{w}_3\bar{w}_4-\bar{w}_5\bar{w}_6)+(\bar{w}_1\bar{w}_4-\bar{w}_2\bar{w}_3)(w_3w_4-w_5w_6) \\
E &=& (w_1w_4-w_2w_3)(\bar{w}_3\bar{w}_4-\bar{w}_7\bar{w}_8)+(\bar{w}_1\bar{w}_4-\bar{w}_2\bar{w}_3)(w_3w_4-w_7w_8) \\
G &=& w_1w_3\bar{w}_1\bar{w}_3+w_2w_4\bar{w}_2\bar{w}_4-w_5w_8\bar{w}_6\bar{w}_7-w_6w_7\bar{w}_5\bar{w}_8 \\
H &=& -(w_3w_4-w_5w_6)(\bar{w}_3\bar{w}_4-\bar{w}_5\bar{w}_6)\\
I &=& -(w_3w_4-w_7w_8)(\bar{w}_3\bar{w}_4-\bar{w}_7\bar{w}_8) 
\eeqr

Specializing to homogeneous variables, we have the following new free-energy expression
\beqr
-\beta f_{8E} &=&  \frac{1}{16\pi^2}\int_0^{2\pi}\int_0^{2\pi} \ln\big[A + 2B\cos(\theta_1) +2C\cos(\theta_2) +2D\cos(\theta_1 - \theta_2) +2E\cos(\theta_1+\theta_2) \nonumber\\
&&\qquad\qquad\qquad\qquad + 2G\cos(2\theta_2) + 2H\cos(\theta_1-2\theta_2) + 2I\cos(\theta_1+2\theta_2) \big] d\theta_1d\theta_2\quad 
\eeqr
where
\beqr
A &=& (w_1^2+w_3^2)^2+2w_1^2w_3^2+(w_2^2+w_4^2)^2+2w_2^2w_4^2+4w_5w_6w_7w_8 \\
B &=& -(w_1w_4-w_2w_3)^2+2w_1w_2w_3w_4-2w_5w_6w_7w_8 \\
C &=& -2w_1w_3(w_1^2+w_3^2)+2w_2w_4(w_2^2+w_4^2) \\
D &=& 2(w_1w_4-w_2w_3)(w_3w_4-w_5w_6) \\
E &=& 2(w_1w_4-w_2w_3)(w_3w_4-w_7w_8) \\
G &=& w_1^2w_3^2+w_2^2w_4^2-2w_5w_6w_7w_8 \\
H &=& -(w_3w_4-w_5w_6)^2 \\
I &=& -(w_3w_4-w_7w_8)^2 
\eeqr

\end{document}